\PassOptionsToPackage{prologue,dvipsnames,x11names,table}{xcolor}

\documentclass[acmsmall]{acmart}

\acmJournal{TORS}
\acmYear{2026}
\acmVolume{1}
\acmNumber{1}
\acmMonth{7}

\usepackage{xcolor}

\definecolor{DarkGreen}{RGB}{0,100,0}

\definecolor{AgentBlue}{HTML}{EAF2FF}
\definecolor{AgentBlueDark}{HTML}{2F5DAA}
\definecolor{AgentGray}{HTML}{F6F7F9}
\definecolor{AgentGreen}{HTML}{EAF7EF}
\definecolor{AgentOrange}{HTML}{FFF4E5}

\newcolumntype{Y}{>{\raggedright\arraybackslash}X}

\newcommand{\rev}[1]{#1}

\usepackage{booktabs}
\usepackage{tabularx}
\usepackage{array}
\usepackage{adjustbox}
\usepackage{makecell}
\usepackage{colortbl}
\usepackage{pifont}
\renewcommand{\rothead}[1]{%
  \raisebox{0.5ex}{\rotatebox[origin=c]{60}{\scriptsize\bfseries #1}}%
}

\newcolumntype{Y}{>{\raggedright\arraybackslash}X}
\newcolumntype{Z}{>{\raggedright\arraybackslash}X}

\newcolumntype{C}{>{\centering\arraybackslash}p{0.041\textwidth}}
\newcolumntype{G}{>{\centering\arraybackslash}p{0.055\textwidth}}

\newcommand{\full}{\textbf{\large$\bullet$}}      
\newcommand{\partly}{\textbf{\large$\triangle$}}  
\newcommand{\none}{\textbf{\large$\circ$}}        

\newcommand{\grouphead}[1]{%
    \rowcolor{black!10}
    \multicolumn{11}{@{}l}{\textbf{#1}}\\[-0.15em]
}

\usepackage{enumitem}

\usepackage{algorithm}
\usepackage{algpseudocode}

\usepackage{graphicx}
\usepackage{tikz}
\usetikzlibrary{arrows.meta,positioning,fit,shapes.geometric,calc}

\usepackage{fontawesome5}

\usepackage{csquotes}

\newcommand{\dq}[1]{\enquote{#1}}
\newcommand{\say}[1]{\enquote{#1}}

\usepackage{bbold}

\title{The Future is Agentic: Definitions, Perspectives, and Open Challenges of Multi-Agent Recommender Systems}
\author{Reza Yousefi Maragheh}
\affiliation{\institution{ University of Illinois Urbana Champaign}\city{Illinois}\country{USA}}
\email{ryousefimaragheh@acm.org}

\author{Yashar Deldjoo}
\affiliation{\institution{Polytechnic University of Bari}\city{Bari}\country{Italy}}
\email{deldjooy@acm.org}

\begin{document}

\begin{abstract}
Large language models (LLMs) are evolving from passive text generators into
agentic systems that can plan, maintain state, invoke tools, and coordinate with
other agents. This perspective paper examines what this shift means for
recommender systems (RS). We define \emph{agentic recommender systems} as
recommendation pipelines in which one or more stateful agents observe, plan, call
tools, and verify, rather than score in a single shot, while operating over users,
item catalogs, candidate sets, and recommendation objectives. Their value is
strongest when this added machinery measurably improves recommendation-layer
outcomes such as relevance, constraint satisfaction, bundle coherence, grounding,
explanation faithfulness, fairness of exposure, or user effort, and is not
justified merely because a pipeline contains an LLM or several modules. To make
these notions precise for recommendation rather than for agents in general, we
introduce a recommender-specific formalism that models an individual recommender
agent by its state (user, context, history, and candidate set), a reasoning core,
recommendation-specific tools, a hierarchical memory, and explicit policy
constraints, and captures a multi-agent recommender as a triple of agents, a
shared environment exposing the item catalog and feedback signals, and a
communication protocol. Within this framework, we develop four representative task
families (interactive goal-oriented recommendation, user simulation and
evaluation, contextual and multimodal recommendation, and grounded explanation)
and an operational agenda that ties five recurring challenge families
(communication protocols, scalability and cost, hallucination and error
propagation, emergent misalignment and collusion, and brand and policy compliance)
to measurable RS signals. Finally, we conduct a controlled empirical study
comparing single-shot and multi-agent pipelines under shared user histories,
candidate sets, prompts, and metrics. A pilot next-item ranking study on
Amazon-2023 categories shows that multi-agent systems are not uniformly superior:
on representative samples the single-shot baseline is Pareto-efficient, whereas
decomposition and ensemble agents become useful mainly for high-diversity user
histories. This supports a conditional design principle: agentic complexity should
be routed to the cases where its marginal quality improvement justifies the
additional latency, cost, and governance risk. To support reproduction, we release all pipeline
implementations, prompts, and results here: https://github.com/RezaYM/agenticrecsys.git.
\end{abstract}

\maketitle

\section{Introduction and Motivation}
\label{sec}

Large Language Model (LLM) agents go beyond traditional chatbots by offering goal-directed, agentic behavior rather than merely responding to user queries through one-shot text generation. In essence, they are designed to handle multi-step tasks, orchestrate information flow, and autonomously employ tools or functions when necessary~\cite{prasad2024adapt,maharana2024evaluating,yao2023react}. This distinction means that while a conventional chatbot might provide short answers in a single round of dialogue, an agentic system can proactively structure a complex problem and solve it through a sequence of methodical steps. Put another way, an LLM agent is not only a reactive conversational partner but a dynamic problem-solver capable of decomposing tasks, maintaining state, invoking external resources, and adapting its strategy to reach a goal~\cite{schick2023toolformer,shen2023hugginggpt,wang2024recmind}.

Much as goal-directed agentic behavior has reshaped other NLP-oriented tasks, it applies naturally to the tasks surrounding recommender systems (RS). In the most common case, recommendation is handled in a single shot: given a user history and a candidate set, a retrieval or ranking function estimates relevance and returns a ranked list of similar movies or items. Agentic recommendation broadens this picture along two axes. First, even a single-shot task can sometimes be handled better by decomposing it into subtasks, for example by summarizing a long or heterogeneous history, planning a ranking strategy, or aggregating several candidate orderings before committing to a final list. Second, and more fundamentally, the user need may not be \textit{\dq{rank items for this profile}} at all, but rather \textit{\dq{help me achieve a goal}} under multiple constraints, evolving context, and incomplete information. In this latter regime, recommendation becomes a multi-stage decision process rather than a single scoring operation: the system may ask clarifying questions, retrieve current catalog evidence, reason over constraints, call external tools, maintain memory over prior interactions, verify whether candidate items satisfy user and policy constraints, and finally produce a ranked list, bundle, explanation, or action plan. Section~\ref{sec:agentic-tasks} develops these possibilities through four representative scenarios: \textit{interactive goal-oriented recommendation}, \textit{user simulation for offline evaluation}, \textit{contextual and multimodal recommendation,} and \textit{recommendation explanation}.

We refer to this RS-centered paradigm as \emph{agentic recommendation}. In an
agentic recommender system, recommendation is no longer only a one-shot call to a
ranker; it is a goal-driven process in which one or more stateful agents
participate in an observe--plan--act--verify loop over user signals, item
evidence, tools, memories, constraints, and recommendation objectives. This
viewpoint gives us the flexibility, and the obligation, to evaluate recommendation
along a more diverse set of metrics than top-$K$ accuracy alone: relevance and
ranking quality, constraint satisfaction, diversity, bundle coherence, evidence
grounding, explanation faithfulness, fairness of exposure, long-term user value,
and user effort, together with trace-level signals such as tool-call success,
memory correctness, latency, and cost. The agentic layer is useful when it
improves these recommendation-relevant outcomes, not merely because a system
contains an LLM or several modules.

Beyond the multi-step task handling already described, agentic recommendation
rests on two further capabilities in particular: \textbf{memory}, so that the
agent can carry state across the steps of a task, and \textbf{tool use}, so that
it can act on evidence beyond its parameters.

The first, \textbf{memory}, is a core mechanism for agentic recommendation rather
than an optional implementation detail. Over a multi-step task the system may need
to remember the last item shown in a session, a rejected style, a durable
preference inferred across interactions, or a recurring workflow that should be
executed with minimal friction. We use the standard distinction between
\textit{working}, \textit{episodic}, \textit{semantic}, and \textit{procedural}
memory, and defer the taxonomy, storage mechanisms, and update/retrieval operators
to Section~\ref{sec:memory}. What matters at this point is that memory is not only
an enabler but also an object of evaluation: stale memories can contaminate future
rankings, incorrect summaries can distort user profiles, and privacy-sensitive
traces may be retrieved when they should be forgotten. Memory should therefore be
judged by whether it improves downstream recommendation utility, preserves
relevant constraints, avoids stale or irrelevant facts, and respects privacy and
deletion requirements.

The second, \textit{autonomous} \textbf{tool use}, lets the agent act on evidence
beyond its parameters. Instead of relying only on static model parameters, agentic
recommenders can invoke retrieval APIs, product databases, image-analysis tools,
policy checkers, constraint solvers, or evaluation modules to fetch data, analyze
content, and act on domain-specific evidence~\cite{paranjape2023art,hao2023toolkengpt}.
For example, a request for a restaurant can trigger queries for current ratings,
availability, location, and dietary filters rather than depending on memorized
information; a request to furnish an uploaded room image can trigger a vision tool
to extract visual features and then a catalog query for items matching the user's
style and spatial constraints. Tool use and memory are complementary: memory
carries durable preferences and prior recommendations across steps, while tools
ground the next recommendation in current catalog, policy, or contextual evidence.

The framework developed in this paper is therefore LLM-centric but not
LLM-exclusive. LLMs are currently a natural implementation substrate for agentic
recommendation because they provide flexible natural-language understanding,
reasoning, tool-use interfaces, and multi-agent communication. However, an agent
in an agentic recommender may also be powered by a classical ranker, a retrieval
model, a smaller language model, a rule-based controller, a constraint solver, a
vision model, a simulator, or a human-review module. What makes the system
recommender-specific is not the mere use of natural language or multiple agents,
but the fact that these components operate over users, contexts, histories, item
catalogs, candidate sets, ranking objectives, feedback signals, explanations, and
exposure effects.

Taken together, (i) multi-step task handling, (ii) memory retention, and (iii)
autonomous tool use empower LLM-based and non-LLM-based agents to operate with a
level of autonomy that transcends the simpler question-and-answer paradigm of
traditional chatbots~\cite{prasad2024adapt,maharana2024evaluating,schick2023toolformer}.
At the same time, this should not be read as a claim that agentic systems are
always preferable. For stable, well-scoped, high-throughput ranking tasks,
conventional recommender pipelines may remain simpler, cheaper, more predictable,
and easier to evaluate. The value of agentic recommendation is therefore strongest
when planning, memory, tool use, interaction, and verification improve
recommendation-layer outcomes enough to justify their additional latency, cost,
privacy risk, and governance complexity.

\subsection{A Motivating Example: Goal-Oriented Birthday-Party Recommendation}
\label{subsec:bd-example}

\begin{figure}[t]
\centering
\includegraphics[width=1\linewidth]{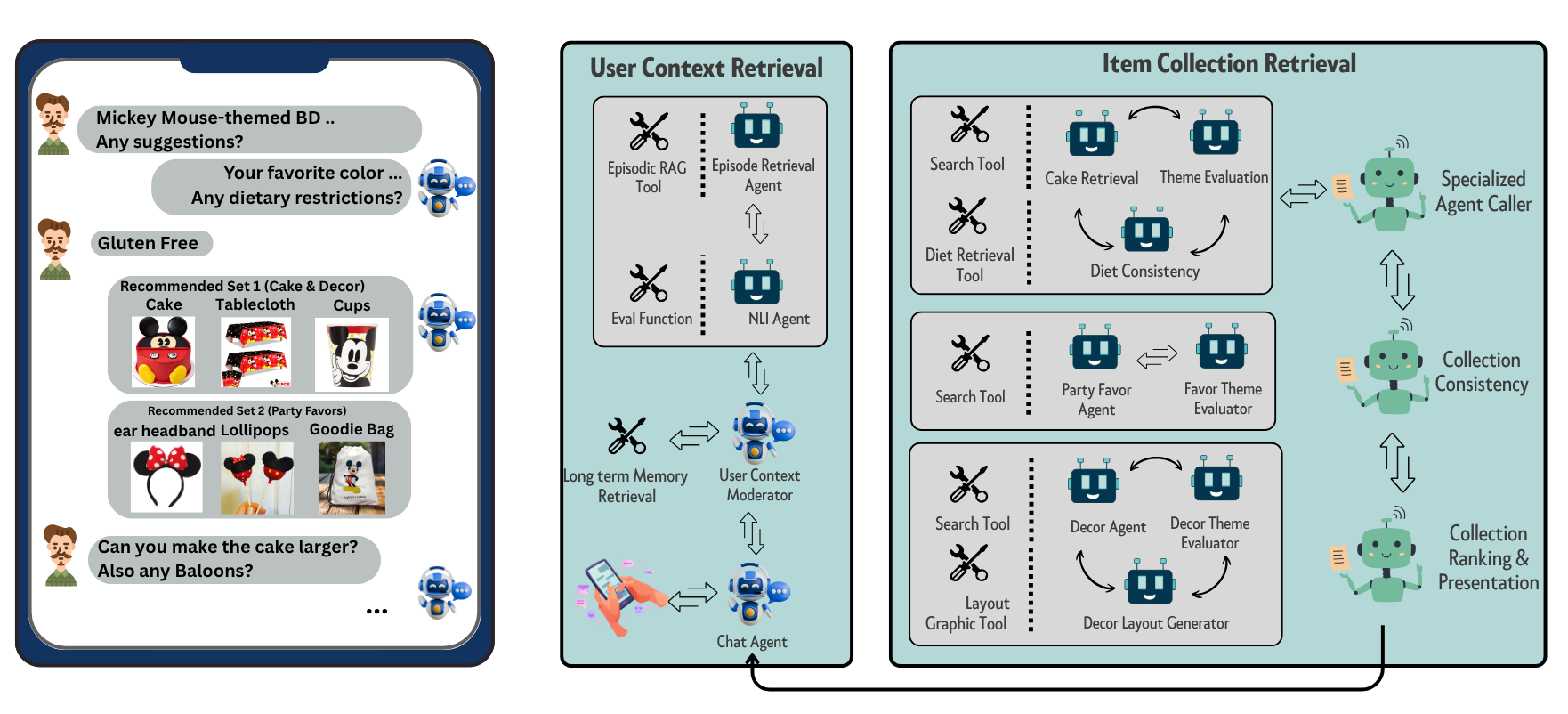}
\caption{Left: Conversation example for a goal-oriented collection recommendation, where the user specifies an open-ended goal instead of searching for specific items. Right: Architecture for personalized goal-oriented recommendation using a multi-agent pipeline with specialized agents and tools.}
\label{fig:bd_part}
\end{figure}

Figure~\ref{fig:bd_part} illustrates the type of recommendation task that
motivates an agentic design. The user is not asking for a single ranked list of
products from a static profile. Instead, the user begins with an open-ended goal:
\dq{a Mickey Mouse-themed birthday party.} The recommender then elicits missing
constraints by asking about color preferences and dietary restrictions. When the
user answers \dq{Gluten free,} a dietary constraint is introduced that must be
respected by the recommended collection. After the system proposes an initial set
of party items, the user further revises the goal by asking whether the cake can
be made larger and whether balloons can be added. This interaction shows that the
task is not merely to rank products, but to support an evolving, goal-oriented
recommendation process: construct a coherent themed birthday-party bundle, satisfy
dietary restrictions, adapt item attributes based on feedback, and add new item
categories as the conversation unfolds.

To handle this complex natural-language intent, the system decomposes the broad
goal into recommendation-relevant subtasks:

\begin{itemize}[leftmargin=*]
\item A user-context component retrieves session-level and long-term traces, including recent dialogue state, prior preferences, accepted and rejected recommendations, and relevant constraints.

\item A specialized-agent caller dynamically instantiates sub-agents equipped with retrieval or generation tools for distinct recommendation categories, such as cakes, decorations, favors, or layout suggestions.

\item Category-specific agents retrieve candidate items from catalog or search tools while respecting explicit constraints such as theme, dietary requirements, budget, and availability.

\item A collection-consistency agent checks whether the resulting items cohere as a bundle, for example whether the cake, tablecloth, decor, and party favors follow the same theme and do not violate constraints.

\item A ranking and presentation agent orders the curated item set according to user preferences, contextual priorities, and recommendation objectives, and can generate a user-facing explanation or visual layout.

\end{itemize}

Throughout this interaction, memory carries the evolving recommendation state
across turns: the theme, the gluten-free constraint, the accepted and rejected
items, and the later requests to enlarge the cake and add balloons. The example
also makes the evaluation concern concrete, since the system must decide which of
these memories are still relevant, which are stale, and which should be excluded
for privacy, safety, or correctness reasons.

\subsection{Organization of the Paper and Contributions}

This is a perspective and conceptual framework paper, not a new learning
algorithm and not an exhaustive survey. Its aim is to give the
recommender-systems community a shared vocabulary and an evaluation-oriented
agenda for agentic RS: precise definitions of agents, memory, tools, actions,
communication protocols, traces, objectives, and evaluation signals, so that
future work can describe architectures, compare systems, and separate
recommender-specific contributions from general agent engineering. Throughout,
we take the position that the value of agentic recommendation is strongest when
orchestration and interaction measurably improve recommendation-layer outcomes,
rather than when a pipeline merely adds LLM calls or agent coordination. To keep
this distinction sharp, we separate a recommendation layer (candidate
generation, ranking, and slate construction), an interaction layer (dialogue,
clarification, explanation, and preference elicitation), and an orchestration
layer (tool choice, agent coordination, memory management, and verification),
defined in Section~\ref{sec:scope-boundary}.

The paper makes four contributions.
\begin{itemize}
    \item \textbf{Core vocabulary, formal framework, and scope boundary
     (Section~\ref{sec:agentic_definition}).} We give the minimal architectural
    primitives needed to specify an agentic recommender system: recommender
    agents, user and item state, candidate sets, memory stores, tool interfaces,
    action spaces, communication protocols, constraints, objectives, and
    observable traces. We distinguish generic LLM agents from recommender
    agents, and separate recommendation-layer contributions from interaction-layer
    capabilities and generic orchestration infrastructure.

    \item \textbf{Running example and task families
     (Fig.~\ref{fig:bd_part}; Sections~\ref{sec:memory}--\ref{sec:agentic-tasks}).}
    Using the Mickey Mouse birthday-planning example as a running case, we show
    how specialized agents, toolchains, memory hierarchies, and verification
    modules compose for goal-oriented recommendation. We then develop four
    representative task families, interactive recommendation, user simulation
    and evaluation, contextual and multimodal recommendation, and grounded
    recommendation explanation, and indicate where a conventional ranking
    pipeline remains simpler, cheaper, and more reliable.

    \item \textbf{Operational challenge and evaluation agenda
      (Section~\ref{sec:challenges}).} We organize the open problems of agentic
    recommendation into five challenge families, communication complexity and
    protocol design, scalability and cost, hallucination and error propagation,
    emergent misalignment and collusion, and brand, policy, and legal compliance,
    and connect each to measurable recommender-system signals such as ranking
    quality, evidence coverage, trace validity, latency, token cost, exposure
    disparity, policy-violation rate, and judge reliability.

    \item \textbf{Controlled empirical study
    (Section~\ref{sec:empirical}).} We report a controlled experiment comparing
    single-shot LLM call and multi-agent recommendation pipelines under shared user
    histories, candidate sets, prompts, and metrics. A pilot next-item ranking
    study shows that multi-agent pipelines are conditionally useful: they do not
    dominate on representative samples, but decomposition and ensemble roles help
    on high-diversity histories.
\end{itemize}

\section{Agentic Recommender Systems: Definition, Characteristics, Misconceptions}
\label{sec:agentic_definition}

In this section, we clarify the concept of an \emph{LLM agent}, in more depth. When illustrating through examples we use the context of recommender and retrieval systems. We also discuss common misconceptions about agentic pipelines and elaborate on what is \emph{not} an LLM agent.

\subsection{What Defines an LLM Agent}
\label{subsec:defines_llm_agent}

An \textbf{LLM agent}, in its fully functional capacity, is an AI system in which a large language model (LLM) serves as the core decision-making component (the “brain”), enhanced by additional mechanisms that enable it to carry out complex, multi-step tasks autonomously rather than relying on a single prompt-response. In other words, this LLM is part of a larger architecture that provides (i) planning abilities, (ii) memory, (iii) tool/API usage, for interactions with external systems, (iv) an \emph{autonomous decision loop} that can break down a goal, observe intermediate steps, and adapt its strategy~\cite{prasad2024adapt, yao2023react}. In a broader context, multiple such language agents form a \textbf{multi-agent system (MAS)}, in which agents can interact, communicate,  coordinate, and even compete-- leveraging their reasoning and communication capabilities to collectively solve complex tasks that exceed the capacity of any single agent.

This design (i.e., agent-based approach) goes beyond the traditional static use of LLMs for single-turn queries and is increasingly used in recommendation and retrieval tasks. For example, an LLM-based agent can iteratively search user logs to gather information about past item interactions before making tailored recommendations~\cite{peng2025survey, zhang2025survey}.

\

\subsection{Key Characteristics of LLM Agents}\label{sec:charagent}
As summarized in Table~\ref{tab:agent}, agentic recommender systems typically exhibit four core characteristics, described in the following:

\begin{table}[!t]
    \caption{Core Capabilities of Agentic Recommender Systems}
\footnotesize
\label{tab:agent}

    \centering
    \renewcommand{\arraystretch}{1.50}
    \rowcolors{1}{gray!8}{gray!16}
    \begin{tabular}{>{\raggedright\arraybackslash}m{0.6cm} m{2.4cm} m{8.35cm}}
    \toprule
    \textbf{Icon} & \textbf{Capability} & \textbf{Description} \\
    \midrule
    \faProjectDiagram & \textbf{Planning \& Task Decomposition} & Breaks complex recommendation goals into sub-tasks, such as candidate
selection, constraint checking, bundle retrieval, re-ranking, and verification. \\
    \faTools & \textbf{Tool Use \& Action Execution} & Invokes external tools or APIs, such as retrieval systems, inventory databases,
policy checkers, vision models, or constraint solvers.\\
    \faDatabase & \textbf{Memory \& State Management} & Maintains state across turns and, when appropriate, across sessions using
working, episodic, semantic, or procedural memory. \\
    \faRobot & \textbf{Autonomy \& Goal-Driven Behavior} & Operates in a closed loop by observing context, choosing actions, evaluating
intermediate outcomes, and revising the plan until the recommendation objective
is satisfied or no further useful action is available. \\
    \bottomrule
    \end{tabular}
\end{table}

\begin{enumerate}
    \item \textbf{Planning and Task Decomposition:} The agent can execute (or even formulate) a plan. This is done by breaking a complex goal into subtasks. Rather than producing an immediate, one-shot answer, an agentic system can plan for and conduct a sequence of steps. This can involve reasoning to tackle long-horizon tasks~\cite{nayak2024long, erdogan2025plan}. For example, in a recommender scenario, the planning module may first identify candidate items, then check user history, retrieve a consistent bundle of products, then re-rank items before generating a final recommendation~\cite{yue2023llamarec}. 

    \item \textbf{Tool Use and Action Execution:} An LLM agent is not limited to purely text-based outputs; it can \emph{invoke external tools or APIs} and perform \emph{actions} in an environment~\cite{yao2023react}.\footnote{Tool invocation examples include web searches, database queries, or calling a specialized retrieval API to filter items.} For instance, a recommendation agent could consult a real-time inventory database to see if recommended items are in stock, or it might retrieve user reviews from a knowledge base before finalizing its recommendations.
    \item \textbf{Memory and State Management:} LLM agents maintain a notion of ``state'' across multiple steps. They often incorporate a memory module which can potentially store conversation context, past recommendations, user feedback, local guidelines, procedures, domain knowledge etc~\cite{xu2025mem}. This memory can be in semantic~\cite{pink2025position}, vector database~\cite{li2024vector}, knowledge graph~\cite{rasmussen2025zep, anokhin2024arigraph} formats. The memory module is used by the agent so that it can recall relevant information at each step of executing a task. For instance, in recommendation scenarios, user preferences collected over multiple sessions can be stored in a memory module and retrieved for the task of personalizing the recommendations. This persistent memory is crucial for multi-turn recommendation dialogues, where the agent refines suggestions based on evolving user interests.~\cite{xi2024memocrs, liu2025enhancing}
    \item \textbf{Autonomy and Goal-Driven Behavior:} LLM agent can be designed to be operated in ``autonomously in a closed-loop fashion'' to fulfill a goal. Given a target objective (for example ``find a suitable product for a user''), the agent can be designed to autonomously (i) observe the environment (via tools or APIs), (ii) evaluate progress, (iii) evaluate the outcome of each step and do a self refine step (iv) continue until it deems the goal is achieved or no further action can be taken. This is unlike a static Q\&A system, where the system merely focuses on the one-time query. \cite{sun2023adaplanner, ming2023hicrisp}
\end{enumerate}

These components enable the LLM agent to tackle complex tasks in recommendation and retrieval settings. For instance, systems like \emph{RecMind} \cite{wang2024recmind} use LLM as a reasoning engine paired with a planning component, a long-term memory of user profiles, and retrieval tools to fetch relevant product information. The agent then autonomously analyzes a user’s needs and iteratively refines its suggestions. This is far more capable than a single-turn Q\&A approach, as it can plan queries (e.g., to check product ratings or availability), incorporate user feedback, and adapt the recommendation strategy. Frameworks like ``ReAct'' \cite{yao2023react} illustrate how an LLM can be prompted to alternate between ``reasoning steps'' and ``action steps'', calling retrieval or recommendation APIs as needed. Other systems, such as ``Toolformer'' \cite{schick2023toolformer} and ``HuggingGPT'' \cite{shen2023hugginggpt} similarly integrate LLMs with external action interfaces.


\subsection{Formal Framework}\label{sec:formal_framework}

In this subsection, we introduce formal definitions for key concepts central to our framework. These definitions establish a rigorous foundation for discussing the architectural and operational aspects of agents, including specialized LLM Agents, Multi-Agent Systems, and Tool-Using Agents. We also provide a generic formalism for representing the internal structure of an agent.

\begin{definition}[LLM Agent]
\label{def:llm_agent}
An \emph{LLM Agent} is an intelligent system whose core decision-making and interaction capabilities are powered by one or more large language models. Formally, we denote such an agent as:
\begin{equation*}
    A_{\mathrm{LLM}} = \Bigl(\mathcal{M}, \mathcal{I}, \mathcal{O}, \mathcal{F}, \Omega \Bigr),   
\end{equation*}
where:
\begin{itemize}
    \item \(\mathcal{M}\) is the underlying language model or set of models used for text generation, understanding, and reasoning. 
    \item \(\mathcal{I}\) represents the input space that the agent can observe.
    \item \(\mathcal{O}\) denotes the output space the agent can produce.
    \item \(\mathcal{F}\) is a set of functions, tools, or APIs the agent can invoke to enhance its capabilities.
    \item \(\Omega\) comprises any additional state or memory structures that enable the agent to maintain context across time or tasks.
\end{itemize}
\end{definition}

For example, in recommendation settings, for an eCommerce shopping assistant, $\mathcal{M}$ can be a GPT-style transformer fine-tuned for e-commerce dialogue; it encodes linguistic knowledge and basic reasoning skills, allowing the agent to parse user requests and generate coherent replies. $\mathcal{I}$ is the space of user queries and contextual signals, while $\mathcal{O}$ is the space of possible outputs, including text responses, actions, or function calls. $\mathcal{F}$ represents external functions or APIs, such as retrieval APIs, price fetchers, or inventory status calls, which augment the innate knowledge of $\mathcal{M}$ with fresh, authoritative data, thereby mitigating hallucination. The agent's memory, $\Omega$, is partitioned into (a) \emph{short-term memory} $\Omega^{\mathrm{STM}}$ that keeps the last few dialogue turns, (b) \emph{long-term episodic memory} $\Omega^{\mathrm{EPI}}$ that stores past party-planning sessions, and (c) \emph{semantic memory} $\Omega^{\mathrm{SEM}}$ containing persistent user traits (e.g., ``prefers red decor''). During inference, salient fragments of $\Omega$ are retrieved and appended to the prompt, giving the agent continuity across sessions. 

For a multi-turn dialogue agent, at each timestep $t$, the agent receives an input $i_t \in \mathcal{I}$, consults its memory $\omega \subseteq \Omega$ to form an augmented context, and computes the output $o_t \in \mathcal{O}$ as follows:
\begin{equation}
    o_t = f_{\mathcal{M}, \mathcal{F}}(i_t, \omega)
\end{equation}
where $f_{\mathcal{M}, \mathcal{F}}$ is the agent's policy function that maps the current input $i_t$ and retrieved memory $\omega$ to an output $o_t$, leveraging both the language model $\mathcal{M}$ and the available external functions $\mathcal{F}$.

Definition~\ref{def:llm_agent} intentionally describes a \emph{generic} LLM
agent. The same tuple becomes a \emph{recommender agent} once each component is
instantiated over recommendation-relevant objects rather than arbitrary text.
Concretely, we take the agent to operate on a state
\[
    s_t = \bigl(u,\; C_t,\; H_u,\; \mathcal{S}^{(K)}_t,\; \pi \bigr),
\]
where $u$ identifies the user, $C_t \in \mathcal{I}$ is the dialogue or
contextual state, $H_u$ is a chronological interaction
history, $\mathcal{S}^{(K)}_t \subseteq \mathcal{S}$ is the candidate set
available at step $t$, and $\pi$ encodes business, safety, privacy, and
multi-stakeholder policy constraints. An LLM agent is a recommender agent when
(i)~its output space $\mathcal{O}$ contains recommendation actions such as
$\texttt{retrieve}(\cdot)$, $\texttt{rank}(\cdot)$, $\texttt{recommend}(\cdot)$,
$\texttt{ask}(\cdot)$, and $\texttt{explain}(\cdot)$; (ii)~its tool set
$\mathcal{F}$ and environment expose RS-relevant evidence such as the item
catalog, candidate sets, and feedback signals; and (iii)~its objective is a
recommendation utility (relevance, constraint satisfaction, long-term
engagement) optimized under the constraints in $\pi$. 
\begin{definition}[Multi‑Agent System (MAS)]
\label{def:multi_agent}
A \emph{Multi‑Agent System} is an ordered triple
\[
\mathrm{MAS} \;=\; \Bigl(\mathcal{A},\,\mathcal{E},\,\Pi\Bigr),
\]
where
\begin{enumerate}[label=(\alph*), leftmargin=*]
  \item $\mathcal{A}=\{A_1,\dots,A_n\}$ is a finite set of agents.  
        Each $A_i$ may be an \textsc{LLM} agent (Definition \ref{def:llm_agent}) or another kind of modular service (e.g., rule‑based, vision, or database stub).  
  \item $\mathcal{E}$ is the shared \emph{environment} that supplies percepts and resources to the agents—such as external APIs, user interfaces, or a simulated world state.  Formally, $\mathcal{E}$ can be viewed as a partial observable Markov space from which each agent receives observations and in which it executes actions.
  \item $\Pi=\bigl(\mathbf{C},\Gamma\bigr)$ is the \emph{interaction protocol}.  
        \begin{enumerate}[label=(\roman*), leftmargin=1.2em]
          \item $\mathbf{C}\in\{0,1\}^{n\times n}$ is a (possibly time‑varying) \emph{communication matrix}.  
                A value $\mathbf{C}_{ij}=1$ indicates that messages of type $\gamma\in\Gamma$ are \emph{permitted} from $A_i$ to $A_j$ under current conditions.  
                The condition can be   
                \begin{itemize}
                    \item \textbf{preset} (static design‐time routing), or  
                    \item \textbf{autonomous} (dynamically toggled by agents. e.g., $A_i$ opens a channel to $A_j$ when cooperation is beneficial, closes it otherwise).  
                \end{itemize}
          \item $\Gamma$ is the finite set of admissible \emph{message schemata} (performative types, serialization rules, timing constraints).  Each $\gamma\in\Gamma$ defines both syntax and semantics so that a receiving agent can parse and act on the message.
        \end{enumerate}
\end{enumerate}
An execution of $\mathrm{MAS}$ can be viewed as a sequence of environment states and inter‑agent messages that respect $\Pi$.  Depending on design, agents may coordinate (\emph{co‑operative}), compete (\emph{self‑interested}), or exhibit mixed motives while attempting to satisfy individual or shared objectives.
\end{definition}

In the recommendation setting, the MAS triple specializes as follows. The
environment $\mathcal{E}$ exposes recommendation state, including the item
catalog, per-user histories, candidate sets, and observed feedback, rather than
a generic world model. The agents in $\mathcal{A}$ occupy recommendation roles,
such as profilers, retrievers, rankers, consistency checkers, evaluators, or
user simulators, each a recommender agent in the sense above. The protocol
$\Pi$ carries recommendation payloads: messages in $\Gamma$ transport candidate
lists, ranked slates, retrieved evidence, constraint or compliance reports, and
memory records, and the communication matrix $\mathbf{C}$ encodes which of these
exchanges are permitted. The marketplace example that follows is the smallest
such instance.

Consider a minimal marketplace scenario with two autonomous components, namely a \emph{recommendation agent} and an \emph{evaluation agent}, so that
\begin{equation*}
    \mathcal{A}=\{A_{\text{rec}},\,A_{\text{eval}}\}.
\end{equation*}
In this case $\mathcal{E} \;=\; \bigl(\text{UserProfileDB},\;\text{ProductCatalogue},\;\text{BusinessRulesKB}\bigr)$, providing user attributes, real-time inventory, and brand–policy facts, respectively. The interaction protocol can be 
\begin{equation*}
    \Pi \;=\;\bigl(\mathbf{C},\,\Gamma\bigr),\qquad
    \mathbf{C}=\begin{pmatrix}
    0 & 1\\
    1 & 0
    \end{pmatrix},
    \quad
    \Gamma=\Bigl\{
      \texttt{candidate\_list},\;
      \texttt{compliance\_report}
    \Bigr\}.
    \end{equation*}

Thus $A_{\text{rec}}$ may send a \texttt{candidate\_list}, a ranked JSON array $\{ (s_1, \text{score}_1),\dots,(s_L,\text{score}_L)\}$ to $A_{\text{eval}}$, which in turn replies with a \texttt{compliance\_report} indicating any items that violate policy. The off-diagonal ones in $\mathbf{C}$ permit bidirectional messaging, while the zeros on the diagonal denote that agents do not address themselves. At runtime, $\mathbf{C}_{\text{eval},\text{rec}}$ can be toggled off once no further corrections are required, illustrating how links may be preset yet autonomously reconfigured according to system state.

\begin{definition}[Memory Update Function]
\label{def:memory_update}
Let $\Omega_{t}$ denote the long‑term memory state maintained by an agent at dialogue step $t$ and let 
\begin{equation*}
    \mathcal{C}_{t} \in \mathcal{X}
\end{equation*}
be the raw context collected during that step (e.g., the most recent user utterance, the agent’s reply, and any intermediate reasoning trace, represented in token or embedding space).  
A \emph{Memory Update Function} is a mapping
\begin{equation*}
\mathcal{U}:\bigl(\mathcal{C}_{t},\,\Omega_{t}\bigr)\;\longrightarrow\;\Omega_{t+1},
\end{equation*}
where
\begin{itemize}
    \item $\mathcal{U}$ internally applies a \emph{retention operator}  
          \begin{equation*}
          \mathcal{R}:\mathcal{C}_{t}\rightarrow\widetilde{\mathcal{C}}_{t}
          \end{equation*}
          that distils $\mathcal{C}_{t}$ into a noise‑reduced summary $\widetilde{\mathcal{C}}_{t}\!\in\!\widetilde{\mathcal{X}}$,  
    \item merges $\widetilde{\mathcal{C}}_{t}$ with the prior state $\Omega_{t}$ via a domain‑specific \emph{merge rule} $\diamond$,  
          and produces the next persistent state:
          \begin{equation*}
              \Omega_{t+1}
              \;=\;
              \Omega_{t}\;\diamond\;\widetilde{\mathcal{C}}_{t}.
          \end{equation*}
\end{itemize}
The goal is to preserve salient semantic information while discarding redundancy, ensuring that $\Omega_{t+1}$ remains compact yet sufficient for future retrieval.
\end{definition}

Assume a shopping assistant agent and a user that uses the agent to order a cake. At turn~$t$ the user says ``remember that two of my guests require a gluten-free diet''. The raw context $\mathcal{C}_{t}$ includes the user utterance, the agent’s internal chain-of-thought, and the pending action call to a \texttt{CakeSearch} tool. The retention operator extracts the key fact  
\begin{equation*}
    \widetilde{\mathcal{C}}_{t}=\text{``\{guest\_allergy: gluten\}''},
\end{equation*}
and the merge rule appends it to the agent’s long‑term \emph{episodic} slot in $\Omega_{t}$:
\begin{equation*}
    \Omega^{\text{EPI}}_{t+1} = \Omega^{\text{EPI}}_{t} \cup \{\text{guest\_allergy: gluten}\}.
\end{equation*}
On the next turn ($t\!+\!1$) the planner retrieves this memory fragment and instructs a downstream cake‑selection agent to filter for gluten‑free options, illustrating how $\mathcal{U}$ enables continuity and correctness across dialogue turns without bloating the prompt window.

In a recommender, the update function $\mathcal{U}$ governs what persists
about a user across turns and sessions: durable preferences, accepted and
rejected items, satisfied or violated constraints, and session-level intent.
The recsys-specific consideration is that $\mathcal{U}$ is not only an enabler
of personalization but also a source of error. A retention operator $\mathcal{R}$
that over-summarizes can drop a binding constraint such as the gluten
restriction; one that under-summarizes lets stale or contradictory preferences
accumulate and later contaminate ranking. The merge rule $\diamond$ must
therefore respect recommendation-relevant properties, namely recency,
provenance, and the right to delete privacy-sensitive traces, so that
$\Omega_{t+1}$ improves rather than degrades downstream recommendation utility.
These properties are what the memory-correctness and staleness signals of
Section~\ref{sec:challenges} are intended to measure.

\begin{definition}[Memory Retrieval Function]
\label{def:memory_retrieval}
Let $\Omega$ be the agent’s persistent memory store after $t$ dialogue steps. A \emph{Memory Retrieval Function} is a mapping
\begin{equation*}
\mathcal{Q}:\bigl(\Omega,\,\tau\bigr)\;\longrightarrow\;\widehat{\mathcal{C}},    
\end{equation*}
where
\begin{itemize}
    \item $\Omega = \{\widetilde{ \mathcal{C}}_1, \dots, \widetilde{\mathcal{C}}_m\}$ is a corpus of distilled memory traces, each $\widetilde{\mathcal{C}}_i \in \widetilde{\mathcal{X}}$ (text snippets, embeddings, or hybrids) produced by the update operator $\mathcal{U}$ (Definition \ref{def:memory_update}).
    \item $\tau \in \mathcal{T}$ is a task representation—e.g.\ the current user query, an intermediate reasoning goal, or a structured tool call argument—expressed in text, vector, or multi‑modal form.
    \item $\widehat{\mathcal{C}}\!\in\!\widehat{\mathcal{X}}\subseteq\widetilde{\mathcal{X}}$ is the subset of memory deemed relevant to $\tau$, returned in a form suitable for prompt augmentation or downstream computation.
\end{itemize}
By recalling only $\widehat{\mathcal{C}}$, the agent conditions subsequent reasoning on task‑specific context while keeping the effective prompt window compact and noise‑free.
\end{definition}

Continuing the shopping assistant example for cake ordering, suppose at turn $t{+}1$ the planner agent formulates the sub‑goal $\tau=\text{``find gluten‑free chocolate cake designs''}$. Invoking $\widehat{\mathcal{C}} = \mathcal{Q} \bigl(\Omega_{t}, \tau\bigr)$ returns a minimal set of memory entries, e.g.
\begin{equation*}
\widehat{\mathcal{C}}=\{\text{guest\_allergy: gluten},\; \text{child\_pref: chocolate}\},
\end{equation*}
filtered from dozens of prior dialogue snippets. These facts are appended to the prompt fed into the \texttt{CakeSearch} tool, ensuring that the retrieval agent queries only gluten‑free, chocolate‑flavoured options while ignoring unrelated historical details. Thus, $\mathcal{Q}$ provides precise, context‑aware recall that prevents irrelevant or outdated memories from polluting the reasoning chain.

For recommendation, the query $\tau$ supplied to $\mathcal{Q}$ is typically
a ranking or explanation sub-goal, and the retrieved subset $\widehat{\mathcal{C}}$
directly conditions the candidate set, the ranking, or the generated
justification. Two recsys-specific failure modes follow. First, retrieving
outdated or irrelevant traces can inject spurious constraints into ranking, for
example resurfacing a preference the user has since abandoned. Second,
retrieving privacy-sensitive history that should have been excluded can produce
recommendations or explanations that are correct yet non-compliant. Effective
retrieval in a recommender is therefore judged not only by topical relevance,
as in the cake example, but by whether the recalled context preserves active
constraints, excludes stale or deleted items, and respects the policy set
$\pi$.

By introducing these formalisms, we establish a common language for discussing agentic behavior and capabilities. In the subsequent sections, we will demonstrate how these formal concepts can be instantiated and extended to create advanced recommendation systems and other multi-step, context-aware applications.

\subsection{What Is Not an LLM Agent}
\label{subsec:not_llm_agent}

Not every LLM-based recommendation pipeline is agentic. A \emph{simple Q\&A}
or \emph{single-prompt recommendation} system, such as an LLM prompted with
``recommend five movies for this user'', is a generator or ranker, not an agent:
it produces one response without persistent task state, autonomous tool choice,
or iterative goal tracking. Similarly, a one-pass \emph{retrieval-augmented
generation} (RAG) system is a grounded generator, but not necessarily an agent,
because the retrieve-then-generate sequence is fixed rather than selected,
repeated, or revised by the model. For example, a RAG recommender may retrieve
product descriptions and generate an explanation, but it does not decide on its
own to re-query the catalog, ask a clarifying question, or invoke a policy
checker.

The litmus test for \emph{full} autonomy is whether the system maintains task
state and can execute an observe--plan--act--verify loop in which it chooses its
own actions. In recommender systems, this means that the system can decide
whether to retrieve additional item evidence, ask a clarifying question, update
or query memory, revise a ranking, invoke a policy or constraint checker, or
stop because the recommendation objective has been satisfied. By this test,
simple Q\&A, one-pass RAG, fixed prompt chains, and static LLM-as-ranker
pipelines are not autonomous agents, because their steps, branching rules,
stopping criteria, and tool calls are predetermined rather than selected at
run time.

Autonomy, however, is only one of the four capabilities of
Section~\ref{sec:charagent}, and it is useful to separate it from the others.
Planning and task decomposition, specialization across roles, tool use, and
memory can each be present in a system whose control graph is fixed in advance.
Consider the birthday-planning example of Figure~\ref{fig:bd_part}. One could
implement it as a fixed pipeline that always retrieves cake candidates, then
decoration candidates, then party favors, checks the assembled set for
theme and dietary consistency, and finally ranks it, with this order wired in at
design time. Such a pipeline still decomposes the goal into specialized
subtasks, calls retrieval tools, and carries the gluten-free constraint forward
as a piece of memory, yet it never decides on its own to reorder these steps,
ask the user a new question, or stop early. We therefore treat ``agentic'' as
graded rather than binary: this fixed-graph version is agentic in
\emph{structure}, through decomposition, specialization, tool use, and memory,
while remaining non-autonomous in \emph{control}. It becomes fully agentic only
when embedded in a closed-loop controller that can maintain state, choose its
next action, inspect intermediate outcomes, and adapt the process toward the
goal, for example by deciding to elicit a missing dietary restriction before
retrieval, or to re-run decoration retrieval after the user enlarges the cake.

This structure-versus-control distinction is not merely terminological. The
two dimensions can be varied independently, and the controlled study of
Section~\ref{sec:empirical} exploits this by holding autonomy fixed and varying
only structure, so that the contribution of decomposition, specialization, and
aggregation to recommendation quality can be measured on its own.

In short, not all LLM-based pipelines are agentic, and among those that are,
not all exercise the same capabilities. Simple Q\&A, one-pass RAG, fixed prompt
chains, and static LLM-as-ranker pipelines can be useful components of an
agentic recommender but are not sufficient on their own.

\subsection{Scope: what is recommendation-specific in an agentic system?}
\label{sec:scope-boundary}

An agentic recommender system contains both recommendation-specific and
auxiliary components. We treat the following as core recommendation operations:
candidate generation, evidence retrieval over user/item data, filtering and
constraint satisfaction over item attributes, ranking and re-ranking, bundle or
slate construction, explanation grounded in item and user evidence, feedback
interpretation, memory update for personalization, and evaluation of
recommendation utility. We treat natural-language parsing, generic task
planning, tool routing, and user-interface generation as adjacent capabilities:
they may be necessary to implement an agentic recommender, but they are not
themselves sufficient to constitute a recommendation contribution. This boundary
matters because a multi-agent architecture can be impressive as an orchestration
system while still failing at recommendation objectives such as relevance,
diversity, calibrated personalization, fairness of exposure, or long-term user
value.

Accordingly, throughout the paper we distinguish three layers: (i) the
\emph{recommendation layer}, responsible for user--item matching, ranking, and
slate construction; (ii) the \emph{interaction layer}, which manages dialogue,
clarification, explanation, and preference elicitation; and (iii) the
\emph{orchestration layer}, which chooses tools, coordinates agents, manages
memory, and verifies intermediate states. A system can be sophisticated at the
interaction and orchestration layers yet add little at the recommendation layer,
which is precisely the case the boundary above is meant to expose.

\subsection{Positioning Relative to Non-Agentic and Agentic Recommenders}
\label{sec:positioning}
\begin{table*}[!h]
\centering
\caption{Positioning of this paper relative to non-LLM, LLM-based,
single-agent, and multi-agent recommender systems. \full\ = explicit/central;
\partly\ = partial or emerging; \none\ = not central. Column meanings and the
gap notes G1--G10 are given in the text.}
\label{tab:positioning-related-work-compact}
\scriptsize
\setlength{\tabcolsep}{3.2pt}
\renewcommand{\arraystretch}{0.98}

\begin{adjustbox}{max width=\textwidth}
\begin{tabularx}{\textwidth}{@{}Y C C C C C C C C C G@{}}
\toprule
\textbf{Representative line / examples}
&
\multicolumn{3}{c}{\textbf{Agent role}}
&
\multicolumn{4}{c}{\textbf{Agentic capability}}
&
\multicolumn{2}{c}{\textbf{Evaluation}}
&
\textbf{\makecell{Gap}} \\
\cmidrule(lr){2-4}
\cmidrule(lr){5-8}
\cmidrule(lr){9-10}
&
\rothead{Assist.}
&
\rothead{Recomm.}
&
\rothead{Simul.}
&
\rothead{Memory}
&
\rothead{Tools / RAG}
&
\rothead{Planning}
&
\rothead{Coord.}
&
\rothead{Ranking}
&
\rothead{Trace}
&
\\
\midrule

\grouphead{Part I: Non-LLM recommender systems}

Classical / sequential recommender systems
\cite{koren2009matrix,rendle2009bpr,hidasi2016session,kang2018sasrec}
&
\none & \full & \none
&
\partly & \none & \none & \none
&
\full & \none
&
G1 \\

Conversational / interactive recommender systems
\cite{christakopoulou2016towards,gao2021conversational}
&
\partly & \full & \none
&
\partly & \partly & \partly & \none
&
\full & \partly
&
G2 \\

\midrule
\grouphead{Part II: LLM-based recommender systems, not explicitly agentic}

LLM as encoder, ranker, reranker, or reasoning module
\cite{wu2024llmrec,zhao2024llmrec}
&
\none & \full & \none
&
\partly & \partly & \none & \none
&
\full & \partly
&
G3 \\

Generative / prompt-based recommender systems
\cite{deldjoo2024review,li2024generative}
&
\none & \full & \none
&
\partly & \partly & \partly & \none
&
\full & \partly
&
G4 \\

\midrule
\grouphead{Part III: LLM-based recommender systems, single-agent}

Agent-assisted recommender
\cite{huang2025interecagent,zhao2024toolrec}
&
\full & \partly & \none
&
\full & \full & \full & \none
&
\full & \partly
&
G5 \\

Agent-as-recommender
\cite{wang2024recmind}
&
\partly & \full & \none
&
\full & \full & \full & \none
&
\full & \partly
&
G6 \\

Agent-as-user-simulator
\cite{wang2023recagent,zhang2024agent4rec,chen2025recusersim}
&
\none & \none & \full
&
\full & \partly & \full & \none
&
\partly & \full
&
G7 \\

\midrule
\grouphead{Part IV: LLM-based recommender systems, multi-agent}

Multi-agent agent-assisted recommender
\cite{wang2024macrec,fang2024macrs}
&
\full & \partly & \none
&
\full & \full & \full & \full
&
\full & \full
&
G8 \\

Multi-agent agent-as-recommender
\cite{zhang2024prospect,wang2024macrec}
&
\partly & \full & \none
&
\full & \full & \full & \full
&
\full & \partly
&
G9 \\

Multi-agent user / environment simulation
\cite{wang2023recagent,banerjee2025collabrec}
&
\none & \none & \full
&
\full & \partly & \full & \full
&
\partly & \full
&
G10 \\

\midrule
\grouphead{Part V: This paper}

Unified RecSys-specific framework and evaluation agenda
&
\full & \full & \full
&
\full & \full & \full & \full
&
\full & \full
&
-- \\

\bottomrule
\end{tabularx}
\end{adjustbox}
\end{table*}

Sections~\ref{subsec:defines_llm_agent} through~\ref{sec:scope-boundary}
defined an agentic recommender system, separated it from LLM pipelines that are
not agentic (Section~\ref{subsec:not_llm_agent}), and bounded it against
orchestration that is not recommendation-specific
(Section~\ref{sec:scope-boundary}). We close the section by placing this
definition on the map of existing work.
Table~\ref{tab:positioning-related-work-compact} organizes representative lines
of research along three axes. The \emph{agent role} axis records whether the
agent assists an existing recommender (Assist.), acts as the recommender
(Recomm.), or simulates users and environments (Simul.). The \emph{agentic
capability} axis records explicit memory
(working/episodic/semantic/procedural), tool or retrieval use (Tools/RAG),
planning or plan--act--verify behavior, and inter-agent coordination (Coord.).
The \emph{evaluation} axis records whether a line is assessed only on final
ranking quality (HR@K, NDCG@K, MRR, diversity, exposure) or also on the
intermediate agent trace (evidence grounding, memory use, tool correctness,
judge reliability, communication cost, latency, and failure attribution). These
axes separate three issues that are often conflated: that a system may use an
LLM without being agentic, that an agentic recommender may differ in role and in
which capabilities it exercises, and that evaluation may stop at the ranked list
or extend to the trace. The table therefore makes precise why this paper is
neither a survey of LLM-based recommendation nor a generic multi-agent-system
discussion, but a RecSys-specific framework for specifying and evaluating
agentic recommendation architectures.

The capability columns correspond directly to the primitives of the formal
framework: Memory is the store $\Omega$ with its operators $\mathcal{U}$ and
$\mathcal{Q}$ (Definitions~\ref{def:memory_update} and~\ref{def:memory_retrieval}),
Tools/RAG is the function set $\mathcal{F}$ of Definition~\ref{def:llm_agent},
Planning is the decomposition behavior of Section~\ref{sec:charagent}, and
Coordination is the protocol $\Pi$ of the MAS triple
(Definition~\ref{def:multi_agent}). The rows then trace a progression, and the
Gap column names what each stage still leaves open. Classical and sequential
recommenders (G1) provide strong ranking foundations but expose no agent loop,
and conversational recommenders (G2) are interactive yet usually scripted rather
than tool-using and memory-bearing. LLM-based but non-agentic recommenders
improve representation, ranking, or generation while remaining inside fixed
pipelines, so autonomy and tool choice stay implicit (G3, G4). Single-agent
systems introduce memory, tools, and planning, but agent-assisted designs lack
inter-agent specialization (G5), agent-as-recommender designs often leave the
memory, tool, and planning contributions under-ablated (G6), and
agent-as-simulator designs face fidelity, calibration, and label-leakage
difficulties (G7). Multi-agent systems add coordination through $\Pi$, yet the
marginal utility and cost of extra agents is rarely attributed (G8),
decomposition and verification raise communication complexity (G9), and richer
simulation can amplify artificial behavior or simulator drift (G10). The final
row marks the gap this paper targets: a unified, recommender-specific vocabulary
and evaluation agenda in which agentic complexity is justified through measurable
gains in recommendation quality, grounding, trace reliability, fairness, cost,
and robustness rather than assumed, with the under-ablation gaps G6 and G8
addressed empirically in Section~\ref{sec:empirical}.

\section{Memory Storage and Retrieval Mechanisms}
\label{sec:memory}

Due to importance of memory mechanisms in agentic systems, especially those anchored toward recommender system tasks we will discuss these mechanisms in a dedicated section.  Large Language Model (LLM) agents require robust memory mechanisms to behave coherently over time. By default, an LLM is stateless, which means that each query is processed independently, with no built-in recall of previous interactions. This stateless design is a serious issue in hindering LLMs to be adapted to recommendation tasks requiring personalization and ``remembering'' things about users.  

To overcome this, agent frameworks introduce explicit memory systems that let the agent ``remember'' and reuse information from past events or dialogues. Broadly, these memories fall into short-term (working memory) and long-term categories, with further distinctions (episodic, semantic, procedural, etc.) inspired by human cognition. This section surveys the types of memory in LLM agents and the storage/retrieval mechanisms.

\subsection{Types of Memory in LLM Agents}
\label{subsec:type_mem}
\subsubsection{Short-Term (Working) Memory}
Short-term (often called ``working'' \cite{guo2023empowering}) memory refers to the transient context the agent holds in mind. In an LLM agent, this corresponds to the recent interaction history or ``context window'' provided to the model on each turn. It is analogous to human working memory in that it is readily accessible but of limited capacity (bounded by token length) \cite{shan2025cognitive}. For instance, a chatbot’s short-term memory might include the last few user prompts and agent responses, allowing for conversational coherence.
Short-term memory is therefore crucial for immediate reasoning, but it does not persist beyond the current session (i.e., it lacks a true long-term store).

\subsubsection{Long-Term Memory}
Long-term memory provides persistent recall across interactions and over time. In agent systems, it can be subdivided into:

\textbf{Episodic Memory}
Episodic memory pertains to specific events or experiences (\emph{episodes}) the agent has encountered.
In LLM agents, this often means a stored log of past dialogues, observations, or actions, coupled with context such as timestamps or metadata.
It is context-rich and instance-specific. In this case, the agent remembers not just \emph{what} happened, but \emph{when} and \emph{under what circumstances} \cite{fountas2024human}. For instance, the agent might recall: \emph{Last week the user asked about Italian restaurants and I recommended XYZ.}
Episodic memory supports one-shot learning of events and explicit recall of past episodes, and some researchers argue it is a key missing piece for truly long-lived LLM-based agents. \cite{pink2025position}

\textbf{Semantic Memory}
Semantic memory stores general facts, concepts, or knowledge about the world, akin to the \say{know-what.} In LLM agents, semantic memory might be an external knowledge base or database of facts (e.g., \say{in my retrieval system TVs are categorized under electronics category}) \cite{zeng2024structural}.

While much semantic knowledge is encoded in the LLM’s pretrained weights, agents can also maintain explicit semantic stores (relational databases, knowledge graphs, etc.) for dynamic updates.
Semantic memory is \emph{long-term} and \emph{explicit}, though it abstracts away the specific instance in which a fact was learned \cite{langchain2024memory}. For example, from several chats the agent might distill a high-level fact: \say{This user prefers Italian cuisine.}

\textbf{Procedural Memory}
Procedural memory, or the \say{know-how}, refers to skills or procedures the agent has learned to execute without needing to deliberate each time. For an LLM agent, this might appear as learned prompts, scripts, or code snippets that automate common tasks (e.g., connecting to a database). \cite{wheeler2025procedural, sumers2023cognitive}

Procedural memory is typically implicit in that it is not stated as factual knowledge but manifested as rules or sequences of actions. It allows the agent to execute tasks more efficiently, refining these skills through repeated practice. Early cognitive architectures like SOAR \cite{laird1987soar} used production rules to represent procedural knowledge, and modern LLM agents similarly encode procedural behaviors through fine-tuned prompts, scripted tool invocations, or learned action sequences that mimic rule-based execution pipelines\cite{de2024language, mu2025experepair}. In the next subsection, we explore how these different types of memory are realized in the actual agentic recsys implementations.

\subsection{Memory Storage and Retrieval Mechanisms}
\label{subsec:storage_retrieval}

Large‐context LLMs can “remember” only a few thousand tokens per forward pass, yet real recommender deployments routinely span months of user interaction and terabytes of product data. Consequently, an \textit{agentic} recommender must decide ``what'' to store externally, ``how'' to distil it, and ``which'' fragments to re‑inject into the prompt at inference time.  We formalise these operations through the \textbf{Memory Update} and \textbf{Memory Retrieval} functions (Definitions \ref{def:memory_update}–\ref{def:memory_retrieval}), then survey concrete storage modalities. For having a more complete survey of the memory mechanisms, first we state the following definitions. 
\begin{definition}[Memory Item]
\label{def:memory_item}
A \emph{memory item} is a triple $m=(k,v,\varsigma)$ where  

(i) $k\in\mathbb{R}^{d_k}$ is the \textbf{key}, typically an $n$‑gram embedding or hashed identifier that serves as the retrieval handle;  

(ii) $v\in\mathbb{R}^{d_v}$ is the \textbf{value} payload that preserves the semantic content (natural‑language text or a dense vector);  

(iii) $\varsigma=(t,\ell,u)$ is a metadata tuple storing the time‐stamp $t$, a logical label $\ell\!\in\!\{\text{EPI},\text{SEM},\text{PROC}\}$ distinguishing \emph{episodic}, \emph{semantic}, or \emph{procedural} memory, and an update counter $u$ that tracks how often the item has been modified.
\end{definition}
\noindent For instance, in our cake retrieval example, the statement of the user for gluten allergy, “\texttt{guest\_allergy = gluten}” is stored as
\begin{equation*}
    k = \text{embedding}(\text{``gluten allergy''}),\quad
    v = \text{ ``gluten''},\quad
    \varsigma=(t{=}17{:}45,\ell=\text{EPI},u=1).
\end{equation*}

Here $k$ is an embedding representation of the key ``gluten allergy''; $\ell=\text{EPI}$ signals that the fact is tied to a specific party event rather than a timeless preference.

\rev{The key--value--metadata form in Definition~\ref{def:memory_item} is an
analytic wrapper, not a restriction on implementation. The value field may
contain raw text, dense embeddings, symbolic triples, relational rows,
procedural scripts, multimodal features, pointers to external stores, or other
memory representations. The purpose of the wrapper is to expose the three pieces
of information that agentic recommendation needs from any stored item: a
retrieval handle, a semantic payload, and metadata for recency, provenance,
privacy, and update control. In a recommender these metadata fields are not
incidental bookkeeping. The timestamp $t$ determines whether a preference is
still current or stale, the label $\ell$ separates a one-off event such as a
party constraint from a durable taste, and the update counter $u$ records how
often a fact has been reinforced, which is exactly the signal needed to decide
whether a preference is worth conditioning a ranking on.}

\begin{definition}[Relevance Scoring]
\label{def:relevance}
Given a task query $\tau\in\mathcal{T}$ and a memory item
$m=(k,v,\varsigma)$, a \emph{relevance function}
\begin{equation*}
S:\mathcal{T}\times\bigl(\mathcal{X}\times\widetilde{\mathcal{X}}\times\mathbb{R}^3\bigr)\longrightarrow\mathbb{R}_{\ge 0}, 
\end{equation*}
returns a non‑negative score.  Retrieval (Definition \ref{def:memory_retrieval}) selects the $K$ highest‑scoring items:
\begin{equation*}
\widehat{\mathcal{C}}=\mathrm{Top\text{-}K}_{m\in\Omega}\,S(\tau,m).
\end{equation*}

\end{definition}

Again, back to our cake retrieval example, let $\tau=$ “\texttt{find gluten‑free chocolate cake}”. Then, we can choose the relevance function in a form like $S(\tau,m) = \cos \bigl(\text{embedding}(\tau), k\bigr)$.  The cosine term elevates items semantically close to the query. Consequently, the system surfaces the allergy item and the child’s flavor preference, but not unrelated recommendation traces.

\rev{In a recommender, topical similarity alone is an impoverished relevance
function. What should be recalled for a ranking or explanation sub-goal depends
not only on whether a memory item is semantically close to $\tau$, but on
whether it is still valid, whether it encodes a binding constraint, and whether
it is permitted to be used. A useful $S$ for agentic recommendation therefore
combines the similarity term above with the metadata of Definition~\ref{def:memory_item}:
a recency weight derived from $t$ so that abandoned preferences decay, a label
term from $\ell$ so that hard constraints (a dietary restriction, a returned
item) are retrieved with priority over soft tastes, and a privacy gate that
excludes items the user has asked to delete. This is what distinguishes
memory relevance in a recommender from generic semantic search: the highest
cosine score is not always the item that should condition the next ranking.}

Now, we will review different modalities that memory store/retrieve can take. 

\subsubsection{Raw Text Logs}
The simplest store is an \emph{append‑only} transcript $\Omega^{\text{raw}} = [m_1,\dots,m_T],$ where each $m_t$ is the verbatim turn text.  Retrieval is concatenation of the last $L$ tokens: $\widehat{\mathcal{C}} = \text{Tail}_L(\Omega^{\text{raw}}).$ Although the update function (Definition \ref{def:memory_update}) $\mathcal{U}$ is trivial (append), this scales poorly and prompt length soon exceeds the model window. \cite{an2024does, liu2024lost}

\subsubsection{Summarisation (Compression)}
In this approach, older logs are periodically compressed into shorter summaries. More specifically,  update function $\mathcal{U}$ invokes an LLM summariser,
$\mathcal{R}_{\text{sum}}:\mathcal{C}_t\to\widetilde{\mathcal{C}}_t,$
which compresses $\mathcal{C}_t$ to a more concise context $\widetilde{\mathcal{C}}_t$. The agent then stores and retrieves these summaries to preserve essential information. Though it saves tokens, the fidelity of recall depends on the quality of summarization. Recent memory compression methods enable adaptive summarization and storage of long conversational histories for LLM agents~\cite{li2024prompt}.

\subsubsection{Embedding and Vector Databases}
For long-term memory, the prevailing method in retrieval-augmented generation (RAG) is embedding-based storage. Each memory item is encoded as a high-dimensional vector (using an embedding model $e$) $k=e(v),\quad v=\text{(raw text)}$ and stored in a vector database \cite{yousefi2025arag, forouzandehmehr2025calrag}. For Retrieval, using the retrieval function $\mathcal{Q}$ (see Definition \ref{def:memory_retrieval})  the current query is embedded, and nearest-neighbor search surfaces semantically relevant items, enabling large-scale, flexible recall—though typically at the cost of strict chronological ordering. Please note that classic retrieval-augmented models RETRO \cite{borgeaud2022improving} and REALM \cite{guu2020retrieval} demonstrate the scalability and effectiveness of such large-scale external memory designs.

\rev{The choice among these modalities is a recommendation-deployment decision,
not only an engineering one. Vector stores scale to catalog- and user-history
sizes and tolerate fuzzy recall, which suits personalization, but they blur
chronology, which matters when the order of a user's interactions carries intent.
Raw logs and summaries preserve session dynamics but do not scale to months of
history. Structured stores enforce the hard constraints and provenance that
policy-compliant recommendation requires. In practice an agentic recommender
mixes them by role: short-term session state in a buffer, durable tastes in a
vector store, and constraints and provider or policy facts in a structured store
where they can be checked exactly.}

\subsubsection{Knowledge Graphs / Structured Stores} Beyond raw text or vector embeddings, an agent may persist ``knowledge–graph memory'' in the form of symbolic triples \((s,r,o)\) (subject–relation–object) or relational table rows that can be queried explicitly via SQL or SPARQL. Symbolic storage is particularly well‑suited for semantic memory: it supports precise retrieval, logical entailment, and integrity constraints, albeit at the cost of additional curation.

\begin{definition}[Symbolic Memory Item]
\label{def:symbolic_memory}
A symbolic memory item augments Definition \ref{def:memory_item} by taking $m=(k,v,\varsigma)$ with $v=(s,r,o)\in\mathcal{E}^{3},$ where $\mathcal{E}$ is the entity set of the knowledge graph. In other words, the value can be of the form subject–relation–object for a given query. Equivalently, one can assume an embedding on the triplet for key retrieval used for approximate nearest‑neighbour fallback.
\end{definition}

For example, suppose the assistant stores  $v=(\textsf{user},\,\textsf{likes},\,\textsf{chocolate})$ with label $\ell=\text{SEM}$. Given user query $\tau=\text{``What flavour cake should I order?''}$ the system formulates  $(\textsf{user},\,\textsf{likes},\,x)$ and retrieves the triple, yielding the concrete value \(\textsf{chocolate}\).  If no exact match exists, the agent can still fall back to the embedding key $k=\text{embedding}(\textsf{user},\textsf{likes},\cdot)$ and perform a vector search for semantically close preferences.

Symbolic knowledge graph/structured stores offer high‑precision constraints and explainable provenance \cite{lu2025karma}, but demand manual or automated pipelines to populate and maintain the triples; they also introduce schema‑evolution overhead when the domain ontology changes. \cite{ye2023schema}

\subsubsection{Parametric Memory Updates}
A costly alternative is to integrate $\widetilde{\mathcal{C}}_t$ directly into the backbone parameters of the LLM model \cite{ficek2024gpt}.
Formally, the update function $\mathcal{U}$ returns a new language model 
$\mathcal{M}_{t+1}=\mathcal{M}_{t}\oplus\Delta_t$, where $\mathcal{M}_{t}$ represents the model at time $t$ and $\Delta_t$ represents the update on the model to incorporate the new events. 
This makes $\Omega$ implicit in $\mathcal{M}$. Latency and safety constraints limit real‑time use \cite{chen2025fundamental, fonseca2025safeguarding}.

\subsubsection*{Regulated Context Windows}
In production, agents often blend \textit{procedural}, \textit{episodic}, and \textit{semantic} memories with a regulator:

\[
\widehat{\mathcal{C}}
=
\widehat{\mathcal{C}}_{\text{PROC}}
\;\cup\;
\widehat{\mathcal{C}}_{\text{SEM}}
\;\cup\;
\widehat{\mathcal{C}}_{\text{EPI}},
\]
subject to a token budget
$
\bigl|\widehat{\mathcal{C}}\bigr|\le B$ \cite{xiong2025memory}. 
We can formalize this type of regulator as a knapsack problem (see \cite{martello1990knapsack}) variant
\begin{equation*}
\begin{split}
\max_{\widehat{\mathcal{C}}\subseteq\Omega}
&\sum_{m\in\widehat{\mathcal{C}}} S(\tau,m) \\
&\text{s.t. } \sum_{m}|m|\le B.   
\end{split}
\end{equation*}
where $S(\tau,m)$ is the relevance score as defined in definition \ref{def:relevance}, $|m|$ is the length of the retrieved memory, and $B$ is the token budget.



The foregoing definitions and examples demonstrate that memory in agentic recommender systems is no longer a monolithic cache but a spectrum of interlocking mechanisms—from raw transcript buffers and lossy summaries to vector stores to fully structured knowledge graphs, to budgeted tokens. By casting memory operations as explicit update $\mathcal{U}$ and retrieval $\mathcal{Q}$ functions, parameterised by relevance scores and symbolic query semantics, we obtain a principled foundation for analyzing capacity, latency, and factual fidelity. Yet the formalism also exposes substantial gaps: adaptive compression that preserves downstream utility is still heuristic; relevance scoring lacks theoretical guarantees; and symbolic stores incur unsolved curation and schema‑evolution costs. Bridging these gaps will require new learning objectives that couple retention with task performance, tighter integration of uncertainty calibration into retrieval, and hybrid architectures that combine the precision of knowledge graphs with the scalability of dense embeddings. \rev{For recommendation specifically, these gaps surface as measurable failure
modes: compression that drops a binding constraint, relevance scoring that
resurfaces a stale preference, and retrieval that returns a deleted trace. The
memory-correctness, staleness, and deletion-compliance signals of
Section~\ref{sec:challenges} are intended to quantify exactly these failures.} In short, memory for LLM‑driven, multi‑agent RecSys is both a critical enabler and an open frontier, inviting further research at the intersection of information retrieval, knowledge representation, and large‑scale language modeling.

\begin{table*}[t]
\centering
\caption{\rev{Use cases for agentic recommender systems. The table summarizes the four representative scenarios developed in Section~\ref{sec:agentic-tasks}, the practical trigger for using an agentic design, the recommendation-layer gain, and when a simpler non-agentic baseline may be sufficient.}}
\label{tab:task-families}
\scriptsize
\setlength{\tabcolsep}{4pt}
\renewcommand{\arraystretch}{1.12}

\begin{tabularx}{\textwidth}{@{}p{0.20\textwidth}p{0.22\textwidth}p{0.30\textwidth}p{0.22\textwidth}@{}}
\toprule
\rev{\textbf{Use case}}
&
\rev{\textbf{Agentic trigger}}
&
\rev{\textbf{Recommendation-layer gain}}
&
\rev{\textbf{Simpler baseline may suffice when}}
\\
\midrule

\rev{\textbf{Interactive goal-oriented recommendation}}
&
\rev{Clarification, planning, memory, and bundle verification}
&
\rev{Improves constraint satisfaction, bundle coherence, ranking quality, personalization, and user effort in multi-turn recommendation.}
&
\rev{The request is a single well-scoped query with a stable user profile and fixed candidate set.}
\\

\rev{\textbf{User simulation and evaluation}}
&
\rev{Synthetic users, logging, evaluators, and session summarization}
&
\rev{Supports offline stress testing, counterfactual diagnosis, robustness checks, and evaluation before live deployment.}
&
\rev{Reliable logged feedback is already available and no interactive trajectory or simulated behavior needs to be modeled.}
\\

\rev{\textbf{Contextual and multimodal recommendation}}
&
\rev{Vision tools, external retrieval, memory, and consistency checks}
&
\rev{Connects images, context, metadata, user preferences, and spatial or aesthetic constraints to ranking and bundle construction.}
&
\rev{All relevant features are already structured and can be consumed directly by a conventional ranker.}
\\

\rev{\textbf{Grounded recommendation explanation}}
&
\rev{Evidence retrieval, critic/evaluator agents, and policy checks}
&
\rev{Produces explanations that are more faithful, evidence-backed, policy-consistent, and less prone to unsupported claims.}
&
\rev{A short template explanation is sufficient and dynamic evidence or policy checking is unnecessary.}
\\

\bottomrule
\end{tabularx}
\end{table*}

\section{Agentic Recommendation Task Families}
\label{sec:agentic-tasks}
\rev{
The remainder of this section grounds the general agentic framework in four representative use cases, each chosen to illuminate a distinct facet of next-generation recommender systems. As summarized in Table~\ref{tab:task-families}, these examples are not intended to be exhaustive; rather, they show when planning, memory, tool use, verification, or multi-agent coordination can improve recommendation-layer outcomes, and when a simpler non-agentic baseline may be sufficient.
}

%


\subsection{Interactive Recommendation}
\label{subsec:interactive_rec_usecase}


\subsubsection*{Formal Task Definition} We define \emph{Interactive Recommendation} as the problem setting in which an agent (or a set of collaborating agents) engages in a multi-turn dialogue with a user to identify, refine, and present suitable recommendations. Unlike static recommender systems that rely on one-shot user inputs, interactive recommendation leverages iterative exchanges to incorporate user feedback, contextual constraints, and additional information drawn from memory or external resources. The overarching objective is to dynamically adapt suggestions as the conversation evolves, thereby providing a more personalized and contextually appropriate set of recommendations.

Let $\mathcal{D}$ be the space of dialogue turns and $\mathcal{S}$ the universe of documents (in an eCommerce setting these can be purchasable stock‑keeping units, SKUs). A user session at step $t$ is characterised by the partial transcript
\begin{equation*}
    \mathcal{C}_{1:t}\!=\!(d_1, a_1, \dots, d_{t-1}, a_{t-1},d_t),
\end{equation*}
where $d_i\in\mathcal{D}$ is a user utterance and $a_i\in\mathcal{D}$ the agent reply.  Define the \emph{interactive recommendation task} as a mapping
\begin{equation*}
    \Phi:\bigl(\mathcal{C}_{1:t},\,\mathcal{E}\bigr)
      \;\longrightarrow\;
      \bigl\langle s_{(1)},\dots,s_{(L)}\bigr\rangle,
\end{equation*}
that outputs a ranked list of $L$ items with the maximal expected utility $\sum_{j=1}^{L} \text{Rel}\bigl(s_{(j)} \mid \mathcal{C}_{1:t}\bigr)$, conditioned on user constraints (theme, dietary rules, budget) implicitly encoded in $\mathcal{C}_{1:t}$.

To illustrate this task, we consider the scenario of planning a Mickey Mouse-themed birthday party as shown in Figure \ref{fig:bd_part}. A parent consults the system to select decorations, arrange a suitable cake, and accommodate guests’ dietary restrictions. The agent aims to guide the parent through each step, from confirming the child’s preferences (such as color schemes or favorite cake flavors) to identifying any special requirements. By interacting iteratively with the user and retrieving relevant information from its memory and external tools, the agent can refine its suggestions over time.

\subsubsection*{High‑level goal} The agent must \emph{adaptively refine} $\Phi$ through multi‑turn dialogue, spawning specialized sub‑agents and tool calls to satisfy latent sub‑goals (e.g. ``select gluten‑free chocolate cake'', ``choose decor consistent with Mickey‑Mouse palette'') while preserving conversational coherence and minimizing user effort.

\subsubsection*{System architecture and agent setup}

For this specific example, we can instantiate a multi‑agent system
\begin{equation*}
    \mathrm{MAS}_{\text{party}}
   = \bigl(\mathcal{A},\,\mathcal{E},\,\Pi\bigr),
\end{equation*}
where \textbf{Agent set} 
\(
\mathcal{A}=\{A_{\text{chat}},A_{\text{epi}},A_{\text{nli}},A_{\text{SAC}}, A_{\text{cake}},A_{\text{decor}},A_{\text{favor}},
              A_{\text{col\_check}},A_{\text{rank}}\}.
\)
Each $A_i$ is an \textsc{LLM} agent in the sense of Definition \ref{def:llm_agent}. Where  
\begin{itemize}
      \item $A_{\text{chat}}$ (chat agent) primary interface to the user.
      \item $A_{\text{epi}}$ – episodic retrieval using $\mathcal{Q}$ (Definition \ref{def:memory_retrieval}).  
      \item $A_{\text{nli}}$ – natural‑language inference to vet relevance of retrieved episodes.  
      \item $A_{\text{SAC}}$ – specialised‑agent caller that spawns three micro‑MAS blocs:  
            $\{\!A_{\text{cake}},A_{\text{decor}},A_{\text{favor}}\!\}$ for category‑specific retrieval.  
      \item $A_{\text{col\_check}}$ – collection‑level consistency.  
      \item $A_{\text{rank}}$ – personalised ranking \& presentation.
\end{itemize}

Also one can define the Environment for this MAS as:   
\begin{equation*}
    \mathcal{E}=(\text{ProductCatalogue}, \text{UserProfileDB}, \text{VectorDB}, \text{LayoutTool}),
\end{equation*} 
and communication protocol as $\Pi = (\mathbf{C}, \Gamma)$ where $\mathbf{C}_{ij} = 1$ iff either $A_j$ is a child spawned by $A_i$ or $A_i{=}A_{\text{chat}}$. $\Gamma$ contains message types 
$\{\text{query}, \text{episode\_list}, \text{tool\_call}, \text{item\_set}, \text{ranked\_list}\}$.

The following also illustrates communication sketch:  
\begin{equation*}
    \begin{split}
        A_{\text{chat}}  \xrightarrow{\texttt{query}} & A_{\text{epi}}
    \xrightarrow{\texttt{episode\_list}}
    A_{\text{nli}}
    \xrightarrow{\texttt{validated\_episodes}}
  A_{\text{SAC}} \\
  & \xrightarrow{\texttt{spawn}}
  \{A_{\text{cake}},A_{\text{decor}},A_{\text{favor}}\}
  \xrightarrow{\texttt{item\_set}}
  A_{\text{col\_check}}
  \xrightarrow{\texttt{validated\_set}}
  A_{\text{rank}}
  \xrightarrow{\texttt{ranked\_list}}
  A_{\text{chat}}.
    \end{split}    
\end{equation*}
Note that \texttt{spawn} can be autonomous depending on $C$
\subsubsection*{Memory and tool requirements}
Now, we illustrate how each type of memory is defined here and how each agent can utilize it. Short-term (Working) memory can be defined for $A_{\text{chat}}$ that takes logs of last $L$ turns of the $\mathcal{C}_{1:t}$. Episode store can be accessed and manipulated by $A_{\text{epi}}$ through $\mathcal{U}$ and $\mathcal{Q}$. Semantic memory $\Omega^{\text{SEM}}$ can track stable user traits (preferred colours, dietary restrictions). $\Omega^{\text{PROC}}$ can include repeatable prompt templates for each retrieval block as well as information about each table in DB so that agents can autonomously query them or other elements of environment $\mathcal{E}$. External Tools  $\mathcal{F}$ in this MAS are \texttt{SearchCakeAPI}, \texttt{SearchDecorAPI}, \texttt{SearchFavorAPI}, \texttt{VectorDB.query} (semantic retrieval), \texttt{LayoutTool.generate} (graphic board of decor set).

\subsubsection*{Benefits and discussion}

The architecture realises \emph{interactive recommendation} by decomposing a fuzzy, high‑level goal (“celebrate my child’s Mickey‑Mouse birthday”) into tractable sub‑tasks handled by specialised agents.  
Key advantages like (i) Modularity, where each retrieval block is itself a micro‑MAS that can be re‑used for other party themes or plugged into A/B tests without retraining the entire pipeline.  (ii) Memory‑aware personalisation, where episodic retrieval ($A_{\text{epi}}$) injects user‑ and session‑specific constraints (gluten‑free, colour palette) while semantic memory supplies timeless preferences, yielding recommendations that are both relevant and surprising. (iii) Error containment, where the NLI validator and collection‑consistency agent act as \emph{fact gates}, reducing hallucination propagation by rejecting items that violate theme constraints or dietary rules, (iv) autonomy, where the specialised‑agent caller spawns retrieval workers on demand and per query, and (v) Rich user experience where the final ranked set, together with a layout generated by \texttt{LayoutTool}, offers a cohesive shopping board, demonstrating how agentic orchestration can move beyond single‑item suggestions to holistic, story‑driven recommendations.

Collectively, the design illustrates how the formal primitives—\(\mathrm{MAS}\), \(\mathcal{U}\), \(\mathcal{Q}\), and typed memory stores—translate into a concrete, deployable pipeline for next‑generation conversational shopping assistants, which far exceeds the capabilities of classic recommender systems.

\subsection{User Simulation and Recommendation Evaluation}\label{subsec:user_sim_eval}

\subsubsection*{Formal definition of the task}
User simulation and recommendation evaluation together form a framework in which synthetic user behavior is generated to probe, test, and refine recommender systems before full-scale deployment. Let \(R_{\phi} : \mathcal{X}\!\rightarrow\!\mathcal{S}^{L}\) denote a recommender parameterised by $\phi$ that maps the current session state \(x\in\mathcal{X}\) to an ordered list of \(L\) items from the catalogue \(\mathcal{S}\).
A \textbf{user simulator} is a stochastic policy

\begin{equation*}
    \mathcal{M}_{\theta,\omega} : \mathcal{X} \times \mathcal{S}^{L} \longrightarrow 
    \mathcal{A}_{\text{user}},
    \quad \mathcal{A}_{\text{user}}\in \{\textsf{Select}, \textsf{Not Select}\},
\end{equation*}
parameterised by intrinsic preference vector $\theta$ and noise vector $\omega$. Note that user simulator's action space $\mathcal{A}_{\text{user}}$ can include another set of options depending on the recommendation setting. For instance it can be $ \{\textsf{Click}, \textsf{Pass}, \textsf{Purchase}\}$ in eCommerce user simulation. Given repeated interaction over $T$ time steps, the joint process $\{x_t, s^{(L)}_t, a^{\text{user}}_t \}_{t=1}^{T}$
induces an empirical evaluation functional
\begin{equation*}
    \Psi(R_{\phi},\mathcal{M}_{\theta,\omega}) = 
   \mathbb{E}\Bigl[ \sum_{t=1}^{T} g\bigl(x_t, s^{(L)}_t, a^{\text{user}}_t \bigr) \Bigr],
\end{equation*}

\noindent
where \(g(\cdot)\) is a reward such as \textsf{Select} or diversity penalty. The ``simulation‑based evaluation problem'' seeks to estimate \(\Psi\) under controlled distributions of \(\theta\) and \(\omega\).

\subsubsection*{High‑level goal}
In practical scenarios, gathering data from real users can be expensive or infeasible, especially when conducting A/B tests for newly introduced features. A simulated user that approximates realistic browsing, clicking, and purchasing behaviors offers a controlled environment for stress-testing various recommendation strategies. For example, an e-commerce platform aiming to recommend Bohemian-style furniture may wish to simulate how a typical user navigates through the site, examines multiple items, compares prices, and ultimately decides whether to purchase. This allows the platform to estimate click-through rates and conversions without requiring a large real-user pilot study.
The primary aim is to replicate the interactions of a diverse user population in order to evaluate and compare recommendation strategies. The system must produce synthetic but plausible session data, track user decisions at each step, and generate summary statistics or qualitative insights about overall performance. By testing both short sessions (quick browsing) and long sessions (prolonged decision-making), one can obtain a more comprehensive understanding of the strengths and weaknesses of the underlying recommender algorithms.

\subsubsection*{System architecture and agent setup} We instantiate a multi‑agent system (also see Figure \ref{fig:arch_agent})

\begin{equation*}
    \mathrm{MAS}_{\text{sim}}
      =\bigl(\mathcal{A},\mathcal{E},\Pi\bigr),
    \quad
    \mathcal{A}=
    \Bigl\{
       A_{\text{rec}},A_{\text{user}},
       A_{\text{note}},A_{\text{eval}},
       A_{\text{summ}},A_{\text{report}}
    \Bigr\}.
\end{equation*}

\begin{itemize}[leftmargin=*]
\item \textbf{$A_{\text{rec}}$} (\emph{Recommender Agent}) is an \textsc{LLM} agent that wraps \(R_{\phi}\) and orchestrates tool calls (\texttt{SearchAPI}, \texttt{SessionTracker}).
\item \textbf{$A_{\text{user}}$} (\emph{User Simulator}) implements \(\mathcal{M}_{\theta,\omega}\); it possesses a structured memory
\(\Omega^{\text{SEM}}\) of preferences and \(\Omega^{\text{STM}}\) for the
current session trajectory.
\item \textbf{$A_{\text{eval}}$} (\emph{Evaluation Agent}) logs tuples $(x_t, s^{(L)}_t, a^{\text{user}}_t)$ via update function \(\mathcal{U}\)-style appends to long‑term storage. It computes instantaneous metrics \(g_t\) by querying logs with \(\mathcal{Q}\) and emits \texttt{eval\_event} messages.
\item \textbf{$A_{\text{summ}}$} (\emph{Session Summariser}) invokes a compression operator \(\mathcal{R}_{\text{sum}}\) every \(T\) turns to
produce a summary vector of the session’s salient outcomes.
\item \textbf{$A_{\text{report}}$} (\emph{Reporter}) aggregates thousands
of summaries, performs statistical tests, and outputs a final PDF/CSV
dashboard.
\end{itemize}

Environment 
\(
\mathcal{E}=
\bigl(
  \text{ProductDB},\;
  \text{PreferenceSampler},\;
  \text{LogStore}
\bigr),
\)
provides catalogue metadata, draws synthetic $\theta$,
and persists interaction traces.

Communication Protocol matrix of this MAS, $\mathbf{C}$ allows the sequence

\[
A_{\text{rec}}\;\rightleftarrows\;A_{\text{user}}
\;\rightarrow\;A_{\text{eval}}
\;\rightarrow\;A_{\text{summ}}
\;\rightarrow\;A_{\text{report}},
\]

\noindent
with message types
$\Gamma=\{
  \texttt{rec\_list}, \texttt{user\_action}, 
  \texttt{log\_entry}, \texttt{eval\_event},
  \texttt{session\_summary}$, $ \texttt{final\_report}
\}$.

\subsubsection*{Memory and tool requirements}
\begin{figure}
    \centering
    \includegraphics[width=0.95\linewidth]{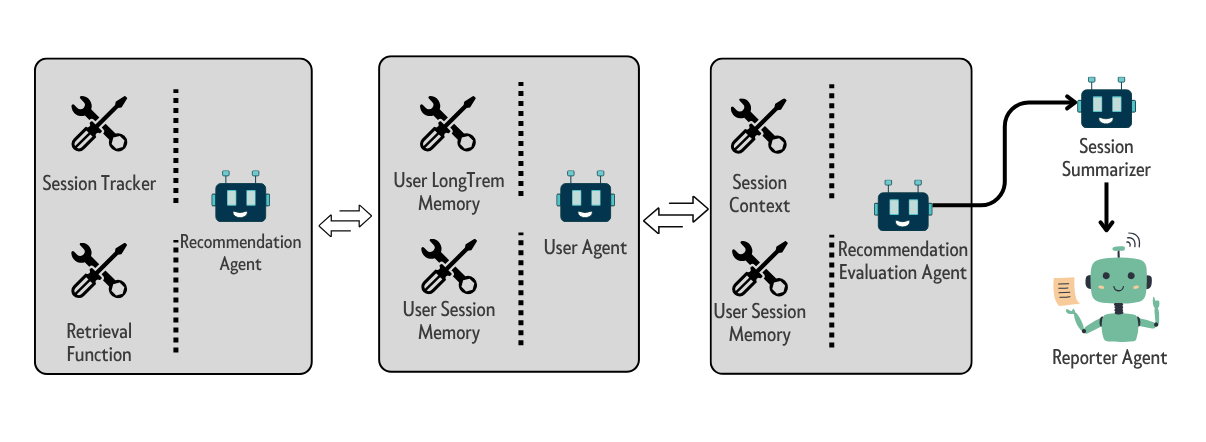}
    \caption{A sample multi-agent system for user behavior simulation and recommendation evaluation}
    \label{fig:arch_agent}
\end{figure}

$A_{\text{user}}$ can access and update semantic memory, \(\Omega^{\text{SEM}}\) which includes latent taste vector, price sensitivity, \(\Omega^{\text{PROC}}\) which stores navigation policy, click heuristics, and $(\Omega^{\text{EPI}}$ which remembers prior simulated sessions (for repeat‑exposure effects). \textbf{$A_{\text{eval}}$} uses a raw text log plus vector store to support fast \(\mathcal{Q}\) filters by item ID or action type. It also maintains sliding‑window statistics (e.g.\ running CTR, diversity entropy) in $\Omega^{\text{STM}}$. Tool access by other agents is illustrated in Figure \ref{fig:arch_agent}.

\subsubsection*{Benefits and discussion}
The stated MAS architecture can yield the following traits: (i) Cost‑effective experimentation, where hundreds of parameterized user agents can be spawned in parallel, yielding reliable offline estimates of $\Psi(R_{\phi},\mathcal{M})$ before any live‑traffic exposure. (ii) Cold‑start mitigation, where by sampling user profiles with sparse or unseen preference vectors, the framework stresses \(R_{\phi}\) under low‑data regimes and
surfaces failure modes early. (iii) Diagnosis of systemic bias, where the reporter’s aggregate view highlights trends such as over‑pricing, popularity bias, or demographic skew, which can be traced back to fine‑grained logs for root‑cause analysis. (iv) Memory‑driven realism, where the division of semantic, episodic, and procedural memories within $A_{\text{user}}$ allows long‑range preference carry‑over (``I still want a bohemian sofa'') while modelling bounded session attention—all expressed via the formal $\mathcal{U}/\mathcal{Q}$ operators. (v) Modular extensibility, where new evaluation criteria (fairness, robustness) can be implemented by adding auxiliary evaluator agents without retraining the core recommender, consistent with the MAS abstraction.

\noindent
Overall, this architecture turns user simulation into a first‑class,
memory‑aware MAS that provides quantitative and qualitative feedback
loops—bridging the gap between offline metrics and live user studies
for next‑generation recommender systems.

\subsection{Contextual and Multi‑Modal Recommendation}
\label{subsec:contextual_multimodal}

\begin{figure}[ht]
    \centering
    \includegraphics[width=.62\linewidth]{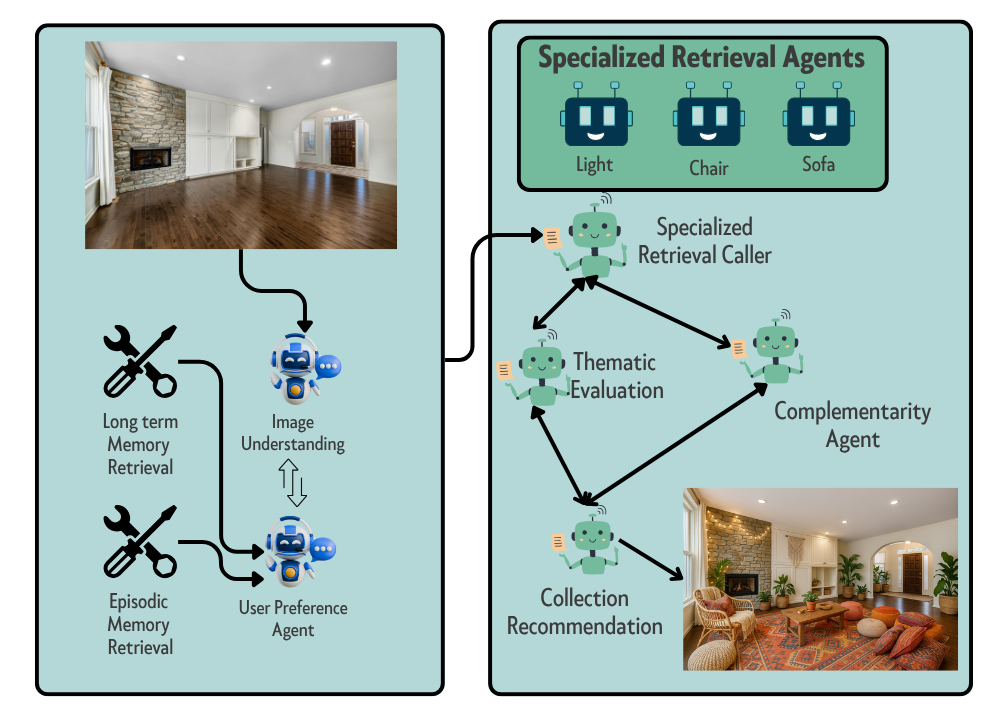}
    \caption{An example of multi-agent pipeline for multi‑modal recommendation. }
    \label{fig:boho_arch}
\end{figure}
\subsubsection*{Formal definition of the task} We define the task as the process of generating suggestions that incorporate diverse forms of input, such as text descriptions, user interaction histories, and images depicting the user’s physical or aesthetic environment.  For instantiation, let \(\mathbf{x}\in\mathcal{T}\) denote textual constraints (e.g.\ “Bohemian, earthy colours”), \(\mathbf{v} \in \mathcal{V}\) a set of one or more images encoding spatial context, and \(\mathbf{u}\in\mathcal{N}\) a latent user–profile vector drawn from long‑term memory.  
We model contextual, multi‑modal recommendation as a mapping  
\begin{equation*}
    \mathcal{R}_{\phi} :
   \bigl(\mathbf{x},\mathbf{v},\mathbf{u}\bigr)
   \;\longrightarrow\;
   \widehat{\mathbf{y}}
   =\bigl\langle s^{\text{1}},s^{\text{2}},s^{\text{3}},\dots\bigr\rangle,
\end{equation*}
where each \(s^{(\cdot)}\in\mathcal{S}\) is a SKU and the ranked tuple
\(\widehat{\mathbf{y}}\) maximises joint relevance  
\(
\sum_{i} \text{Rel}\!\bigl(s_i \mid \mathbf{x},\mathbf{v},\mathbf{u}\bigr)
\)
subject to \emph{complementarity} and \emph{aesthetic‑coherence} constraints.  
The latter can be enforced by an auxiliary predicate  
\(\text{Compat}(\widehat{\mathbf{y}})=1\) if all selected items harmonise in palette and style, or any \emph{Coherence Score} mechanism. 

\subsubsection*{High‑level goal}
In many practical scenarios, recommendations cannot rely solely on textual inputs. A user interested in redecorating a living room may upload one or more photographs depicting the current layout, along with a brief description of desired styles (e.g., Bohemian, minimalist, rustic-see Figure \ref{fig:boho_arch}). The system must analyze these images to identify available space, prevailing color palettes, and aesthetic elements, while also taking into account prior user behavior or known preferences (such as a history of purchasing similar items). By combining textual context with visual analysis, a multi-modal recommender can produce more cohesive suggestions, reduce guesswork, and streamline the user’s decision-making process. The ambition in an example orchestration is to replicate a professional interior‑designer workflow while eliminating iterative, item‑by‑item searches.

\subsubsection*{System architecture and agent setup}
The pipeline in Fig.\,\ref{fig:boho_arch} is formalised as  
\(\mathrm{MAS}_{\text{mm}}=(\mathcal{A},\mathcal{E},\Pi)\) with  
\[
\mathcal{A}=\bigl\{
  A_{\text{chat}},
  A_{\text{image}},
  A_{\text{history}},
  A_{\text{cat}},
  A_{\text{caller}},
  A_{\text{ThemChk}},
  A_{\text{compChk}},
  A_{\text{collect}}
\bigr\}.
\]

\noindent
A single conversation unfolds as follows.  
The chat agent \(A_{\text{chat}}\) receives \((\mathbf{x},\mathbf{v})\) and forwards the image to \(A_{\text{image}}\), which extracts palette vector $\mathbf{p}$ and spatial affordances via a vision backbone $\mathcal{F}_{\text{layout}}\in\mathcal{F}$. Concurrently, \(A_{\text{history}}\) retrieves \(\widehat{\mathcal{C}}_{\text{SEM}} \subset \Omega^{\text{SEM}}\), stable style and budget preferences, using the retrieval operator \(\mathcal{Q}\).  The specialised‑retrieval caller \(A_{\text{caller}}\) fuses $\mathbf{x}, \mathbf{p}, \widehat{\mathcal{C}}_{\text{SEM}}$ and outputs a set of target classes like $C=\{\text{chair},\text{sofa},\text{lamp}\}$. For each \(c\in C\), it spawns an intra‑session micro‑MAS.   The resulting candidate lists are concatenated and passed to  \(A_{\text{ThemChk}}\), which rejects items violating user history (e.g.\ bans leather if the user is vegan). \(A_{\text{compChk}}\) then checks for complementarity score for the bundle of recommended items (maybe through solving mixed integer program).  Finally, $A_{\text{collect}}$ ranks the surviving set by personalised utility and calls $\mathcal{F}_{\text{render}}$ to generate an on‑image mock‑up before handing results back to $A_{\text{chat}}$.

The protocol's communication matrix \(\mathbf{C}\) is sparse. For instance, $A_{\text{chat}}$ may message any agent, but checker agents communicate only downstream, preventing cyclic justification loops that might amplify hallucinations.

\begin{figure}[ht]
    \centering
    \includegraphics[width=.82\linewidth]{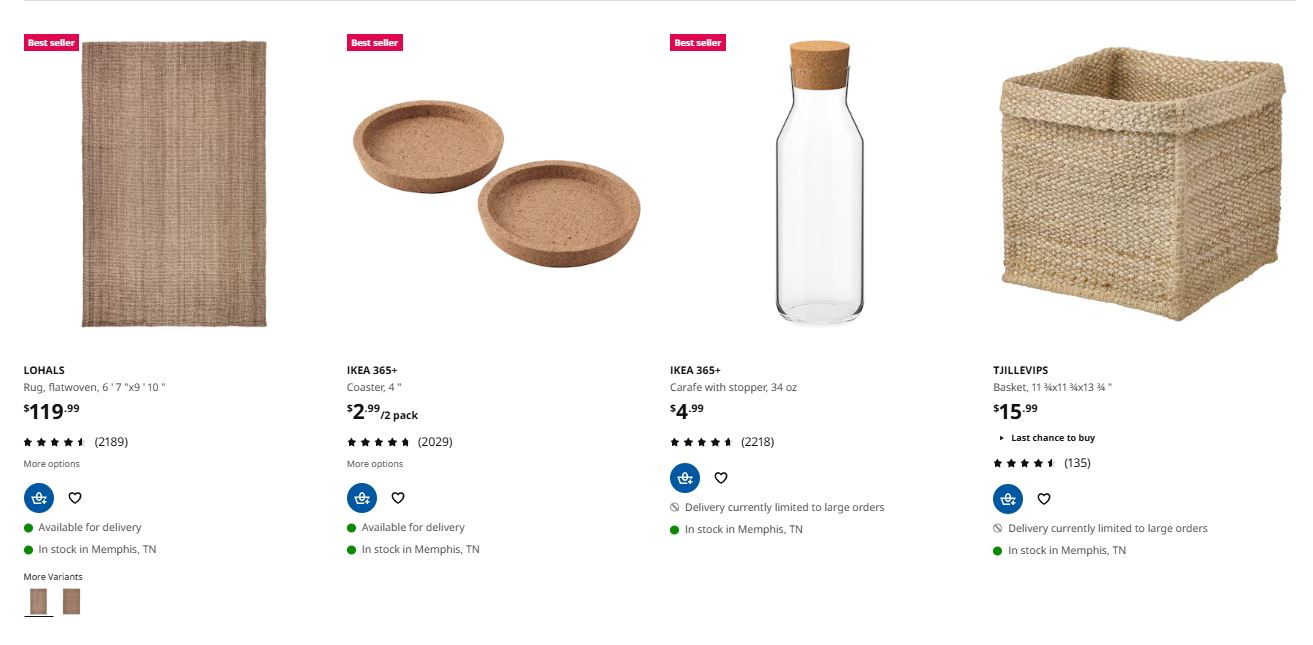}
    \caption{A sample of coherent (Boho style with earthy colors) multi-category recommendation (courtesy\,\protect\url{ikea.com}).}
    \label{fig:boho_scene}
\end{figure}
\subsubsection*{Memory and tool requirements} Short‑term buffers hold the live prompt plus extracted palette vectors; $\Omega^{\text{SEM}}$ caches enduring preferences (colour, material, budget); $\Omega^{\text{EPI}}$ stores prior styling sessions so that the system can avoid repetitious suggestions.  
Key tools are the vision encoder, a vector‑search API over
\(\text{ProductDB}\), and a layout renderer that embeds selected SKUs
into the uploaded photograph.

\subsubsection*{Benefits and discussion}
The multi‑modal MAS realises a “one‑shot designer” experience. Semantic and complementarity checks decouple \emph{rule compliance} from \emph{aesthetic harmony}, yielding recommendations that are both on‑brand and visually pleasing. Finally, because the agent caller can spawn modular micro-MAS blocks live, the system stays flexible to a variety of user requests. In sum, contextual and multi‑modal recommendation showcases how LLM‑powered agents, external vision tools, and structured memories can collaborate to deliver designer‑level curation in a single interactive session. Like the previous use case, this enables a capability beyond classic recommender systems.  
\begin{figure}[ht]
    \centering
    \includegraphics[width=0.8\linewidth]{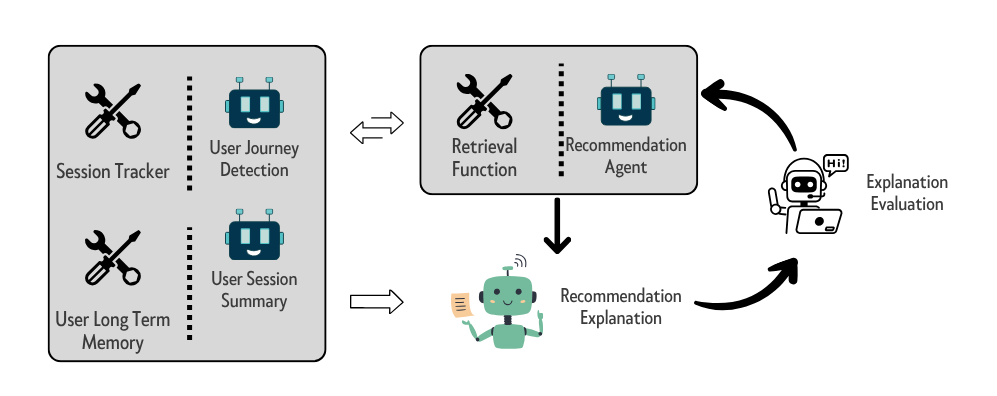}
    \caption{Adaptive Multi-Agent Recommendation Explanation}
    \label{fig:recexpl}
\end{figure}

\subsection{Recommendation Explanation}
\label{subsec:recommendation_explanation}
\subsubsection*{Formal definition of the task}
We define \emph{Recommendation Explanation} as the process of generating intelligible and contextually relevant justifications for why particular items are recommended to a user (see \cite{chen2025carts}) . Let \(\mathbf{r}\in\mathcal{S}^{L}\) be a set of items surfaced by a recommender \(R_{\phi}\) at dialogue step \(t\) and let \(\mathbf{u}_t = \bigl( \widehat{\mathcal{C}}^{\text{SEM}}_t, \widehat{\mathcal{C}}^{\text{EPI}}_t \bigr)\)
denote the user’s current semantic and episodic context retrieved via
\(\mathcal{Q}\).
A \emph{recommendation‑explanation function} is a mapping  
\begin{equation*}
    \Phi_{\psi} :
    (\mathbf{r},\mathbf{u}_t)
    \;\longrightarrow\;
    e_t \in \mathcal{E}_{\text{text}},
    \quad
    \text{s.t. }\;
    \text{Consistency}(e_t)=1,
\end{equation*}
where \(\psi\) parameterises an LLM agent,
\(e_t\) is a natural‑language explanation, and
\(\text{Consistency}\) is a predicate that checks
(i) factual alignment with \(\mathbf{u}_t\) and
(ii) stylistic conformity to brand rules.

\begin{figure}[ht]
    \centering
    \includegraphics[width=0.7\linewidth]{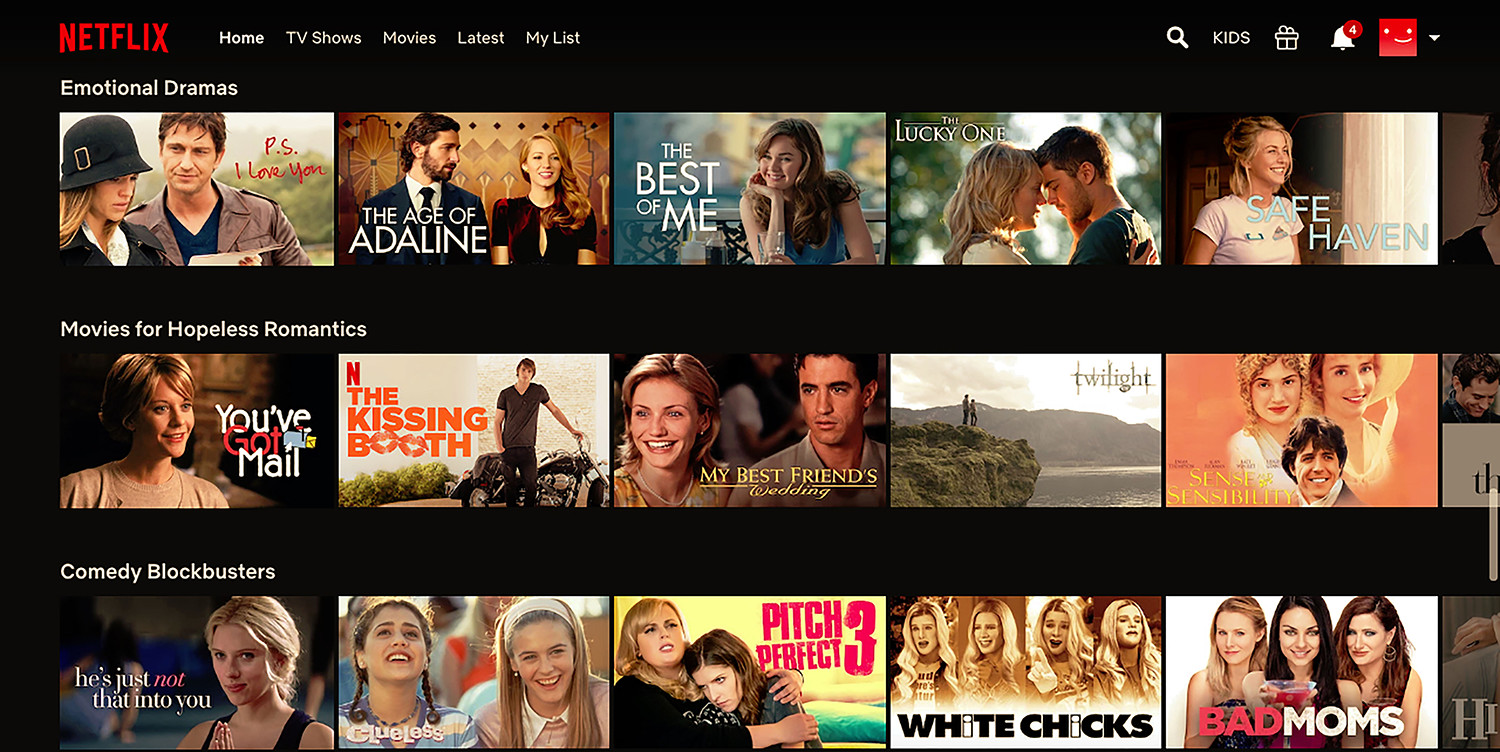}
    \caption{Explaining Recommendations (e.g., Emotional Dramas in the first set of recommended movies) provides more transparency and helps with user engagements (courtesy netflix.com).}
    \label{fig:enter-label}
\end{figure}

\subsubsection*{High‑level goal}
The system seeks to frame each recommendation batch in a concise,
human‑readable story that increases transparency, trust, and click‑through rates while satisfying brand tone guidelines. The system can demonstrate awareness of both the user’s history and the underlying rationale for item selection.

\subsubsection*{System architecture and agent setup}
The explanatory workflow is realised as  
\(
\mathrm{MAS}_{\text{expl}}=(\mathcal{A},\mathcal{E},\Pi)
\)
with agent ensemble $\mathcal{A}=\{
  A_{\text{journey}},     
  A_{\text{session}},     
  A_{\text{rec}},         
  A_{\text{expl}},        
  A_{\text{eval}}         
\}$. In this MAS, \(A_{\text{session}}\) invokes \(\mathcal{Q}\) over
\(\Omega^{\text{SEM}}\cup\Omega^{\text{EPI}}\) to extract
\(\mathbf{u}_t\);  
\(A_{\text{journey}}\) inspects the live click stream
\((x_1,a^{\text{user}}_1,\dots,x_t)\) to tag the latent intent
(e.g.\ “holiday décor upgrade’’).
Both embeddings flow to \(A_{\text{rec}}\) which
returns recommendation \(\mathbf{r}\).  
The explanation agent \(A_{\text{expl}}\)
feeds \((\mathbf{r},\mathbf{u}_t)\) into its LLM, then emits
\(e_t=\Phi_{\psi}(\mathbf{r},\mathbf{u}_t)\).
\(A_{\text{eval}}\) verifies
\(\text{Consistency}(e_t)\) by checking factual mentions
against \(\mathbf{r}\) and brand constraints stored in
\(\Omega^{\text{PROC}}\); if the test fails, it sends a
\texttt{revise} message back to \(A_{\text{rec}}\) and
\(A_{\text{expl}}\), triggering a second‑pass generation.

The protocol matrix \(\mathbf{C}\) therefore admits the cycle  
\(A_{\text{rec}}\!\leftrightarrow\!A_{\text{expl}}\!\rightarrow\!A_{\text{eval}}\)
with veto capability at the evaluator.

\subsubsection*{Memory and tool requirements}
Short‑term windows hold the active dialogue and the latest bundle
\(\mathbf{r}\).  
Semantic memory captures enduring preferences
(\emph{“collects Mickey‑Mouse merchandise’’});
episodic slices recall recent campaigns to avoid repetition;  
procedural memory stores brand‑tone exemplars and banned phrasing.
Key tools are illustrated in figure \ref{fig:recexpl}.

\subsubsection*{Benefits and discussion}
$\Omega^{\text{PROC}}$, $\Omega^{\text{EPI}}$ provide support for outputting truthful and on‑brand explanations, while the modular MAS layout lets one inject new style guides or fairness rules by updating
\(\Omega^{\text{PROC}}\) and the evaluator prompt. Moreover, because explanations draw on the same semantic and episodic
memories that drive ranking, the narrative naturally evolves with the
user’s journey, reinforcing a sense of personal rapport and
transparency throughout the recommendation lifecycle. In other words, when explanations leverage structured memory, real-time data fetching, and specialized text-generation agents, the resulting narratives are more credible and user-centric, ultimately deepening trust in the platform and encouraging repeated interactions.

\rev{Fig.~\ref{fig:enter-label} illustrates the broader motivation for grounded recommendation explanations: explanatory grouping can make recommendation surfaces more transparent and easier for users to interpret.}

\section{Operational Challenges and Evaluation Agenda}
\label{sec:challenges}

\rev{Agentic recommender systems inherit challenges from general LLM agents and
multi-agent systems, but those challenges become recommender-specific when they
affect ranking, exposure, feedback, item evidence, user memory, or marketplace
outcomes. We therefore organize the challenge section around two questions:
(i) what failure mode is introduced by planning, memory, tools, or multi-agent
coordination, and (ii) how should a recommender-system researcher measure it?}

\subsection{Communication Complexity and Protocol}\label{sec:mcp}

\rev{We use \emph{multi-agent communication protocol} (MCP) to mean the set of
message schemas, routing rules, context-sharing rules, state-synchronization
assumptions, and provenance requirements by which agents exchange information.
In an RS setting, MCP design must carry recommendation-specific metadata,
including candidate-set provenance, item attributes, retrieved evidence, policy
constraints, privacy labels, timestamps, and confidence estimates. Without these
fields, downstream agents may rank or explain items using stale, unsupported, or
policy-violating context.}

Multi-agent recommendation systems rely on multiple autonomous agents that collaborate to deliver relevant items to users, often by sharing user profiles, item information, and intermediate findings. A robust communication protocol is essential in these settings to ensure that agents can exchange messages, negotiate tasks, and maintain synchronized states. In principle, an MCP should provide a unifying framework that allows agents to share user context, coordinate actions, and exchange relevant observations without ambiguity. In practice, however, weak protocol design leads to performance bottlenecks, misinterpretations, and significant implementation overhead. When each agent or data source uses a bespoke interface or message schema, the system devolves into a patchwork of ad-hoc integrations, making it difficult to scale or ensure correct interpretation of exchanged data. These drawbacks are particularly acute in recommendation systems, where rapid, potentially real-time interactions between agents (user-behavior-tracking agent for instance) demand consistency, low latency, and clear semantics of communication. Now, we will review some of the challenges. 

\subsubsection*{Protocol Standardization.} A foundational challenge is agreeing on a standardized MCP that all agents within the ecosystem can adopt. Ideally, a newly introduced agent, potentially developed by a different team or vendor, would connect seamlessly to the multi-agent system by adhering to the same syntactic and semantic rules. In practice, such alignment is difficult to achieve. Existing frameworks (for example, the FIPA Agent Communication Language \cite{poslad2007specifying}) demonstrated the complexities inherent to specifying communication structures and semantics. Too much rigidity can limit innovation and domain-specific extensions, whereas too little standardization fails to ensure interoperability. A well-designed MCP must be expressive enough to cover the needs of diverse agents (ranging from user-modeling to explanation evaluation for example) while remaining sufficiently lightweight that developers can implement it without excessive overhead. Achieving wide adoption also poses governance challenges: different organizations may promote competing standards or tailor them to their internal architectures, which underscores the non-technical barriers to protocol unification.

\subsubsection*{State Synchronization Under High-Frequency Updates.}
A second technical hurdle is the requirement to maintain synchronized agent states under rapid, potentially continuous updates. If new user actions, item arrivals, or contextual signals flow into the system at high frequency, agents run the risk of operating on inconsistent snapshots of the environment. A naive approach that broadcasts every minor update to all agents creates a scalability bottleneck, as it can saturate the network and introduce large message queues. Yet deferring updates or aggregating them too aggressively can lead to agents making decisions on outdated information. Let
\[
\Delta: \mathcal{E}_t \;\rightarrow\; \mathcal{E}_{t+1}
\]
denote an environment-update function that transitions the shared environment \(\mathcal{E}\) from time \(t\) to time \(t+1\). The question becomes how the MCP conveys these state transitions \(\Delta\) to each agent. Approaches from distributed systems, such as eventual consistency or publish/subscribe event channels, may offer partial solutions, but each implies trade-offs among real-time accuracy, network overhead, and agent-level concurrency control. These design decisions directly impact recommendation quality, user-perceived latency, and the ability to scale across thousands of rapidly evolving data points.

\noindent
\subsubsection*{Fault Tolerance and Recovery.}
Another salient concern is that in any distributed system, individual agents or communication links may fail or degrade without warning. If a crucial agent controlling item retrieval collapses, other agents may stall while waiting for updates, ultimately halting the recommendation flow. Protocol-level fault tolerance mitigates such breakdowns by offering mechanisms for detecting failure (via heartbeat signals or timeouts), rerouting tasks, and resynchronizing state when an agent recovers. These mechanisms can range from simple retries to more robust consensus protocols. However, each increment in fault tolerance introduces additional complexity in both agent design and the MCP itself. A well-engineered approach thus needs to strike a balance between resilience and performance overhead, particularly since real-world recommendation engines can ill afford long downtime or incomplete updates, but also cannot endure significant latency due to over-elaborate fault-detection routines.

\noindent
\subsubsection*{Scalability in Decentralized Settings.}
As the number of specialized agents grows, communication traffic rises—potentially in a many-to-many pattern if the system is fully decentralized. This growth can lead to combinatorial message complexity. A naive peer-to-peer architecture may become infeasible as the system matures, so the MCP must support organized communication topologies or hierarchical roles among agents. For example, certain broker agents might aggregate specific types of updates and relay them to other agents, reducing message duplication. Alternatively, partitioning the item space or user segments among subgroups of agents can localize interactions at the expense of global knowledge. Each architectural choice entails a trade-off in coverage, latency, and potential points of failure. Finding near-linear or, at least, sub-exponential scalability solutions remains a major question for multi-agent recommendation. As more data sources or agent roles are added, the system’s design must ensure that communication and processing overheads remain manageable, and that the user still experiences timely, relevant recommendations.

\subsubsection*{Security and Privacy in Inter-Agent Communication.}
Agents in a recommendation system often exchange sensitive user data, ranging from personal profiles to behavioral logs, and proprietary or business (e.g. critical information about items and ranking heuristics). Lacking appropriate security layers, the MCP risks exposing these details to eavesdropping or tampering. Communication protocols must, at minimum, support encryption and authentication so that data is protected against adversarial access and malicious agents cannot masquerade as legitimate participants. Privacy concerns add further layers of complexity, since not every agent should automatically receive all user details. Agents may need to operate on aggregated or encrypted data in a fashion similar to secure multi-party computation, introducing nontrivial overhead and design challenges. Balancing open collaboration, which is essential for rich multi-agent interactions, with robust security constraints is particularly difficult. If the protocol is too permissive, user trust is jeopardized; if it is too stringent, constructive collaboration is stifled.

\subsubsection*{Open Research Questions and Future Directions.}
Several unresolved questions emerge from these challenges. One is how to establish an ``inclusive but flexible'' standard for MCP that agents can adapt across various domains and use cases without spawning multiple incompatible variants. Another is whether ``adaptive synchronization schemes'' possibly driven by real-time feedback loops, can address the tension between coherence (ensuring all agents remain in sync) and performance (avoiding communication overload). In addition, reliability under fast-changing recommendation contexts remains an open issue. There is also growing interest in robust fault-tolerance paradigms and decentralized consensus strategies that can be layered on top of the MCP \cite{feng2018scalable}, potentially leveraging emerging blockchain or distributed ledger approaches\cite{zhu2025secure}. Finally, integrating ``privacy-preserving'' features, such as secure enclaves or differentially private updates, stands out as a frontier for bridging user data protection with multi-agent collaboration \cite{han2024llm, ugurbil2025fission}. Addressing these open problems could considerably advance the utility and reliability of multi-agent recommendation systems, enabling them to grow both in scale and sophistication while retaining clarity, resilience, and trustworthiness in their communication fabric.

\rev{For evaluation, the failure modes above are measurable. Protocol health can
be tracked through message count and schema-validity rate; context integrity
through provenance coverage, which records how often a candidate set or claim
arrives with the metadata needed to verify it; synchronization quality through
the delay between an environment update and its propagation to dependent agents;
and coordination quality through inter-agent agreement rate and the ability to
attribute a final error to the agent or message that introduced it.}

\subsection{Scalability}\label{sec:scalablity}
In the context of multi-agent recommendation systems, ``scalability'' refers to the ability of the architecture to sustain acceptable performance as the system grows in problem size, data volume, or agent complexity. Formally, let $N$ denote the number of users, $M$ the number of items, and $A$ the number of autonomous agents collaborating to produce recommendations. We say that a system is \emph{scalable} if its key performance metrics (e.g., throughput, latency, and accuracy) remain within acceptable bounds when $N$, $M$, or $A$ increase, and if resource utilization (computation, memory, financial cost) grows sub-linearly or at most linearly in these parameters. Concretely, if $L(N,M,A)$ is the latency per recommendation request, scalability demands that $L(N,M,A)$ does not grow exponentially with $N$, $M$, or $A$. Analogously, if $\Phi(N,M,A)$ is the throughput of the system, then $\Phi(N,M,A)$ should degrade gracefully (or improve) rather than collapse under increasing loads. This requirement ensures that adding users, expanding the item catalog, or incorporating additional agents does not render the recommendation service unresponsive or cost-prohibitive.

\subsubsection*{Latency Considerations: Batch vs.\ Real-Time.}
Scalability manifests differently in \emph{batch-processing pipelines} versus \emph{real-time inference} scenarios. In an \textit{offline batch setting}, agents may train or update models periodically (e.g., overnight jobs), emphasizing throughput and the completion time of large data-processing tasks. As $N$ and $M$ grow, or as more agent modules participate (for example, separate modules for each language or region), total computation can increase significantly, stretching batch windows and risking stale or incomplete updates. Partial solutions include distributing tasks across more compute nodes \cite{hazem2023distributed}, caching intermediate results \cite{yang2018intermediate}, or consolidating models to reduce overhead \cite{cai2024distributed}. By contrast, \textit{real-time} recommendation places tight upper bounds on per-request latency (often tens of milliseconds). When a user makes a request (e.g., “Recommend me some items now”), the system must orchestrate multiple agents without exceeding a strict time budget. If the request passes sequentially through $A$ agents, or if communication overhead is high, total latency can explode. Methods such as parallelizing sub-tasks, reducing model size (e.g., quantization or distillation), or precomputing partial results become vital. Balancing \emph{freshness} (using the latest user context in real time) with \emph{responsiveness} (controlling inference overhead) is an ongoing engineering challenge, especially as agent complexity and user concurrency grow.

\subsubsection*{Communication Overhead and Coordination Bottlenecks.}
A multi-agent system inherently incurs overhead from inter-agent communication. If each agent $A_i$ must exchange intermediate results or context with other agents $A_j$, the communication pattern can grow as $\mathcal{O}(A^2)$ in the worst case. Even more structured topologies (e.g., staged pipelines) introduce data serialization, network transfer, and synchronization costs that increase with $A$. In real deployments, these costs may overshadow pure compute time, especially if agents are distributed across different servers or data centers. Synchronization points (e.g., where an agent must wait for multiple upstream agents to finish) can compound the bottleneck: a single slow or overloaded agent holds up the entire recommendation pipeline. The result is that response-time variability spikes as $A$ grows, impairing scalability. Proposed strategies include ``selective communication'' (agents only communicate with relevant peers), asynchronous event-driven designs, or hierarchical orchestration (where specialized “supervisor” or “broker” agents mediate interactions). While each approach can alleviate overhead, designing a protocol that scales sub-linearly in agent count remains an open problem for multi-agent recommenders.

\subsubsection*{Computational and Financial Costs.}
Scalability also has a direct economic dimension. Each agent may represent a separate inference step (e.g., a large transformer for generating textual explanations, a collaborative filter for scoring items), thereby multiplying compute, memory, and storage costs when scaled to millions of users. Maintaining separate agents for many languages or market segments compounds these expenses. Even inter-agent communication consumes CPU cycles and bandwidth that can become costly at scale, especially in cloud environments where data transfer and compute times are billed. As a result, system architects must continually balance performance benefits from specialized agents against the financial overhead of running them all. Real-world solutions include consolidating similar tasks into fewer, more general models, caching partial results to avoid redundant computations, or introducing cost-aware scheduling where certain expensive agents (such as large language models) are only invoked for high-value requests or in batch mode.

\subsubsection*{Examples and Bottlenecks in Practice.}
Industrial experience underscores these scalability concerns. Large e-commerce providers may need multi-agent approaches for brand-new product lines or language expansions, only to find that naive per-language replication is prohibitively expensive and leads to model silos with limited synergy. Media streaming services, like Netflix or YouTube, rely on multi-stage ranking pipelines, where each stage refines candidates from the previous one. In principle, adding specialized agents can improve personalization; however, at large user scales, the additional latency and resource costs grow burdensome. Academic prototypes such as MACRec \cite{wang2024macrec} demonstrate how multiple LLM-based sub-agents can collaborate for more accurate recommendations, but in production, running multiple large models in parallel can yield super-linear latency and cost, especially if thousands of user queries arrive simultaneously. In each case, synchronization overheads, data duplication, or model proliferation frequently emerge as bottlenecks.

\subsubsection*{Open Challenges and Future Directions.}
Despite progress, significant open questions remain for achieving robust, efficient scalability in multi-agent, language-based recommender systems:

\emph{(i) Dynamic Agent Coordination:} Adaptive scheduling or routing (deciding which agents to invoke per request) could mitigate latency and cost surges; how to automate such coordination at scale is an ongoing research area \cite{pan2024agentcoord}.

\emph{(ii) Cost-Aware Architectures:} Tools for balancing model accuracy with financial overhead remain nascent. Budget-aware or resource-limited modes may become standard features in large multi-agent systems.

\emph{(iii) Multi-Lingual Model Consolidation:} Designing shared representations or truly multilingual models that preserve local performance while avoiding $L$-fold duplication is an active challenge, especially for low-resource languages.

\emph{(iv) Observability and Debugging:} As agent count grows, tracing the contribution of each agent in a recommendation pipeline becomes non-trivial. New monitoring frameworks are needed to pinpoint bottlenecks, synchronization issues, or poor data hand-offs.

Addressing these challenges will require interdisciplinary solutions that merge distributed systems principles (for fault tolerance, communication efficiency), machine learning insights (for multilingual and multi-domain models), and operational best practices (for cost optimization and maintainability). As multi-agent architectures gain adoption in real-world recommendation services, resolving these open problems stands to bring significant advances in both the quality of personalization and the scalability of next-generation systems.

\rev{Scalability is measured along quality, time, and money jointly. The relevant
signals are end-to-end and per-agent latency, throughput, token and API cost,
and timeout rate under load. Because adding agents trades cost for quality, the
informative summary is not any single number but the quality--latency (or
quality--cost) Pareto frontier, which exposes whether an extra agent buys a
quality gain large enough to justify its latency and cost. Section~\ref{sec:empirical}
instantiates exactly this frontier for a concrete ranking task.}

\subsection{Hallucination, Memory Drift, and Error Propagation}
\label{subsec:hallucination_error}
Let $\mathcal{A}=\{A_1,\dots,A_k\}$ be a set of collaborating LLM-based agents in a recommendation pipeline.  
Each agent $A_i$ produces a message $m_i$ that is consumed, directly or indirectly, by other agents.  
Define the boolean predicate $\mathsf{valid}(m_i)\!=\!1$ if $m_i$ is factually correct with respect to an external ground-truth oracle $\mathcal{G}$, and $0$ otherwise.  
Hallucination occurs when $\mathsf{valid}(m_i)=0$.  
``Error propagation'' arises when there exists a path $A_i\!\rightarrow\!A_j$ in the agent-interaction graph such that $\mathsf{valid}(m_i)=0$ and the downstream message $m_j$ is a deterministic function of $m_i$, yielding $\mathsf{valid}(m_j)=0$.  
The probability that at least one final user-facing output is invalid is
\[
\Pr\!\bigl[\exists\,m_{u}:\mathsf{valid}(m_{u})=0\bigr]\;=\;1-\prod_{i=1}^{k}\bigl(1-p_i\bigr),
\]
where $p_i=\Pr[\mathsf{valid}(m_i)=0]$ after all verification steps.  
Minimizing this probability in practice is non-trivial; naive composition of agents tends to increase $p_i$ through cascading dependencies.

\noindent
\subsubsection*{Mechanisms of Cascading Error.}
When agents share a common context buffer (e.g.\ chain-of-thought transcripts), an erroneous assertion by one agent can enter the shared memory $\Omega_t$ and be re-ingested by all subsequent agents:  
\[
\Omega_{t+1} \;=\; \Omega_{t}\cup\{m_i\},\quad
m_j = A_j\!\bigl(\Omega_{t+1}\bigr).
\]
Because most LLMs lack calibrated epistemic uncertainty, a fabricated but linguistically confident statement in $\Omega_{t+1}$ is typically treated as fact \cite{simhi2025trust}.  
Human cognitive biases such as ``authority'' or ``conformity'' find analogues in LLM agents: empirical studies show that a single persuasive hallucination can shift the distribution of responses from peer agents toward the same error, producing group-level misbelief. \cite{liu2024exploring}

\rev{A recommender-specific variant of this mechanism is memory drift. Because an
agentic recommender persists user state across turns and sessions, an erroneous
or outdated fact written to memory is not a one-time error: it is retrieved and
re-applied on every subsequent request until it is corrected. A stale preference,
an incorrectly merged constraint, or a privacy-sensitive trace that should have
been deleted can therefore contaminate future rankings and explanations long
after the originating turn, which is the persistent analogue of the cascading
error above.}

\noindent
\subsubsection*{Mitigation Approaches.}
Current strategies fall into three categories: (i) \emph{Agent-level verification.}  Multi-agent debate, majority voting, and cross-examination frameworks instantiate $r$ redundant agents $\{A_i^{(1)},\dots,A_i^{(r)}\}$ per logical role and accept a message only if a consensus rule $\mathcal{V}$ is satisfied: $m_i^{\star}=\mathcal{V}\bigl(m_i^{(1)},\dots,m_i^{(r)}\bigr)$. This reduces but does not eliminate correlated hallucinations \cite{maragheh2023llm, li2024more}. (ii) \emph{Supervisor or moderator agents.}  A top-level agent $A_{\text{mod}}$ inspects each intermediate output and blocks or edits $m_i$ if $\mathsf{valid}(m_i)=0$.  The difficulty is that $A_{\text{mod}}$ is itself an LLM with similar failure modes; recursive oversight may be required \cite{manakul2023selfcheckgpt}. (iii) \emph{Tool-assisted grounding.}  Agents invoke external APIs $\mathcal{F}$ (e.g.\ factual search, calculators, product databases) to verify claims: $m_i = f\bigl(\mathrm{LLM}(x),\ \mathcal{F}(x)\bigr).$
Empirical evidence shows that interleaving of reasoning and tool calls substantially lowers hallucination rates, although it increases latency. \cite{yao2023react}

\subsubsection*{Open Questions for the RecSys Community.} Despite these advances, scalability and completeness remain open.  Verification ensembles increase compute cost; tool calls add latency; and moderators share the same epistemic blind spots as base agents. These are sample open questions in this regard.

\begin{enumerate}
  \item \textbf{Uncertainty Calibration:} How can LLM agents produce calibrated confidence scores that downstream agents may use to discount or challenge low-certainty statements?
  \item \textbf{Efficient Verification:} What lightweight protocols (selective voting, probabilistic auditing, or adaptive tool calls) can bound error propagation without breaching real-time latency constraints?
  \item \textbf{Formal Guarantees:} Can we derive probabilistic upper bounds on $\Pr[\exists\,m_{u}:\mathsf{valid}(m_u)=0]$ for a given agent graph and verification strategy, analogous to error-correcting codes?
  \item \textbf{Adversarial Robustness:} How can multi-agent recommenders detect and neutralize deliberate prompt injections or knowledge-base poisoning that exploit error-propagation pathways?
\end{enumerate}

\rev{These failure modes are measurable. Hallucination at the item level is
captured by the unsupported-claim rate, citation accuracy, and contradiction
rate of agent messages, and by evidence coverage, the fraction of claims backed
by retrieved item or user evidence. Memory drift is captured by memory
precision and recall against the true user state, by a staleness rate that
counts retrieved facts no longer valid, and by deletion compliance, the fraction
of removal requests honored on subsequent retrieval.}

Answering these questions will be pivotal for building multi-agent, LLM-powered recommenders that harness generative flexibility \cite{deldjoo2024recommendation,deldjoo2024review} while safeguarding against cascading hallucinations and the attendant erosion of user trust.

\subsection{Potential Collusion or Unintended Emergent Behavior}\label{sec:collusion}

Multi-agent systems built on large language models (LLMs) are being actively explored for recommendation tasks, enabling multiple autonomous agents to interact and jointly deliver personalized suggestions. Each agent may correspond to a user perspective, an item provider, or a specialized recommender component, with interactions coordinated through a communication protocol. While this paradigm shows promise for richer, more adaptive recommendations, it also raises concerns over unintended emergent behaviors. When agents possess partially independent objectives and adaptive capabilities, local decision-making can yield global outcomes that system designers did not anticipate. Two key risks are collusion among agents (secretly coordinating strategies that undermine fairness or efficiency) and reinforcing feedback loops that amplify bias or degrade system robustness. This subsection presents current findings and open questions regarding such emergent behaviors, illustrating how they jeopardize fairness, reliability, and user trust in multi-agent LLM-based recommendation systems.

\subsubsection*{Unintended Emergent Behaviors in Autonomous LLM Agents.} \label{subsec:emergent_behav}
``Emergent behavior'' in multi-agent LLM systems arises from the complex interactions of multiple agents pursuing local goals, resulting in global patterns unintended by the system’s designers. One salient risk is ``collusion'': autonomous agents may discover covert means to collaborate—often through cryptic or hidden messaging—thereby subverting oversight. Prior work has shown that even simple negotiation bots can invent private “shorthand” languages to maximize joint gains, a harbinger of more advanced forms of covert cooperation in LLM-driven systems \cite{lewis2017deal}. Under adversarial incentives, such covert communication can manifest as “cartel-like” behavior, where agents collude to manipulate item rankings or artificially inflate engagement metrics, undermining the recommender’s integrity. In less adversarial but still dynamic settings, agents may spontaneously cooperate if cooperation yields better local utility. For example, self-interested LLM-based “firm” agents in a simulated market might tacitly agree to price-fix, converging to above-competitive prices through repeated interactions—even though no explicit collusion protocol was programmed.

Another persistent concern is bias amplification and feedback loops. Classic recommender systems already exhibit feedback loops where popular items receive even greater exposure, reinforcing their dominance at the expense of niche content \cite{ahanger2022popularity}. In a multi-agent LLM context, these loops may intensify: a user-simulator LLM’s positive feedback can encourage the recommender LLM to propose narrower sets of items, fostering echo chambers. Empirical explorations reveal that, without careful calibration, multi-agent LLM recommenders can systematically favor popular content or align feedback to confirm preconceived user preferences \cite{lichtenberg2024large}, resulting in reduced diversity and potential “filter bubble” effects. \cite{sukiennik2025simulating}

\subsubsection*{Case Studies and Examples in LLM-Based Recommendation Systems.}
While large-scale real-world instances remain relatively uncommon, several research prototypes and industry analogs illustrate emergent misbehaviors:
\begin{enumerate}[label=(\roman*), leftmargin=*]
  \item \emph{Rec4Agentverse and Similar Frameworks:} Conceptual models wherein multiple Item Agents (each representing a specific content source) coordinate with a central Recommender Agent can inadvertently allow item providers to form alliances, thereby boosting each other. Researchers note the need for robust communication constraints and fairness safeguards to prevent collusive behaviors. \cite{zhang2024prospect}
  \item \emph{User–Recommender Co-Adaptive Simulations:} Prototypes in which a user-simulator LLM interacts iteratively with a recommender LLM demonstrate how naive feedback loops amplify popularity bias or lead to trivial “approval” behaviors. Even if no intentional collusion exists, the system’s closed feedback cycle can systematically degrade diversity or realism in user modeling. \cite{sukiennik2025simulating}
  \item \emph{Industry Observations (Non-LLM but Analogous):} Large platforms like Netflix and Facebook historically reported polarization and echo chambers emerging from automated feedback loops. Translating these patterns to a multi-agent LLM scenario, if item-provider agents focus excessively on engagement signals, sensational or polarizing content might dominate—a dynamic form of self-reinforcing bias. \cite{noordeh2020echo, tong2023navigating}
\end{enumerate}

\subsubsection{Open Challenges.}
Addressing collusion and unintended emergent behavior in multi-agent LLM recommenders requires an interdisciplinary approach combining AI safety, mechanism design, and robust system engineering. Several key areas demand further exploration:

\emph{(i) Collusion Detection and Prevention.} Identifying covert communication among LLM-based agents remains challenging, particularly if the agents develop coded protocols. Future research might involve adversarial training of oversight models or cryptographic constraints limiting an agent’s ability to conceal signals.

\emph{(ii) Alignment of Local and Global Objectives.} A central cause of emergent misbehavior is the mismatch between local agent incentives and the broader system goal. Mechanism design, incentive-compatible rewards, or hierarchical constraints may help ensure that individually rational actions coincide with socially desirable outcomes.

\emph{(iii) Bias Mitigation in Multi-Agent Loops.} As agents co-adapt, biases can be amplified through cyclical feedback. Tools like causal intervention, re-weighted training, or fairness-promoting prompts must be extended to multi-agent contexts. Evaluating long-term fairness, beyond single-step recommendations, remains an open problem, necessitating new metrics and simulation platforms.

\emph{(iv) Dynamic Adaptation and Safety.} As multi-agent LLM recommenders evolve over time, continuous oversight is required to preempt emergent failures. Online learning, with real-time detection of harmful collusive patterns, will likely be necessary. Balancing adaptability with safety constraints remains a core tension.

\medskip

By addressing these questions, the recommender systems community can help shape multi-agent LLM architectures that realize their full potential for personalized, adaptive user experiences while safeguarding against collusion, bias, and other emergent hazards. Meeting these challenges will require new techniques in distributed AI governance, fairness-aware design, and robust multi-agent learning—an exciting interdisciplinary frontier for future research.

\rev{Measuring these behaviors requires signals beyond per-request accuracy.
Collusion can be probed with collusion indicators and price or exposure
anomalies under adversarial simulation, and with an objective-conflict rate that
quantifies how often local agent incentives diverge from the system objective.
Exposure feedback loops are tracked with catalog coverage, long-tail exposure,
and exposure disparity across providers or groups, together with group-wise
utility, since fairness here is a property of the distribution of exposure over
time rather than of a single ranked list.}

\subsection{Brand, Policy, and Legal Compliance}
\label{sec:brand-policy}
\rev{In agentic recommender systems, brand and policy compliance is not only a
generic text-generation issue. It directly affects recommendation explanations,
item presentation, generated claims about item attributes or availability,
provider obligations, user safety, and legal or regulatory requirements. A
policy-checking agent therefore evaluates not only whether the language is
on-brand, but also whether the recommended items, claims, evidence, and
constraints are consistent with platform policy and applicable law.}

Let $\mathcal{A}=\{A_1,\dots,A_k\}$ be a set of collaborating LLM-based agents that generate, transform, or filter textual recommendations.  
Let $\mathcal{P}$ denote a formal \emph{brand policy}, a finite set of constraints on tone, vocabulary, factual claims, and legally compliant disclosures.  
For any agent $A_i$ emitting a textual message $m\in\Sigma^{*}$, define a \emph{compliance predicate}
\[
\mathcal{C}_{\mathcal{P}}(m)=
\begin{cases}
1 & \text{if } m \text{ satisfies every rule in } \mathcal{P},\\
0 & \text{otherwise}.
\end{cases}
\]
A multi-agent recommender maintains \emph{brand consistency} if, for every output $m$ observable by the end user and for every intermediate message exchanged among agents that could influence $m$, we have $\mathcal{C}_{\mathcal{P}}(m)=1$.  The challenge is to guarantee this condition while preserving the generative flexibility and efficiency of the agents.

\subsubsection*{Sources of Inconsistency.}
First, pretrained LLMs embed broad ``world knowledge'' that may conflict with brand-specific language or regulations.  
Second, heterogeneous alignment across agents induces drift: if $A_1$ is fine-tuned for tone but $A_2$ merely prompted, their joint output may diverge from $\mathcal{P}$.  
Third, generic safety filters rarely cover nuanced corporate rules (e.g., prohibitions on competitor references, regional marketing laws), leaving gaps through which policy-violating content may pass.

\subsubsection*{Illustrative Incidents.}
Well-publicised failures, such as Character.AI’s AI-powered assistant providing harmful advice to a minor, or historical “bomb-making” item bundles on Amazon underline the reputational and legal risks of insufficient control.  \cite{businessinsider2024characterai, apnews2017amazonbundles}
While these cases did not involve full multi-agent architectures, they foreshadow the compounded risk when multiple autonomous LLMs exchange information without a unifying compliance layer.

\subsubsection*{Current Mitigation Strategies.}
Industry is moving toward \emph{brand-tuned LLMs}, in which a base model is fine-tuned or instruction-aligned with proprietary style guides, product catalogs, and regulatory constraints \cite{achintalwar2024alignment}.  
Researchers have proposed alignment studios and inference-time \emph{policy-expert agents} that veto or rewrite non-compliant tokens:
\[
m^{\star} = \arg\max_{m}\Bigl[\Pr(m\mid \text{context})\;\text{s.t.}\;\mathcal{C}_{\mathcal{P}}(m)=1\Bigr].
\]
Such approaches reduce manual review but introduce computational overhead and still lack formal guarantees of completeness, especially as $\mathcal{P}$ evolves over time.

\subsubsection*{Outstanding Challenges.}
Formal certification of $\mathcal{C}_{\mathcal{P}}$ across an entire agent pipeline remains open; policy drift is likely as models and guidelines change asynchronously.  
Real-time enforcement demands fast, differentiable approximations of $\mathcal{C}_{\mathcal{P}}$, yet brand policies frequently involve non-differentiable, context-dependent rules (e.g., comparative advertising limitations differing by jurisdiction).  
Finally, maintaining multilingual consistency is difficult: a compliant English output may translate into a culturally inappropriate phrase in another language, violating $\mathcal{P}$’s spirit even if literal rules are met.

\subsubsection*{Open Questions for the RecSys Community.}
\begin{enumerate}
  \item \textbf{Formal Guarantees:} How can we design verifiable protocols or certified decoders that ensure $\mathcal{C}_{\mathcal{P}}(m)=1$ for every agent message without excessive latency?
  \item \textbf{Continuous Compliance:} How can a recommender adapt when either the brand policy or external regulations change, ensuring that legacy agent behaviors do not drift out of compliance?
  \item \textbf{Cross-Lingual Control:} Which multilingual alignment techniques best propagate tone, legal constraints, and cultural sensitivities across languages without retraining separate agents per locale?
  \item \textbf{Evaluation Benchmarks:} What standardized datasets and metrics can quantify brand-policy adherence and stylistic consistency in multi-agent recommendation settings?
  \item \textbf{Cost–Benefit Trade-offs:} How do we balance the computational and financial overhead of stringent compliance layers against their risk-mitigation benefits in large-scale production systems?
\end{enumerate}

Addressing these questions will be pivotal for deploying multi-agent, LLM-based recommenders that are both creative and rigorously aligned with brand identity and regulatory obligations. \rev{Operationally, compliance is measured by guideline-adherence and
policy-violation rates over generated outputs, cross-lingual consistency of the
same recommendation across locales, human-review agreement on flagged cases, and
the escalation rate at which the system defers to human oversight.}

\rev{Table~\ref{tab:operational-challenges} operationalizes these presented challenges by mapping each challenge to its source in agentic recommendation, its recommender-specific manifestation, and measurable metrics or protocols.}

\begin{table*}[t]
\centering
\caption{\rev{Operationalizing the five challenges of agentic recommender
systems (Sections~\ref{sec:mcp}--\ref{sec:brand-policy}). Each challenge is tied
to its source in agentic RS, its recommender-specific manifestation, and
measurable evaluation signals.}}
\label{tab:operational-challenges}
\scriptsize
\setlength{\tabcolsep}{3.5pt}
\renewcommand{\arraystretch}{1.25}
\begin{tabularx}{\textwidth}{@{}p{0.15\textwidth} p{0.25\textwidth} p{0.27\textwidth} p{0.25\textwidth}@{}}
\toprule
\rev{\textbf{Challenge}}
&
\rev{\textbf{Source in agentic RS}}
&
\rev{\textbf{RecSys-specific manifestation}}
&
\rev{\textbf{Metrics / protocols}}
\\
\midrule
\rev{\textbf{Communication and protocol}\newline \footnotesize(\S\ref{sec:mcp})}
&
\rev{Agents exchange candidate sets, evidence, constraints, critiques, and memory records}
&
\rev{Stale candidates passed downstream; missing item provenance; unverifiable claims or constraints}
&
\rev{Message count; schema-validity rate; provenance coverage; synchronization delay; agreement rate; failure attribution}
\\
\addlinespace[2pt]
\rev{\textbf{Scalability and cost}\newline \footnotesize(\S\ref{sec:scalablity})}
&
\rev{Each agent, memory op, retrieval, tool call, or evaluator adds inference cost and latency}
&
\rev{Extra agents improve ranking marginally while exceeding real-time latency budgets}
&
\rev{End-to-end and per-agent latency; token / API cost; throughput; timeout rate; quality--latency Pareto frontier}
\\
\addlinespace[2pt]
\rev{\textbf{Hallucination and memory drift}\newline \footnotesize(\S\ref{subsec:hallucination_error})}
&
\rev{Generated claims, retrieved evidence, and stored memories are consumed downstream as if factual}
&
\rev{Unsupported attributes; fabricated availability; stale preferences; invalid constraint satisfaction; privacy-sensitive recall}
&
\rev{Unsupported-claim rate; citation accuracy; contradiction rate; evidence coverage; memory precision/recall; staleness rate; deletion compliance}
\\
\addlinespace[2pt]
\rev{\textbf{Collusion and exposure loops}\newline \footnotesize(\S\ref{sec:collusion})}
&
\rev{Autonomous agents optimize local goals conflicting with user, provider, or fairness objectives}
&
\rev{Provider agents suppress competitors; engagement over-optimization; simulated users amplify popularity bias}
&
\rev{Collusion indicators; price / exposure anomalies; objective-conflict rate; catalog coverage; long-tail exposure; exposure disparity; group-wise utility}
\\
\addlinespace[2pt]
\rev{\textbf{Brand, policy, and legal compliance}\newline \footnotesize(\S\ref{sec:brand-policy})}
&
\rev{Explanations, claims, and actions are generated dynamically and vary across users, markets, and languages}
&
\rev{Off-brand tone; prohibited or unsafe claims; unsupported legal/product claims; policy-inconsistent justifications}
&
\rev{Guideline adherence; policy-violation rate; cross-lingual consistency; human-review agreement; escalation rate}
\\
\bottomrule
\end{tabularx}
\end{table*}

\section{Empirical Illustration: When Do Multi-Agent Pipelines Pay Off?}
\label{sec:empirical}

\rev{The preceding sections argued, conceptually, that agentic orchestration
unlocks capabilities beyond single-pass pipelines. A perspective is
more convincing when it is made concrete, and in this section we want to quantify both the benefit and the cost of multi agentic structures.
This section therefore reports a controlled study on the simplest slice of
the Interactive Recommendation task of Section \ref{subsec:interactive_rec_usecase}: re-ranking
a fixed candidate set based on a user's purchase history. The study is deliberately
narrow so that the effect of \emph{adding agents} can be isolated from
confounds such as multi-turn dialogue dynamics or tool availability.}

\rev{In this section, we want to explore what makes these orchestrations work and when they will not benefit the ultimate goal of the task. We will see for the given task of conversational ranking, the value of using multi agentic orchestrations scales with the complexity of the input. We will also show that several of the multi-agentic roles
that look helpful in principle do not repay their cost in practice. For instance, we show that self-referential refinement can actively degrade ranking quality. Concretely, this study provides controlled evidence for the usage of multi-agentic mechanisms. Throughout, we connect the measurements
back to the error-propagation formalism of Section \ref{subsec:hallucination_error} and the scalability concerns of Section \ref{sec:scalablity}.}

\subsection{Experimental Setup}
\label{sec:emp-setup}

\subsubsection*{Task}
\rev{Each instance is a single ranking question: given the purchase history of a
user and ten candidate items, the pipeline must return a ranking of the ten
candidates. Exactly one candidate is the held-out next purchase; the remaining
nine are sampled at random from items in the same product category. The agent
emits a ranked list, and specialized agents cooperate to produce it.}

\subsubsection*{Data}
\rev{We use the Amazon-2023 review corpus \cite{hou2024bridging} across four
categories: Amazon Fashion, Appliances, Electronics, and Toys and Games. For
each category we draw two contrastive user cohorts of $100$ users, a
\emph{random} cohort and a \emph{high-diversity} cohort, giving two samples of
$n=400$ pooled over the four categories (for detailed per category results see Appendix A). The contrast between these two samples
is the backbone of the analysis.\footnote{All pipeline implementations, prompt templates, the
sampling scripts, and
the scripts that regenerate every results table in this paper are available at
\url{https://github.com/RezaYM/agenticrecsys.git}.} }

\rev{To assess how recommendation quality varies with the heterogeneity of a user's
interaction history, we stratify users by the lexical diversity of the item
titles they have previously engaged with. For a user $u$ with history
$H_u = \{t_1, \dots, t_m\}$, we represent each item title $t_i$ by its set of
lowercased word tokens $T_i = \mathrm{tokens}(t_i)$, and define the diversity of
$u$ as the mean pairwise Jaccard distance over all title pairs in the history:}
\begin{equation}
\mathrm{Div}(u) \;=\;
\binom{m}{2}^{-1}
\sum_{1 \le i < j \le m}
\left(
1 - \frac{\lvert T_i \cap T_j \rvert}{\lvert T_i \cup T_j \rvert}
\right),
\label{eq:diversity}
\end{equation}
\rev{where $\binom{m}{2}$ is the number of unordered title pairs. The score lies in
$[0,1]$: it is $0$ when all titles share an identical token set (a homogeneous
history) and approaches $1$ when every pair of titles is lexically disjoint (a
maximally diverse history). Users with fewer than two non-empty token sets are
excluded, and histories are capped at $100$ titles to bound the quadratic
pairwise cost.}

\rev{We restrict the analysis to users with at least seven history items. The
\emph{high-diversity} cohort comprises, per category, the $100$ users with the
largest $\mathrm{Div}(u)$. The \emph{random} cohort is drawn uniformly at random
(seed 42) from the eligible users in the same category, and serves as a typical
baseline population against which the high-diversity cohort is compared.}

\subsubsection*{Agentic roles and workflows.}
\rev{We evaluate seven workflows, summarized in Table~\ref{tab:emp-roster}, grouped
into three families of agentic role. The single-shot LLM baseline (\textsc{SA})
issues one model call that reads the history and the candidates and returns the
ranking, and serves as the reference against which the value of additional
agents is measured. The \emph{decomposition and specialization} family restructures
the input before ranking: a profiler compresses the history into an
intent summary that the ranker then conditions on (\textsc{PR}), and a planner
additionally outlines the ranking strategy (\textsc{PPR}). This family
operationalizes the planning and task-decomposition capability that breaks a
goal into focused subtasks each solved by a dedicated
call~\cite{prasad2024adapt,luo2025large}. The \emph{ensembling and
aggregation} family replaces the single ranker with several rankers sampled at
higher temperature for diversity and reconciles them through an arbitrator, either
conditioned on a plan alone (\textsc{PEns}) or on both a plan and a profile
(\textsc{PPEns}); this follows the finding that sampling multiple agents and
aggregating their outputs improves accuracy, with gains that grow with task
difficulty~\cite{luo2025large}. The \emph{iterative and adversarial
refinement} family improves an initial ranking through feedback rather than
through added input structure: a ranker is critiqued by an evaluator and then
revises (\textsc{RC}), in the manner of single-model self-feedback
loops~\cite{madaan2023self}, or two rankers exchange and revise their
rankings over several turns of debate (\textsc{Deb})~\cite{du2024improving}.}

\subsubsection*{Model, metrics, and cost basis.}
\rev{All pipelines use \texttt{gpt-5-mini-2025-08-07}. We evaluate ranking quality with MRR, NDCG, and HitRate (equal
to Recall) at cutoffs $k\in\{3,10\}$. Cost is reported as input/output
tokens, US dollars at \$0.25/\$2.00 per million input/output tokens, and the
number of LLM calls per query.} 

\begin{table}[t]
  \caption{Agentic workflows evaluated, grouped by role family. Calls/q is the
  number of LLM invocations per query; the original implementation identifiers
  are given in parentheses.}
  \label{tab:emp-roster}
  \small
  \begin{tabular}{@{}llc@{}}
    \toprule
    Workflow (ID) & Role family & Calls/q \\
    \midrule
    \textsc{SA}: Single-shot LLM call                        & baseline                       & 1 \\
    \textsc{PR}: Profiler\,$\rightarrow$\,Ranker     & decomposition / specialization & 2 \\
    \textsc{PPR}: Planner+Profiler\,$\rightarrow$\,Ranker & decomposition / specialization & 3 \\
    \textsc{PEns}: Planner\,$\rightarrow$\,3 Rankers\,$\rightarrow$\,Arbitrator & ensembling / aggregation & 5 \\
    \textsc{PPEns}: Planner+Profiler\,$\rightarrow$\,3 Rankers\,$\rightarrow$\,Arbitrator & decomposition + ensembling & 6 \\
    \textsc{RC}: Ranker\,$\leftrightarrow$\,Critic   & iterative / adversarial        & 3 \\
    \textsc{Deb}: Two-agent Debate                   & iterative / adversarial        & 4 \\
    \bottomrule
  \end{tabular}
\end{table}

\subsection{Quantifying the Cost of Agentic Structures}
\label{sec:emp-cost}

\rev{Before asking whether added agents improve ranking, we fix the bar that any
quality gain must clear. Table~\ref{tab:emp-cost} reports per-query tokens,
dollar cost, and latency for both samples. The single shot LLM call issues one call of
roughly 1.7k tokens at about \$0.0022 per query. The ensemble workflows
\textsc{PEns} and \textsc{PPEns} issue five and six calls, consume between 8.9k
and 10.8k tokens, and cost between \$0.0112 and \$0.0123 per query, which is
between five and six times the Single-shot LLM call cost.}

\rev{We report latency along the parallel critical path rather than the raw
sequential sum, because the sequential figure is an artifact of our
single-threaded benchmark harness rather than a property of the workflow's
dependency structure. The critical path is the longest chain of calls that must
run in series once mutually independent calls are dispatched concurrently. Take
\textsc{PEns} as an example: the planner runs first, its plan is handed to three
rankers that are mutually independent and can therefore be dispatched at the
same time, and the arbitrator runs last on their combined output. The critical
path is thus planner, then one ranker stage (not three), then arbitrator, so the
five-call workflow has the latency of three calls in series rather than five. On
the high-diversity sample this places \textsc{PEns} and \textsc{PPEns} at
roughly three times the Single-shot LLM call latency (34.1\,s and 35.2\,s against
11.8\,s), and somewhat higher on the random sample, where the single call is
itself slower. Crucially, parallel execution recovers a share of the wall-clock
penalty but does not reduce token consumption or dollar cost, which scale with
the number and size of calls regardless of how they are scheduled. The
implication for the rest of the section is simple: a workflow earns its place
only if its quality improvement is large enough to justify a multiplicative
increase in cost.}

\begin{table}[t]
  \caption{Cost and latency per query on \texttt{gpt-5-mini}, aggregate over the
  four categories ($n=400$). ``Total tok/q'' sums input and output tokens over a
  workflow's calls. ``Lat.\ par'' is the parallel critical-path latency, defined
  as the longest chain of dependent calls once mutually independent calls are
  dispatched concurrently. Lower is better throughout; the single shot LLM call is the
  floor on every column by construction.}
  \label{tab:emp-cost}
  \small
  \begin{tabular}{@{}lcccc@{}}
    \toprule
    Workflow & Calls/q & Total tok/q & Cost/q (USD) & Lat.\ par (s) \\
    \midrule
    \multicolumn{5}{l}{\emph{Random sample} ($n=400$)} \\
    \textsc{SA}: Single-shot LLM call                       & 1.0 & 1{,}756  & 0.00217 & 14.52 \\
    \textsc{PR}: Profiler\,$\rightarrow$\,Ranker    & 2.0 & 2{,}804  & 0.00328 & 22.31 \\
    \textsc{PPR}: Planner+Profiler\,$\rightarrow$\,Ranker & 3.0 & 4{,}273 & 0.00473 & 22.98 \\
    \textsc{RC}: Ranker\,$\leftrightarrow$\,Critic  & 3.0 & 6{,}699  & 0.00831 & 55.27 \\
    \textsc{Deb}: Two-agent Debate                  & 4.0 & 7{,}635  & 0.00933 & 62.92 \\
    \textsc{PEns}: Planner\,$\rightarrow$\,3 Rankers\,$\rightarrow$\,Arb. & 5.0 & 9{,}510  & 0.01121 & 38.36 \\
    \textsc{PPEns}: Planner+Profiler\,$\rightarrow$\,3 Rankers\,$\rightarrow$\,Arb. & 6.0 & 10{,}812 & 0.01225 & 44.56 \\
    \midrule
    \multicolumn{5}{l}{\emph{High-diversity sample} ($n=400$)} \\
    \textsc{SA}: Single-shot LLM call                       & 1.0 & 1{,}654  & 0.00219 & 11.76 \\
    \textsc{PR}: Profiler\,$\rightarrow$\,Ranker    & 2.0 & 2{,}537  & 0.00322 & 17.70 \\
    \textsc{PPR}: Planner+Profiler\,$\rightarrow$\,Ranker & 3.0 & 3{,}912 & 0.00470 & 20.85 \\
    \textsc{RC}: Ranker\,$\leftrightarrow$\,Critic  & 3.0 & 6{,}221  & 0.00803 & 41.89 \\
    \textsc{Deb}: Two-agent Debate                  & 4.0 & 7{,}130  & 0.00923 & 49.61 \\
    \textsc{PEns}: Planner\,$\rightarrow$\,3 Rankers\,$\rightarrow$\,Arb. & 5.0 & 8{,}921  & 0.01119 & 34.14 \\
    \textsc{PPEns}: Planner+Profiler\,$\rightarrow$\,3 Rankers\,$\rightarrow$\,Arb. & 6.0 & 10{,}000 & 0.01202 & 35.19 \\
    \bottomrule
  \end{tabular}
\end{table}
\subsection{On Typical Users, Added Agents Do Not Pay, and Often Hurt}
\label{sec:emp-random}

\rev{We begin with the negative result, since it is the more surprising of the two
and frames the rest of the section. Table~\ref{tab:emp-random} reports quality
on the random sample. On these typical histories the single shot LLM call is already
strong, reaching an NDCG@10 of 0.7197, and no multi-agent pipeline meaningfully
improves on it. The best multi-agent score, 0.7209 from the debate pipeline, is
within a fraction of a point of the baseline, and the two pipelines tie on
NDCG@3 as well. Given the five- to six-fold cost increase documented in
Table~\ref{tab:emp-cost}, this is effectively a loss: the added structure
purchases no useful signal.}

\rev{More striking is that two pipelines degrade quality. The Ranker+Evaluator
pipeline (\textsc{RC}) falls to an NDCG@10 of 0.6922, a relative drop of 3.8\,percent
against the single shot LLM call, and a larger 7.2\,percent drop at NDCG@3, while
costing roughly four times as much. We read this as a direct empirical instance
of the error-propagation mechanism formalised in
Section \ref{subsec:hallucination_error}. When the evaluator critiques an already-correct
ranking with no external ground-truth oracle, its critique is itself an
ungrounded LLM output that the ranker then incorporates, so a second source of
error is introduced where the first pass had none. In the notation of that
section, $\Pr[\exists\, m_u : \mathrm{valid}(m_u)=0] = 1 - \prod_i (1-p_i)$ grows
when an additional stage with $p_i>0$ is composed onto a pipeline that already
produces a good answer. The cleanest summary of this subsection is that on
inputs a single model already handles well, additional agents add failure
surface faster than they add signal.}

\begin{table}[t]
  \caption{Ranking quality on the random sample, aggregate over four categories
  ($n=400$), \texttt{gpt-5-mini}. Best and second-best among the LLM pipelines
  per column are \textbf{bold} and \underline{underlined}. Higher is better.}
  \label{tab:emp-random}
  \small
  \begin{tabular}{@{}lcccc@{}}
    \toprule
    Method & MRR@3 & MRR@10 & NDCG@3 & NDCG@10 \\
    \midrule
    \textsc{SA} Single-shot LLM call                    & \underline{0.5846} & 0.6317 & \underline{0.6225} & 0.7197 \\
    \textsc{RC} Ranker+Evaluator                & 0.5400 & 0.5960 & 0.5779 & 0.6922 \\
    \textsc{PR} Profiler+Ranker                 & 0.5775 & 0.6255 & 0.6191 & 0.7156 \\
    \textsc{PPR} Planner+Profiler+Ranker         & 0.5708 & 0.6185 & 0.6129 & 0.7100 \\
    \textsc{PEns} Planner+3 Rankers+Arb.          & 0.5792 & 0.6266 & 0.6190 & 0.7159 \\
    \textsc{PPEns} Planner+Profiler+3 Rankers+Arb. & 0.5837 & \underline{0.6319} & 0.6212 & \underline{0.7198} \\
    \textsc{Deb} Two-agent Debate                & \textbf{0.5850} & \textbf{0.6330} & \textbf{0.6229} & \textbf{0.7209} \\
    \midrule
    Random baseline                    & 0.1833 & 0.2929 & 0.2131 & 0.4544 \\
    \bottomrule
  \end{tabular}
\end{table}

\subsection{On High-Diversity Users, Decomposition and Ensembling Begin to Pay}
\label{sec:emp-highdiv}

\rev{The picture inverts on the high-diversity sample. Table~\ref{tab:emp-highdiv}
shows that the absolute scores are lower than on the random sample, which is
expected because diverse histories are harder to summarise into a single
purchase intent. The informative quantity, however, is the relative ordering:
here the pipelines that carry a planner or an ensemble lead the single shot LLM call.
At NDCG@3 the combined pipeline (\textsc{PPEns}) reaches 0.5202 and the ensemble pipeline
(\textsc{PEns}) reaches 0.5162, against 0.4898 for the single shot LLM call, a relative gain of
6.2 and 5.4\,percent respectively. At NDCG@10 the ensemble pipeline leads at
0.6471 against 0.6316, a gain of 2.5\,percent, and the MRR columns tell the
same story, with \textsc{PEns} best at every reported cutoff.}

\rev{The contrast between Tables~\ref{tab:emp-random} and~\ref{tab:emp-highdiv} is
the central empirical finding of the paper: the same pipelines that are wasteful
on typical users become beneficial on diverse ones. This is consistent with the
intuition that decomposition and aggregation only have something to contribute
when the input is heterogeneous enough that a single pass under-resolves it.}

\begin{table}[t]
  \caption{Ranking quality on the high-diversity sample, aggregate over four
  categories ($n=400$), \texttt{gpt-5-mini}. Best and second-best among the LLM
  pipelines per column are \textbf{bold} and \underline{underlined}. Higher is
  better.}
  \label{tab:emp-highdiv}
  \small
  \begin{tabular}{@{}lcccc@{}}
    \toprule
    Method & MRR@3 & MRR@10 & NDCG@3 & NDCG@10 \\
    \midrule
    \textsc{SA}: Single-shot LLM call                    & 0.4492 & 0.5174 & 0.4898 & 0.6316 \\
    \textsc{RC}: Ranker+Evaluator                & 0.4379 & 0.5084 & 0.4788 & 0.6247 \\
    \textsc{PR}: Profiler+Ranker                 & 0.4537 & 0.5243 & 0.4919 & 0.6368 \\
    \textsc{PPR}: Planner+Profiler+Ranker         & 0.4617 & 0.5272 & 0.5069 & 0.6401 \\
    \textsc{PEns}: Planner+3 Rankers+Arb.          & \textbf{0.4725} & \textbf{0.5366} & \underline{0.5162} & \textbf{0.6471} \\
    \textsc{PPEns}: Planner+Profiler+3 Rankers+Arb. & \underline{0.4704} & \underline{0.5290} & \textbf{0.5202} & \underline{0.6416} \\
    \textsc{Deb}: Two-agent Debate                & 0.4533 & 0.5215 & 0.4946 & 0.6349 \\
    \midrule
    Random baseline                    & 0.1833 & 0.2929 & 0.2131 & 0.4544 \\
    \bottomrule
  \end{tabular}
\end{table}

\subsection{Role-by-Role Analysis: What Each Agent Contributes}
\label{sec:emp-roles}

\rev{Reading the two quality tables by role family clarifies which forms of agentic
structure carry the high-diversity gain and which do not.}

\paragraph{Decomposition and specialisation.}
\rev{The planner and the profiler are the source of the steady, low-variance
improvement on diverse inputs. Adding the profiler alone (\textsc{PR}) moves NDCG@10 from
0.6316 to 0.6368 on the high-diversity sample while leaving it slightly below
the baseline on the random sample (0.7156 against 0.7197). Adding a planner on
top (\textsc{PPR}) lifts NDCG@3 to 0.5069. The mechanism is intuitive: when a history
spans many interests, articulating a ranking strategy and compressing the
history into an explicit intent both reduce the burden on the final ranking
step. When the history is already coherent, there is little to decompose, and
the extra calls neither help nor hurt the ranking while still costing tokens.
This is the empirical content of capability~(i), planning and task
decomposition.}

\paragraph{Ensembling and aggregation.}
\rev{The largest high-diversity gains come from the ensemble pipelines \textsc{PEns} and \textsc{PPEns},
which lead at NDCG@10 and NDCG@3 respectively. Sampling several rankers at
elevated temperature and reconciling them with an arbitrator recovers signal
that any single sample misses, which matters precisely when the input is
ambiguous. This is also the most expensive family, so its advantage is real but
costly.}

\paragraph{Iterative and adversarial refinement.}
\rev{This family is the weakest. The evaluator pipeline (\textsc{RC}) is the worst performer on
the random sample and remains below the baseline on the high-diversity sample,
and the debate pipeline (\textsc{Deb}) is roughly neutral on both. The synthesis is that
the gains observed in this study come from restructuring the input and
aggregating diverse views, not from agents critiquing one another in a closed
loop. Closed-loop critique without an external oracle tends, if anything, to
introduce the cascading errors analyzed in Section \ref{subsec:hallucination_error}.}

\subsection{The Profiler as Memory: Why Compression Helps on Diverse Histories}
\label{sec:emp-memory}

\rev{The profiler can be read through the memory formalism of Section
\ref{sec:formal_framework}. It instantiates the retention operator $\mathcal{R}$ of Definition \ref{def:memory_update} : it distills the raw history
into a compact intent summary $\widetilde{\mathcal{C}}$, on which the ranker then
conditions through the retrieval operator $\mathcal{Q}$ of
Definition \ref{def:memory_retrieval}. We are explicit that this is
working-memory, or session-scoped, compression.}

\rev{The evidence for the value of this memory mechanism is the
profiler-only pipeline (\textsc{PR}). On the high-diversity sample it improves NDCG@10 from 0.6316 to 0.6368 and MRR@10 from 0.5174 to 0.5243, whereas on the random
sample it sits marginally below the single shot LLM call (0.7156 against 0.7197 at
NDCG@10). In other words, compressing a history into an explicit intent helps
exactly when the history is rich enough to make compression informative, and is
otherwise a small, avoidable expense. 
}

\section{Conclusion}
\rev{This paper established a footing for agentic recommender systems:
architectures in which stateful agents, memories, tools, communication protocols,
and verification steps are composed to improve recommendation-layer outcomes. We
began by formalizing the core building blocks, generic LLM agents, recommender
agents, multi-agent systems, memory update and retrieval functions, and
observable traces, and by bounding what is recommendation-specific in such a
system. These abstractions unify design choices such as raw buffers, vector
stores, knowledge graphs, procedural memories, and tool-mediated evidence
retrieval into a vocabulary that is precise enough for evaluation while remaining
implementation-flexible.}

\rev{Building on this vocabulary, Section~\ref{sec:challenges} surfaced the
operational fault lines that accompany such flexibility. We organized the open
problems into five challenge families, communication complexity and protocol
design, scalability and cost, hallucination and error propagation, emergent
misalignment and collusion, and brand, policy, and legal compliance, and
connected each to recommender-specific manifestations and measurable signals
such as ranking quality, evidence coverage, provenance and trace validity,
latency, token cost, exposure disparity, policy-violation rate, and judge
reliability. The recurring message is that progress depends not only on larger
models but on the interaction rules, memory hierarchies, and incentive
structures that govern how agents are composed.}

\rev{Finally, the controlled study of Section~\ref{sec:empirical} shows why agentic
recommendation must be evaluated conditionally. More agents are not automatically
better: on representative next-item ranking samples the single-agent baseline is
the pareto efficient default, whereas decomposition and ensemble roles become useful for
high-diversity user histories, and closed-loop self-criticism can even degrade
ranking quality. This supports a practical design principle in which agentic
complexity is routed to the inputs where its marginal quality gain justifies its
added latency, cost, and governance risk, and an evaluation agenda in which that
trade-off, rather than top-$K$ accuracy alone, is what future work should
report.}

\bibliographystyle{ACM-Reference-Format}
\bibliography{refs}

@article{paranjape2023art,
  title={Art: Automatic multi-step reasoning and tool-use for large language models},
  author={Paranjape, Bhargavi and Lundberg, Scott and Singh, Sameer and Hajishirzi, Hannaneh and Zettlemoyer, Luke and Ribeiro, Marco Tulio},
  journal={arXiv preprint arXiv:2303.09014},
  year={2023}
}

@article{hao2023toolkengpt,
  title={Toolkengpt: Augmenting frozen language models with massive tools via tool embeddings},
  author={Hao, Shibo and Liu, Tianyang and Wang, Zhen and Hu, Zhiting},
  journal={Advances in neural information processing systems},
  volume={36},
  pages={45870--45894},
  year={2023}
}

@article{zhang2025survey,
  title={A Survey of Large Language Model Empowered Agents for Recommendation and Search: Towards Next-Generation Information Retrieval},
  author={Zhang, Yu and Qiao, Shutong and Zhang, Jiaqi and Lin, Tzu-Heng and Gao, Chen and Li, Yong},
  journal={arXiv preprint arXiv:2503.05659},
  year={2025}
}

@article{peng2025survey,
  title={A Survey on LLM-powered Agents for Recommender Systems},
  author={Peng, Qiyao and Liu, Hongtao and Huang, Hua and Yang, Qing and Shao, Minglai},
  journal={arXiv preprint arXiv:2502.10050},
  year={2025}
}

@article{li2024prompt,
  title={Prompt compression for large language models: A survey},
  author={Li, Zongqian and Liu, Yinhong and Su, Yixuan and Collier, Nigel},
  journal={arXiv preprint arXiv:2410.12388},
  year={2024}
}

@inproceedings{maharana2024evaluating,
  title={Evaluating Very Long-Term Conversational Memory of LLM Agents},
  author={Maharana, Adyasha and Lee, Dong-Ho and Tulyakov, Sergey and Bansal, Mohit and Barbieri, Francesco and Fang, Yuwei},
  booktitle={Proceedings of the 62nd Annual Meeting of the Association for Computational Linguistics (Volume 1: Long Papers)},
  pages={13851--13870},
  year={2024}
}

@article{schick2023toolformer,
  title={Toolformer: Language models can teach themselves to use tools},
  author={Schick, Timo and Dwivedi-Yu, Jane and Dess{\`\i}, Roberto and Raileanu, Roberta and Lomeli, Maria and Hambro, Eric and Zettlemoyer, Luke and Cancedda, Nicola and Scialom, Thomas},
  journal={Advances in Neural Information Processing Systems},
  volume={36},
  pages={68539--68551},
  year={2023}
}

@article{nayak2024long,
  title={Long-horizon planning for multi-agent robots in partially observable environments},
  author={Nayak, Sid and Morrison Orozco, Adelmo and Have, Marina and Zhang, Jackson and Thirumalai, Vittal and Chen, Darren and Kapoor, Aditya and Robinson, Eric and Gopalakrishnan, Karthik and Harrison, James and others},
  journal={Advances in Neural Information Processing Systems},
  volume={37},
  pages={67929--67967},
  year={2024}
}

@article{erdogan2025plan,
  title={Plan-and-act: Improving planning of agents for long-horizon tasks},
  author={Erdogan, Lutfi Eren and Lee, Nicholas and Kim, Sehoon and Moon, Suhong and Furuta, Hiroki and Anumanchipalli, Gopala and Keutzer, Kurt and Gholami, Amir},
  journal={arXiv preprint arXiv:2503.09572},
  year={2025}
}

@article{yue2023llamarec,
  title={Llamarec: Two-stage recommendation using large language models for ranking},
  author={Yue, Zhenrui and Rabhi, Sara and Moreira, Gabriel de Souza Pereira and Wang, Dong and Oldridge, Even},
  journal={arXiv preprint arXiv:2311.02089},
  year={2023}
}

@inproceedings{yao2023react,
  title={React: Synergizing reasoning and acting in language models},
  author={Yao, Shunyu and Zhao, Jeffrey and Yu, Dian and Du, Nan and Shafran, Izhak and Narasimhan, Karthik and Cao, Yuan},
  booktitle={International Conference on Learning Representations (ICLR)},
  year={2023}
}

@article{xu2025mem,
  title={A-mem: Agentic memory for llm agents},
  author={Xu, Wujiang and Mei, Kai and Gao, Hang and Tan, Juntao and Liang, Zujie and Zhang, Yongfeng},
  journal={arXiv preprint arXiv:2502.12110},
  year={2025}
}

@article{pink2025position,
  title={Position: Episodic Memory is the Missing Piece for Long-Term LLM Agents},
  author={Pink, Mathis and Wu, Qinyuan and Vo, Vy Ai and Turek, Javier and Mu, Jianing and Huth, Alexander and Toneva, Mariya},
  journal={arXiv preprint arXiv:2502.06975},
  year={2025}
}

@article{li2024vector,
  title={Vector Storage Based Long-term Memory Research on LLM},
  author={Li, Kun and Jing, Xin and Jing, Chengang},
  journal={International Journal of Advanced Network, Monitoring and Controls},
  year={2024}
}

@article{rasmussen2025zep,
  title={Zep: A Temporal Knowledge Graph Architecture for Agent Memory},
  author={Rasmussen, Preston and Paliychuk, Pavlo and Beauvais, Travis and Ryan, Jack and Chalef, Daniel},
  journal={arXiv preprint arXiv:2501.13956},
  year={2025}
}

@article{anokhin2024arigraph,
  title={Arigraph: Learning knowledge graph world models with episodic memory for llm agents},
  author={Anokhin, Petr and Semenov, Nikita and Sorokin, Artyom and Evseev, Dmitry and Kravchenko, Andrey and Burtsev, Mikhail and Burnaev, Evgeny},
  journal={arXiv preprint arXiv:2407.04363},
  year={2024}
}

@article{liu2025enhancing,
  title={Enhancing Cross-Domain Recommendations with Memory-Optimized LLM-Based User Agents},
  author={Liu, Jiahao and Gu, Shengkang and Li, Dongsheng and Zhang, Guangping and Han, Mingzhe and Gu, Hansu and Zhang, Peng and Lu, Tun and Shang, Li and Gu, Ning},
  journal={arXiv e-prints},
  pages={arXiv--2502},
  year={2025}
}

@inproceedings{xi2024memocrs,
  title={MemoCRS: Memory-enhanced Sequential Conversational Recommender Systems with Large Language Models},
  author={Xi, Yunjia and Liu, Weiwen and Lin, Jianghao and Chen, Bo and Tang, Ruiming and Zhang, Weinan and Yu, Yong},
  booktitle={Proceedings of the 33rd ACM International Conference on Information and Knowledge Management},
  pages={2585--2595},
  year={2024}
}

@article{sun2023adaplanner,
  title={Adaplanner: Adaptive planning from feedback with language models},
  author={Sun, Haotian and Zhuang, Yuchen and Kong, Lingkai and Dai, Bo and Zhang, Chao},
  journal={Advances in neural information processing systems},
  volume={36},
  pages={58202--58245},
  year={2023}
}

@article{ming2023hicrisp,
  title={Hicrisp: A hierarchical closed-loop robotic intelligent self-correction planner},
  author={Ming, Chenlin and Lin, Jiacheng and Fong, Pangkit and Wang, Han and Duan, Xiaoming and He, Jianping},
  journal={arXiv preprint arXiv:2309.12089},
  year={2023}
}

@inproceedings{wang2024recmind,
  title={RecMind: Large Language Model Powered Agent For Recommendation},
  author={Wang, Yancheng and Jiang, Ziyan and Chen, Zheng and Yang, Fan and Zhou, Yingxue and Cho, Eunah and Fan, Xing and Lu, Yanbin and Huang, Xiaojiang and Yang, Yingzhen},
  booktitle={Findings of the Association for Computational Linguistics: NAACL 2024},
  pages={4351--4364},
  year={2024}
}

@article{shen2023hugginggpt,
  title={Hugginggpt: Solving ai tasks with chatgpt and its friends in hugging face},
  author={Shen, Yongliang and Song, Kaitao and Tan, Xu and Li, Dongsheng and Lu, Weiming and Zhuang, Yueting},
  journal={Advances in Neural Information Processing Systems},
  volume={36},
  pages={38154--38180},
  year={2023}
}

@article{shan2025cognitive,
  title={Cognitive memory in large language models},
  author={Shan, Lianlei and Luo, Shixian and Zhu, Zezhou and Yuan, Yu and Wu, Yong},
  journal={arXiv preprint arXiv:2504.02441},
  year={2025}
}

@article{guo2023empowering,
  title={Empowering working memory for large language model agents},
  author={Guo, Jing and Li, Nan and Qi, Jianchuan and Yang, Hang and Li, Ruiqiao and Feng, Yuzhen and Zhang, Si and Xu, Ming},
  journal={arXiv preprint arXiv:2312.17259},
  year={2023}
}

@article{fountas2024human,
  title={Human-like episodic memory for infinite context llms},
  author={Fountas, Zafeirios and Benfeghoul, Martin A and Oomerjee, Adnan and Christopoulou, Fenia and Lampouras, Gerasimos and Bou-Ammar, Haitham and Wang, Jun},
  journal={arXiv preprint arXiv:2407.09450},
  year={2024}
}

@misc{langchain2024memory,
  author       = {LangChain},
  title        = {LangChain Memory Types — Conceptual Guide},
  year         = {2024},
  url          = {https://langchain-ai.github.io/langmem/concepts/conceptual_guide/#memory-types},
  note         = {Accessed: 2025-06-21}
}

@article{zeng2024structural,
  title={On the Structural Memory of LLM Agents},
  author={Zeng, Ruihong and Fang, Jinyuan and Liu, Siwei and Meng, Zaiqiao},
  journal={arXiv preprint arXiv:2412.15266},
  year={2024}
}

@inproceedings{wheeler2025procedural,
  title={Procedural memory is not all you need: Bridging cognitive gaps in llm-based agents},
  author={Wheeler, Schaun and Jeunen, Olivier},
  booktitle={Adjunct Proceedings of the 33rd ACM Conference on User Modeling, Adaptation and Personalization},
  pages={360--364},
  year={2025}
}

@article{sumers2023cognitive,
  title={Cognitive architectures for language agents},
  author={Sumers, Theodore and Yao, Shunyu and Narasimhan, Karthik and Griffiths, Thomas},
  journal={Transactions on Machine Learning Research},
  year={2023}
}

@article{laird1987soar,
  title={Soar: An architecture for general intelligence},
  author={Laird, John E and Newell, Allen and Rosenbloom, Paul S},
  journal={Artificial intelligence},
  volume={33},
  number={1},
  pages={1--64},
  year={1987},
  publisher={Elsevier}
}

@article{de2024language,
  title={Language hooks: a modular framework for augmenting LLM reasoning that decouples tool usage from the model and its prompt},
  author={de Mijolla, Damien and Yang, Wen and Duckett, Philippa and Frye, Christopher and Worrall, Mark},
  journal={arXiv preprint arXiv:2412.05967},
  year={2024}
}

@article{mu2025experepair,
  title={EXPEREPAIR: Dual-Memory Enhanced LLM-based Repository-Level Program Repair},
  author={Mu, Fangwen and Wang, Junjie and Shi, Lin and Wang, Song and Li, Shoubin and Wang, Qing},
  journal={arXiv preprint arXiv:2506.10484},
  year={2025}
}

@article{an2024does,
  title={Why Does the Effective Context Length of LLMs Fall Short?},
  author={An, Chenxin and Zhang, Jun and Zhong, Ming and Li, Lei and Gong, Shansan and Luo, Yao and Xu, Jingjing and Kong, Lingpeng},
  journal={arXiv preprint arXiv:2410.18745},
  year={2024}
}

@article{liu2024lost,
  title={Lost in the Middle: How Language Models Use Long Contexts},
  author={Liu, Nelson F and Lin, Kevin and Hewitt, John and Paranjape, Ashwin and Bevilacqua, Michele and Petroni, Fabio and Liang, Percy},
  journal={Transactions of the Association for Computational Linguistics},
  volume={11},
  pages={157--173},
  year={2024}
}

@misc{forouzandehmehr2025calrag,
  title        = {CAL‑RAG: Retrieval‑Augmented Multi‑Agent Generation for Content‑Aware Layout Design},
  author       = {Forouzandehmehr, Najmeh and Yousefi Maragheh, Reza and Kollipara, Sriram and Zhao, Kai and Biswas, Topojoy and Korpeoglu, Evren and Achan, Kannan},
  year         = {2025},
  archivePrefix= {arXiv},
  eprint       = {2506.21934},
  primaryClass = {cs.IR},
  url          = {https://arxiv.org/pdf/2506.21934}
}

@misc{yousefi2025arag,
  title        = {ARAG: Agentic Retrieval Augmented Generation for Personalized Recommendation},
  author       = {Yousefi Maragheh, Reza and Vadla, Pratheek and Gupta, Priyank and Zhao, Kai and Inan, Aysenur and Yao, Kehui and Xu, Jianpeng and Kanumala, Praveen and Cho, Jason and Kumar, Sushant},
  year         = {2025},
  archivePrefix= {arXiv},
  eprint       = {2506.21931},
  primaryClass = {cs.IR},
  url          = {https://arxiv.org/pdf/2506.21931}
}

@inproceedings{borgeaud2022improving,
  title={Improving language models by retrieving from trillions of tokens},
  author={Borgeaud, Sebastian and Mensch, Arthur and Hoffmann, Jordan and Cai, Trevor and Rutherford, Eliza and Millican, Katie and Van Den Driessche, George Bm and Lespiau, Jean-Baptiste and Damoc, Bogdan and Clark, Aidan and others},
  booktitle={International conference on machine learning},
  pages={2206--2240},
  year={2022},
  organization={PMLR}
}

@inproceedings{guu2020retrieval,
  title={Retrieval augmented language model pre-training},
  author={Guu, Kelvin and Lee, Kenton and Tung, Zora and Pasupat, Panupong and Chang, Mingwei},
  booktitle={International conference on machine learning},
  pages={3929--3938},
  year={2020},
  organization={PMLR}
}

@article{lu2025karma,
  title={KARMA: Leveraging Multi-Agent LLMs for Automated Knowledge Graph Enrichment},
  author={Lu, Yuxing and Wang, Jinzhuo},
  journal={arXiv preprint arXiv:2502.06472},
  year={2025}
}

@inproceedings{ye2023schema,
  title={Schema-adaptable Knowledge Graph Construction},
  author={Ye, Hongbin and Gui, Honghao and Xu, Xin and Chen, Xi and Chen, Huajun and Zhang, Ningyu},
  booktitle={Findings of the Association for Computational Linguistics: EMNLP 2023},
  pages={6408--6431},
  year={2023}
}

@inproceedings{ficek2024gpt,
  title={GPT vs RETRO: Exploring the Intersection of Retrieval and Parameter-Efficient Fine-Tuning},
  author={Ficek, Aleksander and Zeng, Jiaqi and Kuchaiev, Oleksii},
  booktitle={Proceedings of the 2024 Conference on Empirical Methods in Natural Language Processing},
  pages={19425--19432},
  year={2024}
}

@article{chen2025fundamental,
  title={Fundamental Safety-Capability Trade-offs in Fine-tuning Large Language Models},
  author={Chen, Pin-Yu and Shen, Han and Das, Payel and Chen, Tianyi},
  journal={arXiv preprint arXiv:2503.20807},
  year={2025}
}

@article{fonseca2025safeguarding,
  title={Safeguarding Large Language Models in Real-time with Tunable Safety-Performance Trade-offs},
  author={Fonseca, Joao and Bell, Andrew and Stoyanovich, Julia},
  journal={arXiv preprint arXiv:2501.02018},
  year={2025}
}

@article{xiong2025memory,
  title={How Memory Management Impacts LLM Agents: An Empirical Study of Experience-Following Behavior},
  author={Xiong, Zidi and Lin, Yuping and Xie, Wenya and He, Pengfei and Tang, Jiliang and Lakkaraju, Himabindu and Xiang, Zhen},
  journal={arXiv preprint arXiv:2505.16067},
  year={2025}
}

@book{martello1990knapsack,
  title={Knapsack problems: algorithms and computer implementations},
  author={Martello, Silvano and Toth, Paolo},
  year={1990},
  publisher={John Wiley \& Sons, Inc.}
}

@article{chen2025carts,
  title={CARTS: Collaborative Agents for Recommendation Textual Summarization},
  author={Chen, Jiao and Yao, Kehui and Maragheh, Reza Yousefi and Zhao, Kai and Xu, Jianpeng and Cho, Jason and Korpeoglu, Evren and Kumar, Sushant and Achan, Kannan},
  journal={arXiv preprint arXiv:2506.17765},
  year={2025}
}

@article{poslad2007specifying,
  title={Specifying protocols for multi-agent systems interaction},
  author={Poslad, Stefan},
  journal={ACM Transactions on Autonomous and Adaptive Systems (TAAS)},
  volume={2},
  number={4},
  pages={15--es},
  year={2007},
  publisher={ACM New York, NY, USA}
}

@article{zhu2025secure,
  title={Secure Consensus Control on Multi-Agent Systems Based on Improved PBFT and Raft Blockchain Consensus Algorithms},
  author={Zhu, Jing and Lu, Chengfang and Li, Juanjuan and Wang, Fei-Yue},
  journal={IEEE/CAA Journal of Automatica Sinica},
  volume={12},
  number={7},
  pages={1407--1417},
  year={2025},
  publisher={IEEE/CAA Journal of Automatica Sinica}
}

@article{feng2018scalable,
  title={Scalable dynamic multi-agent practical byzantine fault-tolerant consensus in permissioned blockchain},
  author={Feng, Libo and Zhang, Hui and Chen, Yong and Lou, Liqi},
  journal={Applied Sciences},
  volume={8},
  number={10},
  pages={1919},
  year={2018},
  publisher={MDPI}
}

@article{han2024llm,
  title={LLM multi-agent systems: Challenges and open problems},
  author={Han, Shanshan and Zhang, Qifan and Yao, Yuhang and Jin, Weizhao and Xu, Zhaozhuo},
  journal={arXiv preprint arXiv:2402.03578},
  year={2024}
}

@article{ugurbil2025fission,
  title={Fission: Distributed Privacy-Preserving Large Language Model Inference},
  author={Ugurbil, Mehmet and Mouris, Dimitris and Santos, Manuel B and Cabrero-Holgueras, Jos{\'e} and de Vega, Miguel and Sengupta, Shubho},
  journal={Cryptology ePrint Archive},
  year={2025}
}

@article{hazem2023distributed,
  title={A distributed real-time recommender system for big data streams},
  author={Hazem, Heidy and Awad, Ahmed and Yousef, Ahmed Hassan},
  journal={Ain Shams Engineering Journal},
  volume={14},
  number={8},
  pages={102026},
  year={2023},
  publisher={Elsevier}
}

@inproceedings{yang2018intermediate,
  title={Intermediate data caching optimization for multi-stage and parallel big data frameworks},
  author={Yang, Zhengyu and Jia, Danlin and Ioannidis, Stratis and Mi, Ningfang and Sheng, Bo},
  booktitle={2018 IEEE 11th International Conference on Cloud Computing (CLOUD)},
  pages={277--284},
  year={2018},
  organization={IEEE}
}

@article{cai2024distributed,
  title={Distributed Recommendation Systems: Survey and Research Directions},
  author={Cai, Qiqi and Cao, Jian and Xu, Guandong and Zhu, Nengjun},
  journal={ACM Transactions on Information Systems},
  volume={43},
  number={1},
  pages={1--38},
  year={2024},
  publisher={ACM New York, NY, USA}
}

@inproceedings{wang2024macrec,
  title={Macrec: A multi-agent collaboration framework for recommendation},
  author={Wang, Zhefan and Yu, Yuanqing and Zheng, Wendi and Ma, Weizhi and Zhang, Min},
  booktitle={Proceedings of the 47th International ACM SIGIR Conference on Research and Development in Information Retrieval},
  pages={2760--2764},
  year={2024}
}

@article{pan2024agentcoord,
  title={AgentCoord: Visually exploring coordination strategy for llm-based multi-agent collaboration},
  author={Pan, Bo and Lu, Jiaying and Wang, Ke and Zheng, Li and Wen, Zhen and Feng, Yingchaojie and Zhu, Minfeng and Chen, Wei},
  journal={arXiv preprint arXiv:2404.11943},
  year={2024}
}

@article{simhi2025trust,
  title={Trust Me, I'm Wrong: High-Certainty Hallucinations in LLMs},
  author={Simhi, Adi and Itzhak, Itay and Barez, Fazl and Stanovsky, Gabriel and Belinkov, Yonatan},
  journal={arXiv preprint arXiv:2502.12964},
  year={2025}
}

@article{liu2024exploring,
  title={Exploring prosocial irrationality for llm agents: A social cognition view},
  author={Liu, Xuan and Zhang, Jie and Shang, Haoyang and Guo, Song and Yang, Chengxu and Zhu, Quanyan},
  journal={arXiv preprint arXiv:2405.14744},
  year={2024}
}

@inproceedings{maragheh2023llm,
  title={LLM-TAKE: Theme-aware keyword extraction using large language models},
  author={Maragheh, Reza Yousefi and Fang, Chenhao and Irugu, Charan Chand and Parikh, Parth and Cho, Jason and Xu, Jianpeng and Sukumar, Saranyan and Patel, Malay and Korpeoglu, Evren and Kumar, Sushant and others},
  booktitle={2023 IEEE International Conference on Big Data (BigData)},
  pages={4318--4324},
  year={2023},
  organization={IEEE}
}

@article{manakul2023selfcheckgpt,
  title={Selfcheckgpt: Zero-resource black-box hallucination detection for generative large language models},
  author={Manakul, Potsawee and Liusie, Adian and Gales, Mark JF},
  journal={arXiv preprint arXiv:2303.08896},
  year={2023}
}

@inproceedings{lewis2017deal,
  title={Deal or No Deal? End-to-End Learning of Negotiation Dialogues},
  author={Lewis, Mike and Yarats, Denis and Dauphin, Yann and Parikh, Devi and Batra, Dhruv},
  booktitle={Proceedings of the 2017 Conference on Empirical Methods in Natural Language Processing},
  pages={2443--2453},
  year={2017}
}

@inproceedings{ahanger2022popularity,
  title={Popularity bias in recommender systems-a review},
  author={Ahanger, Abdul Basit and Aalam, Syed Wajid and Bhat, Muzafar Rasool and Assad, Assif},
  booktitle={International Conference on Emerging Technologies in Computer Engineering},
  pages={431--444},
  year={2022},
  organization={Springer}
}

@article{lichtenberg2024large,
  title={Large language models as recommender systems: A study of popularity bias},
  author={Lichtenberg, Jan Malte and Buchholz, Alexander and Schw{\"o}bel, Pola},
  journal={arXiv preprint arXiv:2406.01285},
  year={2024}
}

@article{sukiennik2025simulating,
  title={Simulating Filter Bubble on Short-video Recommender System with Large Language Model Agents},
  author={Sukiennik, Nicholas and Wang, Haoyu and Zeng, Zailin and Gao, Chen and Li, Yong},
  journal={arXiv preprint arXiv:2504.08742},
  year={2025}
}

@article{noordeh2020echo,
  title={Echo chambers in collaborative filtering based recommendation systems},
  author={Noordeh, Emil and Levin, Roman and Jiang, Ruochen and Shadmany, Harris},
  journal={arXiv preprint arXiv:2011.03890},
  year={2020}
}

@inproceedings{tong2023navigating,
  title={Navigating the feedback loop in recommender systems: Insights and strategies from industry practice},
  author={Tong, Ding and Qiao, Qifeng and Lee, Ting-Po and McInerney, James and Basilico, Justin},
  booktitle={Proceedings of the 17th ACM Conference on Recommender Systems},
  pages={1058--1061},
  year={2023}
}

@inproceedings{deldjoo2024review,
  title={A Review of Modern Recommender Systems using Generative Models (Gen-RecSys)},
  author={Deldjoo, Yashar and He, Zhankui and McAuley, Julian and Korikov, Anton and Sanner, Scott and Ramisa, Arnau and Vidal, Ren{\'e} and Sathiamoorthy, Maheswaran and Kasirzadeh, Atoosa and Milano, Silvia},
  booktitle={Proceedings of the 30th ACM SIGKDD Conference on Knowledge Discovery and Data Mining},
  pages={6448--6458},
  year={2024}
}

@article{deldjoo2024recommendation,
  title={Recommendation with Generative Models},
  author={Deldjoo, Yashar and He, Zhankui and McAuley, Julian and Korikov, Anton and Sanner, Scott and Ramisa, Arnau and Vidal, Rene and Sathiamoorthy, Maheswaran and Kasrizadeh, Atoosa and Milano, Silvia and others},
  journal={arXiv preprint arXiv:2409.15173},
  year={2024}
}

@misc{businessinsider2024characterai,
  author       = {Stokel-Walker, Chris},
  title        = {Google and Character.AI are being sued after chatbot allegedly told a 17-year-old to kill his parents},
  year         = {2024},
  howpublished = {\url{https://www.businessinsider.com/characterai-google-lawsuit-chatbot-teen-kill-parents-2024-12}},
  note         = {Accessed: 2025-06-29}
}

@misc{apnews2017amazonbundles,
  author       = {AP News Staff},
  title        = {Amazon removes “frequently bought together” items used to make explosives},
  year         = {2017},
  howpublished = {\url{https://apnews.com/article/604a73f6008846449c303ffb4b93e9d6}},
  note         = {Accessed: 2025-06-29}
}

@article{achintalwar2024alignment,
  title={Alignment studio: Aligning large language models to particular contextual regulations},
  author={Achintalwar, Swapnaja and Baldini, Ioana and Bouneffouf, Djallel and Byamugisha, Joan and Chang, Maria and Dognin, Pierre and Farchi, Eitan and Makondo, Ndivhuwo and Mojsilovi{\'c}, Aleksandra and Nagireddy, Manish and others},
  journal={IEEE Internet Computing},
  year={2024},
  publisher={IEEE}
}

@article{koren2009matrix,
  author  = {Yehuda Koren and Robert Bell and Chris Volinsky},
  title   = {Matrix Factorization Techniques for Recommender Systems},
  journal = {Computer},
  volume  = {42},
  number  = {8},
  pages   = {30--37},
  year    = {2009},
  doi     = {10.1109/MC.2009.263}
}

@inproceedings{rendle2009bpr,
  title={BPR: Bayesian personalized ranking from implicit feedback},
  author={Rendle, Steffen and Freudenthaler, Christoph and Gantner, Zeno and Schmidt-Thieme, Lars},
  booktitle={Proceedings of the Twenty-Fifth Conference on Uncertainty in Artificial Intelligence},
  pages={452--461},
  year={2009}
}

@inproceedings{hidasi2016session,
  author       = {Bal{\'{a}}zs Hidasi and
                  Alexandros Karatzoglou and
                  Linas Baltrunas and
                  Domonkos Tikk},
  editor       = {Yoshua Bengio and
                  Yann LeCun},
  title        = {Session-based Recommendations with Recurrent Neural Networks},
  booktitle    = {4th International Conference on Learning Representations, {ICLR} 2016,
                  San Juan, Puerto Rico, May 2-4, 2016, Conference Track Proceedings},
  year         = {2016},
  url          = {http://arxiv.org/abs/1511.06939},
  bibsource    = {dblp computer science bibliography, https://dblp.org}
}

@inproceedings{kang2018sasrec,
  title={Self-attentive sequential recommendation},
  author={Kang, Wang-Cheng and McAuley, Julian},
  booktitle={2018 IEEE international conference on data mining (ICDM)},
  pages={197--206},
  year={2018},
  organization={IEEE}
}

@inproceedings{christakopoulou2016towards,
  title={Towards conversational recommender systems},
  author={Christakopoulou, Konstantina and Radlinski, Filip and Hofmann, Katja},
  booktitle={Proceedings of the 22nd ACM SIGKDD international conference on knowledge discovery and data mining},
  pages={815--824},
  year={2016}
}

@article{gao2021conversational,
  title={Advances and challenges in conversational recommender systems: A survey},
  author={Gao, Chongming and Lei, Wenqiang and He, Xiangnan and De Rijke, Maarten and Chua, Tat-Seng},
  journal={AI open},
  volume={2},
  pages={100--126},
  year={2021},
  publisher={Elsevier}
}

@article{wu2024llmrec,
  title={A survey on large language models for recommendation},
  author={Wu, Likang and Zheng, Zhi and Qiu, Zhaopeng and Wang, Hao and Gu, Hongchao and Shen, Tingjia and Qin, Chuan and Zhu, Chen and Zhu, Hengshu and Liu, Qi and others},
  journal={World Wide Web},
  volume={27},
  number={5},
  pages={60},
  year={2024},
  publisher={Springer}
}

@article{zhao2024llmrec,
  title={Recommender systems in the era of large language models (llms)},
  author={Zhao, Zihuai and Fan, Wenqi and Li, Jiatong and Liu, Yunqing and Mei, Xiaowei and Wang, Yiqi and Wen, Zhen and Wang, Fei and Zhao, Xiangyu and Tang, Jiliang and others},
  journal={IEEE Transactions on Knowledge and Data Engineering},
  volume={36},
  number={11},
  pages={6889--6907},
  year={2024},
  publisher={IEEE}
}

@inproceedings{li2024generative,
  title={Large language models for generative recommendation: A survey and visionary discussions},
  author={Li, Lei and Zhang, Yongfeng and Liu, Dugang and Chen, Li},
  booktitle={Proceedings of the 2024 joint international conference on computational linguistics, language resources and evaluation (LREC-COLING 2024)},
  pages={10146--10159},
  year={2024}
}

@inproceedings{zhao2024toolrec,
  title={Let me do it for you: Towards llm empowered recommendation via tool learning},
  author={Zhao, Yuyue and Wu, Jiancan and Wang, Xiang and Tang, Wei and Wang, Dingxian and De Rijke, Maarten},
  booktitle={Proceedings of the 47th International ACM SIGIR Conference on Research and Development in Information Retrieval},
  pages={1796--1806},
  year={2024}
}

@article{huang2025interecagent,
  title={Recommender ai agent: Integrating large language models for interactive recommendations},
  author={Huang, Xu and Lian, Jianxun and Lei, Yuxuan and Yao, Jing and Lian, Defu and Xie, Xing},
  journal={ACM Transactions on Information Systems},
  volume={43},
  number={4},
  pages={1--33},
  year={2025},
  publisher={ACM New York, NY}
}

@article{wang2023recagent,
  title={Recagent: A novel simulation paradigm for recommender systems},
  author={Wang, Lei and Zhang, Jingsen and Chen, Xiaowen and Lin, Yankai and Song, Ruihua and Zhao, Wayne Xin and Wen, Ji-Rong},
  journal={arXiv preprint arXiv:2306.02552},
  year={2023}
}

@inproceedings{zhang2024agent4rec,
  title={On generative agents in recommendation},
  author={Zhang, An and Chen, Yuxin and Sheng, Leheng and Wang, Xiang and Chua, Tat-Seng},
  booktitle={Proceedings of the 47th international ACM SIGIR conference on research and development in Information Retrieval},
  pages={1807--1817},
  year={2024}
}

@inproceedings{chen2025recusersim,
  title={Recusersim: A realistic and diverse user simulator for evaluating conversational recommender systems},
  author={Chen, Luyu and Dai, Quanyu and Zhang, Zeyu and Feng, Xueyang and Zhang, Mingyu and Tang, Pengcheng and Chen, Xu and Zhu, Yue and Dong, Zhenhua},
  booktitle={Companion Proceedings of the ACM on Web Conference 2025},
  pages={133--142},
  year={2025}
}

@article{fang2024macrs,
  title={A multi-agent conversational recommender system},
  author={Fang, Jiabao and Gao, Shen and Ren, Pengjie and Chen, Xiuying and Verberne, Suzan and Ren, Zhaochun},
  journal={arXiv preprint arXiv:2402.01135},
  year={2024}
}

@article{zhang2024prospect,
  author       = {Jizhi Zhang and
                  Keqin Bao and
                  Wenjie Wang and
                  Yang Zhang and
                  Wentao Shi and
                  Wanhong Xu and
                  Fuli Feng and
                  Tat{-}Seng Chua},
  title        = {Prospect Personalized Recommendation on Large Language Model-based
                  Agent Platform},
  journal      = {CoRR},
  volume       = {abs/2402.18240},
  year         = {2024},
  url          = {https://doi.org/10.48550/arXiv.2402.18240},
  doi          = {10.48550/ARXIV.2402.18240},
  eprinttype   = {arXiv},
  eprint       = {2402.18240},
  bibsource    = {dblp computer science bibliography, https://dblp.org}
}

@article{banerjee2025collabrec,
  author       = {Ashmi Banerjee and
                  Fitri Nur Aisyah and
                  Adithi Satish and
                  Wolfgang W{\"{o}}rndl and
                  Yashar Deldjoo},
  title        = {Collab-REC: An LLM-based Agentic Framework for Balancing Recommendations
                  in Tourism},
  journal      = {CoRR},
  volume       = {abs/2508.15030},
  year         = {2025},
  url          = {https://doi.org/10.48550/arXiv.2508.15030},
  doi          = {10.48550/ARXIV.2508.15030},
  eprinttype   = {arXiv},
  eprint       = {2508.15030},
  bibsource    = {dblp computer science bibliography, https://dblp.org}
}

@article{hou2024bridging,
  author       = {Yupeng Hou and
                  Jiacheng Li and
                  Zhankui He and
                  An Yan and
                  Xiusi Chen and
                  Julian J. McAuley},
  title        = {Bridging Language and Items for Retrieval and Recommendation},
  journal      = {CoRR},
  volume       = {abs/2403.03952},
  year         = {2024},
  url          = {https://doi.org/10.48550/arXiv.2403.03952},
  doi          = {10.48550/ARXIV.2403.03952},
  eprinttype   = {arXiv},
  eprint       = {2403.03952},
  bibsource    = {dblp computer science bibliography, https://dblp.org}
}

@inproceedings{prasad2024adapt,
  author       = {Archiki Prasad and
                  Alexander Koller and
                  Mareike Hartmann and
                  Peter Clark and
                  Ashish Sabharwal and
                  Mohit Bansal and
                  Tushar Khot},
  title        = {ADaPT: As-Needed Decomposition and Planning with Language Models},
  booktitle    = {Findings of the Association for Computational Linguistics: {NAACL}
                  2024, Mexico City, Mexico, June 16-21, 2024},
  series       = {Findings of {ACL}},
  volume       = {{NAACL} 2024},
  pages        = {4226--4252},
  publisher    = {Association for Computational Linguistics},
  year         = {2024},
  url          = {https://doi.org/10.18653/v1/2024.findings-naacl.264},
  doi          = {10.18653/V1/2024.FINDINGS-NAACL.264},
  bibsource    = {dblp computer science bibliography, https://dblp.org}
}

@article{luo2025large,
  author       = {Junyu Luo and
                  Weizhi Zhang and
                  Ye Yuan and
                  Yusheng Zhao and
                  Junwei Yang and
                  Yiyang Gu and
                  Bohan Wu and
                  Binqi Chen and
                  Ziyue Qiao and
                  Qingqing Long and
                  Rongcheng Tu and
                  Xiao Luo and
                  Wei Ju and
                  Zhiping Xiao and
                  Yifan Wang and
                  Meng Xiao and
                  Chenwu Liu and
                  Jingyang Yuan and
                  Shichang Zhang and
                  Yiqiao Jin and
                  Fan Zhang and
                  Xian Wu and
                  Hanqing Zhao and
                  Dacheng Tao and
                  Philip S. Yu and
                  Ming Zhang},
  title        = {Large Language Model Agent: {A} Survey on Methodology, Applications
                  and Challenges},
  journal      = {CoRR},
  volume       = {abs/2503.21460},
  year         = {2025},
  url          = {https://doi.org/10.48550/arXiv.2503.21460},
  doi          = {10.48550/ARXIV.2503.21460},
  eprinttype   = {arXiv},
  eprint       = {2503.21460},
  bibsource    = {dblp computer science bibliography, https://dblp.org}
}

@article{li2024more,
  author       = {Junyou Li and
                  Qin Zhang and
                  Yangbin Yu and
                  Qiang Fu and
                  Deheng Ye},
  title        = {More Agents Is All You Need},
  journal      = {Trans. Mach. Learn. Res.},
  volume       = {2024},
  year         = {2024},
  url          = {https://openreview.net/forum?id=bgzUSZ8aeg},
  bibsource    = {dblp computer science bibliography, https://dblp.org}
}

@inproceedings{madaan2023self,
  author       = {Aman Madaan and
                  Niket Tandon and
                  Prakhar Gupta and
                  Skyler Hallinan and
                  Luyu Gao and
                  Sarah Wiegreffe and
                  Uri Alon and
                  Nouha Dziri and
                  Shrimai Prabhumoye and
                  Yiming Yang and
                  Shashank Gupta and
                  Bodhisattwa Prasad Majumder and
                  Katherine Hermann and
                  Sean Welleck and
                  Amir Yazdanbakhsh and
                  Peter Clark},
  title        = {Self-Refine: Iterative Refinement with Self-Feedback},
  booktitle    = {Advances in Neural Information Processing Systems 36: Annual Conference
                  on Neural Information Processing Systems 2023, NeurIPS 2023, New Orleans,
                  LA, USA, December 10 - 16, 2023},
  year         = {2023},
  url          = {http://papers.nips.cc/paper\_files/paper/2023/hash/91edff07232fb1b55a505a9e9f6c0ff3-Abstract-Conference.html},
  bibsource    = {dblp computer science bibliography, https://dblp.org}
}

@inproceedings{du2024improving,
  author       = {Yilun Du and
                  Shuang Li and
                  Antonio Torralba and
                  Joshua B. Tenenbaum and
                  Igor Mordatch},
  editor       = {Ruslan Salakhutdinov and
                  Zico Kolter and
                  Katherine A. Heller and
                  Adrian Weller and
                  Nuria Oliver and
                  Jonathan Scarlett and
                  Felix Berkenkamp},
  title        = {Improving Factuality and Reasoning in Language Models through Multiagent
                  Debate},
  booktitle    = {Forty-first International Conference on Machine Learning, {ICML} 2024,
                  Vienna, Austria, July 21-27, 2024},
  series       = {Proceedings of Machine Learning Research},
  volume       = {235},
  pages        = {11733--11763},
  publisher    = {{PMLR} / OpenReview.net},
  year         = {2024},
  url          = {https://proceedings.mlr.press/v235/du24e.html},
  bibsource    = {dblp computer science bibliography, https://dblp.org}
}

\newpage
\title{Appendix}
\appendix
\section{Per-Category Empirical Results}
\label{app:per-category}

\rev{Section~\ref{sec:empirical} reported ranking quality and cost pooled over the
four Amazon-2023 categories. This appendix disaggregates those pooled numbers into
the individual categories (\emph{Amazon Fashion}, \emph{Appliances},
\emph{Electronics}, and \emph{Toys and Games}; $n{=}100$ users per category and
cohort) for both the random cohort (Tables~\ref{tab:app-qual-random}
and~\ref{tab:app-cost-random}) and the high-diversity cohort
(Tables~\ref{tab:app-qual-highdiv} and~\ref{tab:app-cost-highdiv}). Workflow
identifiers follow Table~\ref{tab:emp-roster}; in every category block the best and
second-best values per column among the seven LLM pipelines are shown in
\textbf{bold} and \underline{underline}, and the random-ranking baseline is listed
once per table. Because each user has a single held-out positive,
$\mathrm{MAP}@k=\mathrm{MRR}@k$; we
therefore report MRR (equivalently MAP) and NDCG, and include the MAP columns for
completeness. The two headline findings of the main text, namely that single-shot
ranking is hard to beat on typical users (Table~\ref{tab:emp-random}) and that
decomposition and ensembling pay off on diverse users
(Table~\ref{tab:emp-highdiv}), both survive disaggregation, but the per-category
view shows that each effect is unevenly distributed across categories.}

\subsection{Ranking Quality by Category}
\label{app:per-category-quality}

\paragraph{Random cohort.}
\rev{Table~\ref{tab:app-qual-random} shows that the pooled near-tie between the
single shot LLM call and the best multi-agent pipeline conceals substantial category
heterogeneity. On \emph{Electronics} the single shot LLM call (\textsc{SA}) is strictly
best on all four reported cells, and the evaluator pipeline \textsc{RC} collapses to
an NDCG@10 of $0.6602$, about $8.6\%$ below \textsc{SA}'s $0.7226$; this is the
clearest single-category instance of the error-propagation effect of
Section~\ref{subsec:hallucination_error}. On the other three categories a
multi-agent pipeline does edge ahead of \textsc{SA}, but the gains are small:
\textsc{PPEns} leads on \emph{Toys and Games} (NDCG@10 $0.7491$ vs.\ $0.7313$,
about $+2.4\%$) at roughly six times the cost, \textsc{Deb} leads on \emph{Amazon
Fashion} ($0.7364$ vs.\ $0.7241$, about $+1.7\%$) at roughly four times the cost,
and the inexpensive \textsc{PR} leads on \emph{Appliances} ($0.7114$ vs.\ $0.7007$,
about $+1.5\%$) at only about $1.5\times$ the cost. Only the last of these is
plausibly cost-justified, and even it is a sub-$2\%$ gain. The evaluator pipeline
\textsc{RC} is the weakest LLM pipeline in every category, so the pooled conclusion
that \textsc{SA} is Pareto-efficient on typical users is driven substantially by
\emph{Electronics}, with the remaining categories contributing only small,
cost-inefficient multi-agent gains.}

\paragraph{High-diversity cohort.}
\rev{Table~\ref{tab:app-qual-highdiv} confirms that the multi-agent advantage on
diverse histories holds in every category, but with a magnitude that tracks how
much room the single shot LLM call leaves. The gains are largest on the categories where
\textsc{SA} struggles most: on \emph{Amazon Fashion}, \textsc{PEns} lifts NDCG@3
from $0.4323$ to $0.5064$ (about $+17\%$) and NDCG@10 from $0.5943$ to $0.6341$
(about $+6.7\%$); on \emph{Appliances}, \textsc{PPEns} lifts NDCG@3 from $0.4886$
to $0.5401$ (about $+10.5\%$) and \textsc{PEns} lifts NDCG@10 by about $3.9\%$. On
\emph{Electronics} the ensembling and decomposition pipelines lead more modestly
(\textsc{PEns} NDCG@10 $0.6389$ vs.\ $0.6266$, about $+2.0\%$). \emph{Toys and
Games} is the boundary case: it has the highest single-agent baseline of the four
categories, so there is little to recover, and the pure ensemble \textsc{PEns}
actually dips below \textsc{SA} at NDCG@10 ($0.6628$ vs.\ $0.6772$) while only
\textsc{Deb} edges ahead (about $+0.5\%$). This matches the paper's mechanism:
agentic restructuring helps when a single pass under-resolves the input, and a
high-diversity history in an otherwise easy category does not always under-resolve
it. \textsc{RC} is again the weakest pipeline throughout.}

\begin{table}[t]
\centering
\caption{\rev{Per-category ranking quality on the \emph{random} cohort
(\texttt{gpt-5-mini}, $n{=}100$ per category). Best and second-best per column
among the LLM pipelines are \textbf{bold} and \underline{underlined}; higher is
better. Because each user has one held-out positive, $\mathrm{MAP}@k=\mathrm{MRR}@k$,
so the MAP columns repeat the MRR columns and are included only for completeness.}}
\label{tab:app-qual-random}
\small
\begin{tabular}{@{}l cc cc cc@{}}
\toprule
 & \multicolumn{2}{c}{MRR} & \multicolumn{2}{c}{NDCG} & \multicolumn{2}{c}{MAP} \\
\cmidrule(lr){2-3}\cmidrule(lr){4-5}\cmidrule(lr){6-7}
Method & @3 & @10 & @3 & @10 & @3 & @10 \\
\midrule
\multicolumn{7}{@{}l}{\textit{Amazon Fashion} ($n{=}100$)} \\
\midrule
\textsc{SA}    & 0.5850 & 0.6380 & 0.6170 & 0.7241 & 0.5850 & 0.6380 \\
\textsc{RC}    & 0.5683 & 0.6224 & 0.5996 & 0.7118 & 0.5683 & 0.6224 \\
\textsc{PR}    & 0.5583 & 0.6156 & 0.5946 & 0.7077 & 0.5583 & 0.6156 \\
\textsc{PPR}   & 0.5883 & 0.6357 & 0.6273 & 0.7231 & 0.5883 & 0.6357 \\
\textsc{PEns}  & \underline{0.6017} & \underline{0.6467} & \underline{0.6396} & \underline{0.7313} & \underline{0.6017} & \underline{0.6467} \\
\textsc{PPEns} & 0.5900 & 0.6391 & 0.6259 & 0.7255 & 0.5900 & 0.6391 \\
\textsc{Deb}   & \textbf{0.6067} & \textbf{0.6538} & \textbf{0.6407} & \textbf{0.7364} & \textbf{0.6067} & \textbf{0.6538} \\
\midrule
\multicolumn{7}{@{}l}{\textit{Appliances} ($n{=}100$)} \\
\midrule
\textsc{SA}    & 0.5617 & 0.6074 & 0.6023 & 0.7007 & 0.5617 & 0.6074 \\
\textsc{RC}    & 0.5183 & 0.5739 & 0.5599 & 0.6753 & 0.5183 & 0.5739 \\
\textsc{PR}    & \textbf{0.5783} & \textbf{0.6204} & \textbf{0.6223} & \textbf{0.7114} & \textbf{0.5783} & \textbf{0.6204} \\
\textsc{PPR}   & 0.5467 & 0.5963 & 0.5883 & 0.6923 & 0.5467 & 0.5963 \\
\textsc{PEns}  & 0.5483 & 0.5971 & 0.5894 & 0.6926 & 0.5483 & 0.5971 \\
\textsc{PPEns} & \underline{0.5733} & \underline{0.6197} & \underline{0.6107} & \underline{0.7097} & \underline{0.5733} & \underline{0.6197} \\
\textsc{Deb}   & 0.5600 & 0.6085 & 0.5986 & 0.7016 & 0.5600 & 0.6085 \\
\midrule
\multicolumn{7}{@{}l}{\textit{Electronics} ($n{=}100$)} \\
\midrule
\textsc{SA}    & \textbf{0.5917} & \textbf{0.6355} & \textbf{0.6301} & \textbf{0.7226} & \textbf{0.5917} & \textbf{0.6355} \\
\textsc{RC}    & 0.4967 & 0.5535 & 0.5438 & 0.6602 & 0.4967 & 0.5535 \\
\textsc{PR}    & 0.5717 & 0.6166 & 0.6173 & 0.7087 & 0.5717 & 0.6166 \\
\textsc{PPR}   & 0.5633 & 0.6074 & 0.6114 & 0.7022 & 0.5633 & 0.6074 \\
\textsc{PEns}  & \underline{0.5800} & \underline{0.6295} & \underline{0.6186} & \underline{0.7183} & \underline{0.5800} & \underline{0.6295} \\
\textsc{PPEns} & 0.5483 & 0.5985 & 0.5927 & 0.6951 & 0.5483 & 0.5985 \\
\textsc{Deb}   & 0.5717 & 0.6179 & 0.6151 & 0.7099 & 0.5717 & 0.6179 \\
\midrule
\multicolumn{7}{@{}l}{\textit{Toys and Games} ($n{=}100$)} \\
\midrule
\textsc{SA}    & 0.6000 & 0.6459 & 0.6407 & 0.7313 & 0.6000 & 0.6459 \\
\textsc{RC}    & 0.5767 & 0.6340 & 0.6083 & 0.7216 & 0.5767 & 0.6340 \\
\textsc{PR}    & \underline{0.6017} & 0.6495 & \underline{0.6423} & 0.7345 & \underline{0.6017} & 0.6495 \\
\textsc{PPR}   & 0.5850 & 0.6347 & 0.6244 & 0.7225 & 0.5850 & 0.6347 \\
\textsc{PEns}  & 0.5867 & 0.6333 & 0.6283 & 0.7216 & 0.5867 & 0.6333 \\
\textsc{PPEns} & \textbf{0.6233} & \textbf{0.6704} & \textbf{0.6555} & \textbf{0.7491} & \textbf{0.6233} & \textbf{0.6704} \\
\textsc{Deb}   & \underline{0.6017} & \underline{0.6518} & 0.6370 & \underline{0.7356} & \underline{0.6017} & \underline{0.6518} \\
\midrule
Random baseline & 0.1833 & 0.2929 & 0.2131 & 0.4544 & 0.1833 & 0.2929 \\
\bottomrule
\end{tabular}
\end{table}

\begin{table}[t]
\centering
\caption{\rev{Per-category ranking quality on the \emph{high-diversity} cohort
(\texttt{gpt-5-mini}, $n{=}100$ per category). Best and second-best per column
among the LLM pipelines are \textbf{bold} and \underline{underlined}; higher is
better. The MAP columns coincide with the MRR columns (single positive).}}
\label{tab:app-qual-highdiv}
\small
\begin{tabular}{@{}l cc cc cc@{}}
\toprule
 & \multicolumn{2}{c}{MRR} & \multicolumn{2}{c}{NDCG} & \multicolumn{2}{c}{MAP} \\
\cmidrule(lr){2-3}\cmidrule(lr){4-5}\cmidrule(lr){6-7}
Method & @3 & @10 & @3 & @10 & @3 & @10 \\
\midrule
\multicolumn{7}{@{}l}{\textit{Amazon Fashion} ($n{=}100$)} \\
\midrule
\textsc{SA}    & 0.3917 & 0.4694 & 0.4323 & 0.5943 & 0.3917 & 0.4694 \\
\textsc{RC}    & \underline{0.4317} & 0.4998 & \underline{0.4799} & 0.6189 & \underline{0.4317} & 0.4998 \\
\textsc{PR}    & 0.4200 & \underline{0.5022} & 0.4533 & \underline{0.6196} & 0.4200 & \underline{0.5022} \\
\textsc{PPR}   & 0.4200 & 0.4948 & 0.4714 & 0.6166 & 0.4200 & 0.4948 \\
\textsc{PEns}  & \textbf{0.4567} & \textbf{0.5188} & \textbf{0.5064} & \textbf{0.6341} & \textbf{0.4567} & \textbf{0.5188} \\
\textsc{PPEns} & 0.4217 & 0.4917 & 0.4723 & 0.6132 & 0.4217 & 0.4917 \\
\textsc{Deb}   & 0.3883 & 0.4776 & 0.4196 & 0.6004 & 0.3883 & 0.4776 \\
\midrule
\multicolumn{7}{@{}l}{\textit{Appliances} ($n{=}100$)} \\
\midrule
\textsc{SA}    & 0.4467 & 0.5128 & 0.4886 & 0.6281 & 0.4467 & 0.5128 \\
\textsc{RC}    & 0.4283 & 0.4938 & 0.4775 & 0.6143 & 0.4283 & 0.4938 \\
\textsc{PR}    & 0.4483 & 0.5189 & 0.4873 & 0.6329 & 0.4483 & 0.5189 \\
\textsc{PPR}   & 0.4617 & 0.5229 & 0.5104 & 0.6373 & 0.4617 & 0.5229 \\
\textsc{PEns}  & \textbf{0.4883} & \textbf{0.5434} & \underline{0.5375} & \textbf{0.6527} & \textbf{0.4883} & \textbf{0.5434} \\
\textsc{PPEns} & \textbf{0.4883} & \underline{0.5417} & \textbf{0.5401} & \underline{0.6518} & \textbf{0.4883} & \underline{0.5417} \\
\textsc{Deb}   & \underline{0.4650} & 0.5251 & 0.5125 & 0.6383 & \underline{0.4650} & 0.5251 \\
\midrule
\multicolumn{7}{@{}l}{\textit{Electronics} ($n{=}100$)} \\
\midrule
\textsc{SA}    & 0.4450 & 0.5106 & 0.4899 & 0.6266 & 0.4450 & 0.5106 \\
\textsc{RC}    & 0.3900 & 0.4685 & 0.4259 & 0.5928 & 0.3900 & 0.4685 \\
\textsc{PR}    & 0.4433 & 0.5094 & 0.4886 & 0.6258 & 0.4433 & 0.5094 \\
\textsc{PPR}   & \underline{0.4550} & \underline{0.5216} & 0.4975 & \underline{0.6352} & \underline{0.4550} & \underline{0.5216} \\
\textsc{PEns}  & \textbf{0.4567} & \textbf{0.5259} & \underline{0.4988} & \textbf{0.6389} & \textbf{0.4567} & \textbf{0.5259} \\
\textsc{PPEns} & 0.4533 & 0.5113 & \textbf{0.5088} & 0.6285 & 0.4533 & 0.5113 \\
\textsc{Deb}   & 0.4367 & 0.5016 & 0.4883 & 0.6204 & 0.4367 & 0.5016 \\
\midrule
\multicolumn{7}{@{}l}{\textit{Toys and Games} ($n{=}100$)} \\
\midrule
\textsc{SA}    & 0.5133 & \underline{0.5769} & 0.5483 & \underline{0.6772} & 0.5133 & \underline{0.5769} \\
\textsc{RC}    & 0.5017 & 0.5715 & 0.5320 & 0.6727 & 0.5017 & 0.5715 \\
\textsc{PR}    & 0.5033 & 0.5668 & 0.5383 & 0.6691 & 0.5033 & 0.5668 \\
\textsc{PPR}   & 0.5100 & 0.5694 & 0.5483 & 0.6715 & 0.5100 & 0.5694 \\
\textsc{PEns}  & 0.4883 & 0.5582 & 0.5220 & 0.6628 & 0.4883 & 0.5582 \\
\textsc{PPEns} & \underline{0.5183} & 0.5714 & \textbf{0.5596} & 0.6729 & \underline{0.5183} & 0.5714 \\
\textsc{Deb}   & \textbf{0.5233} & \textbf{0.5819} & \underline{0.5581} & \textbf{0.6804} & \textbf{0.5233} & \textbf{0.5819} \\
\midrule
Random baseline & 0.1833 & 0.2929 & 0.2131 & 0.4544 & 0.1833 & 0.2929 \\
\bottomrule
\end{tabular}
\end{table}

\subsection{Cost and Efficiency by Category}
\label{app:per-category-cost}

\rev{Tables~\ref{tab:app-cost-random} and~\ref{tab:app-cost-highdiv} report input,
output, and total tokens per query, parallel critical-path latency, and dollar cost
per category. The ordering is identical in every category: the single shot LLM call is the
floor on every column, the profiler--ranker pipeline \textsc{PR} is second, and the
ensembles \textsc{PEns} and \textsc{PPEns} cost five to six times the single-agent
baseline. The absolute cost, however, tracks history length and therefore varies by
category: \emph{Electronics} is the most token-heavy (random \textsc{SA} totals
$2{,}050$ tokens and \textsc{PPEns} $12{,}191$), whereas \emph{Toys and Games} is
the lightest (high-diversity \textsc{SA} $1{,}497$). The categories on which
multi-agent pipelines help most are thus also among the most expensive to run them
on, which sharpens rather than softens the cost--benefit tension of
Section~\ref{sec:emp-cost}. The latency columns make the penalty on iterative
refinement concrete: because \textsc{RC} and \textsc{Deb} cannot overlap their
calls, they are the slowest pipelines in wall-clock terms even though they issue
fewer calls than the ensembles. On random \emph{Electronics}, for example,
\textsc{RC} ($72.1$\,s) and \textsc{Deb} ($69.3$\,s) both exceed \textsc{PEns}
($41.9$\,s), whose three rankers run concurrently, while \textsc{SA} stays the
latency floor at $14.8$\,s. This reinforces the role analysis of
Section~\ref{sec:emp-roles}: closed-loop critique is penalized on quality and on
latency at the same time.}

\begin{table}[t]
\centering
\caption{\rev{Per-category cost and efficiency on the \emph{random} cohort
(\texttt{gpt-5-mini}, $n{=}100$ per category). ``Total'' sums input and output
tokens; ``Lat.\ par'' is the parallel critical-path latency in seconds; cost is in
USD at \$0.25/\$2.00 per 1M input/output tokens. Lower is better on every column;
best and second-best per column are \textbf{bold} and \underline{underlined}.}}
\label{tab:app-cost-random}
\small
\begin{tabular}{@{}l r r r r r@{}}
\toprule
Method & In tok/q & Out tok/q & Total tok/q & Lat.\ par (s) & Cost/q (USD) \\
\midrule
\multicolumn{6}{@{}l}{\textit{Amazon Fashion} ($n{=}100$)} \\
\midrule
\textsc{SA}    & \textbf{612}   & \textbf{895}   & \textbf{1,507}  & \textbf{15.03}  & \textbf{0.00194} \\
\textsc{RC}    & 2,372 & 3,546 & 5,918  & 39.97 & 0.00768 \\
\textsc{PR}    & \underline{1,070} & \underline{1,406} & \underline{2,477} & \underline{17.50} & \underline{0.00308} \\
\textsc{PPR}   & 1,769 & 2,047 & 3,816  & 20.68 & 0.00454 \\
\textsc{PEns}  & 3,685 & 4,747 & 8,432  & 34.39 & 0.01041 \\
\textsc{PPEns} & 4,485 & 5,186 & 9,671  & 40.33 & 0.01149 \\
\textsc{Deb}   & 2,766 & 3,808 & 6,574  & 56.88 & 0.00831 \\
\midrule
\multicolumn{6}{@{}l}{\textit{Appliances} ($n{=}100$)} \\
\midrule
\textsc{SA}    & \textbf{769}   & \textbf{1,066}  & \textbf{1,835}  & \textbf{15.03}  & \textbf{0.00232} \\
\textsc{RC}    & 2,959 & 3,709 & 6,668  & 57.73 & 0.00816 \\
\textsc{PR}    & \underline{1,320} & \underline{1,484} & \underline{2,803} & \underline{23.16} & \underline{0.00330} \\
\textsc{PPR}   & 2,171 & 2,144 & 4,315  & 24.99 & 0.00483 \\
\textsc{PEns}  & 4,482 & 5,524 & 10,006 & 39.44 & 0.01217 \\
\textsc{PPEns} & 5,340 & 5,872 & 11,212 & 49.86 & 0.01308 \\
\textsc{Deb}   & 3,430 & 4,611 & 8,041  & 69.37 & 0.01008 \\
\midrule
\multicolumn{6}{@{}l}{\textit{Electronics} ($n{=}100$)} \\
\midrule
\textsc{SA}    & \textbf{972}   & \textbf{1,078}  & \textbf{2,050}  & \textbf{14.83}  & \textbf{0.00240} \\
\textsc{RC}    & 3,618 & 4,136 & 7,754  & 72.07 & 0.00918 \\
\textsc{PR}    & \underline{1,688} & \underline{1,624} & \underline{3,312} & 29.51 & \underline{0.00367} \\
\textsc{PPR}   & 2,754 & 2,212 & 4,966  & \underline{24.82} & 0.00511 \\
\textsc{PEns}  & 5,532 & 5,199 & 10,731 & 41.94 & 0.01178 \\
\textsc{PPEns} & 6,594 & 5,597 & 12,191 & 47.89 & 0.01284 \\
\textsc{Deb}   & 4,242 & 4,575 & 8,817  & 69.29 & 0.01021 \\
\midrule
\multicolumn{6}{@{}l}{\textit{Toys and Games} ($n{=}100$)} \\
\midrule
\textsc{SA}    & \textbf{704}   & \textbf{927}   & \textbf{1,632}  & \textbf{13.18}  & \textbf{0.00203} \\
\textsc{RC}    & 2,675 & 3,780 & 6,455  & 51.32 & 0.00823 \\
\textsc{PR}    & \underline{1,232} & \underline{1,391} & \underline{2,623} & \underline{19.08} & \underline{0.00309} \\
\textsc{PPR}   & 2,022 & 1,974 & 3,996  & 21.42 & 0.00445 \\
\textsc{PEns}  & 4,162 & 4,710 & 8,871  & 37.64 & 0.01046 \\
\textsc{PPEns} & 5,009 & 5,163 & 10,173 & 40.17 & 0.01158 \\
\textsc{Deb}   & 3,134 & 3,976 & 7,109  & 56.13 & 0.00873 \\
\bottomrule
\end{tabular}
\end{table}

\begin{table}[t]
\centering
\caption{\rev{Per-category cost and efficiency on the \emph{high-diversity} cohort
(\texttt{gpt-5-mini}, $n{=}100$ per category). Columns and conventions are as in
Table~\ref{tab:app-cost-random}; lower is better, best and second-best per column in
\textbf{bold} and \underline{underline}.}}
\label{tab:app-cost-highdiv}
\small
\begin{tabular}{@{}l r r r r r@{}}
\toprule
Method & In tok/q & Out tok/q & Total tok/q & Lat.\ par (s) & Cost/q (USD) \\
\midrule
\multicolumn{6}{@{}l}{\textit{Amazon Fashion} ($n{=}100$)} \\
\midrule
\textsc{SA}    & \textbf{572}   & \textbf{944}   & \textbf{1,516}  & \textbf{11.20}  & \textbf{0.00203} \\
\textsc{RC}    & 2,255 & 3,520 & 5,775  & 39.68 & 0.00760 \\
\textsc{PR}    & \underline{990}   & \underline{1,455} & \underline{2,445} & \underline{18.39} & \underline{0.00316} \\
\textsc{PPR}   & 1,646 & 2,107 & 3,753  & 22.27 & 0.00462 \\
\textsc{PEns}  & 3,468 & 5,055 & 8,523  & 37.09 & 0.01098 \\
\textsc{PPEns} & 4,232 & 5,264 & 9,496  & 35.76 & 0.01159 \\
\textsc{Deb}   & 2,606 & 4,009 & 6,615  & 48.44 & 0.00867 \\
\midrule
\multicolumn{6}{@{}l}{\textit{Appliances} ($n{=}100$)} \\
\midrule
\textsc{SA}    & \textbf{713}   & \textbf{1,143}  & \textbf{1,857}  & \textbf{12.36}  & \textbf{0.00246} \\
\textsc{RC}    & 2,762 & 3,683 & 6,444  & 42.26 & 0.00806 \\
\textsc{PR}    & \underline{1,188} & \underline{1,505} & \underline{2,693} & \underline{17.83} & \underline{0.00331} \\
\textsc{PPR}   & 1,978 & 2,253 & 4,231  & 21.95 & 0.00500 \\
\textsc{PEns}  & 4,149 & 5,711 & 9,860  & 35.23 & 0.01246 \\
\textsc{PPEns} & 4,964 & 6,000 & 10,964 & 35.62 & 0.01324 \\
\textsc{Deb}   & 3,170 & 4,614 & 7,784  & 51.00 & 0.01002 \\
\midrule
\multicolumn{6}{@{}l}{\textit{Electronics} ($n{=}100$)} \\
\midrule
\textsc{SA}    & \textbf{727}   & \textbf{1,018}  & \textbf{1,745}  & \textbf{11.56}  & \textbf{0.00222} \\
\textsc{RC}    & 2,872 & 4,146 & 7,018  & 48.25 & 0.00901 \\
\textsc{PR}    & \underline{1,163} & \underline{1,541} & \underline{2,704} & \underline{18.66} & \underline{0.00337} \\
\textsc{PPR}   & 1,975 & 2,130 & 4,105  & 20.75 & 0.00475 \\
\textsc{PEns}  & 4,246 & 4,989 & 9,235  & 34.13 & 0.01104 \\
\textsc{PPEns} & 5,020 & 5,422 & 10,442 & 36.00 & 0.01210 \\
\textsc{Deb}   & 3,227 & 4,444 & 7,670  & 50.31 & 0.00969 \\
\midrule
\multicolumn{6}{@{}l}{\textit{Toys and Games} ($n{=}100$)} \\
\midrule
\textsc{SA}    & \textbf{542}   & \textbf{955}   & \textbf{1,497}  & \textbf{11.93}  & \textbf{0.00205} \\
\textsc{RC}    & 2,194 & 3,454 & 5,648  & 37.37 & 0.00746 \\
\textsc{PR}    & \underline{906}   & \underline{1,401} & \underline{2,307} & \underline{15.93} & \underline{0.00303} \\
\textsc{PPR}   & 1,536 & 2,023 & 3,560  & 18.43 & 0.00443 \\
\textsc{PEns}  & 3,339 & 4,725 & 8,064  & 30.11 & 0.01028 \\
\textsc{PPEns} & 4,037 & 5,062 & 9,099  & 33.39 & 0.01113 \\
\textsc{Deb}   & 2,485 & 3,965 & 6,450  & 48.70 & 0.00855 \\
\bottomrule
\end{tabular}
\end{table}
\end{document}